\documentclass[prd,floatfix,onecolumn,amsmath,amssymb,11pt,nofootinbib]{revtex4} 
\usepackage{graphicx} 
\usepackage{epstopdf} 

\usepackage{hyperref} 

\begin{document}   
\newcommand{\figref}[1]{Fig.~\ref{#1}}
\newcommand{\MTTWO}{\ensuremath{m_{T2}}}
\newcommand{\MTGEN}{\ensuremath{M_{T\mathrm{Gen}}}}
\newcommand{\casetwo}{{\tt CASE 2}}
\newcommand{\casefour}{{\tt CASE 4}}
\newcommand{\casethrees}{{\tt CASE 3s}}
\newcommand{\casethreev}{{\tt CASE 3v}}
\newcommand{\casesixs}{{\tt CASE 6s}}
\newcommand{\casesixv}{{\tt CASE 6v}}
\newcommand{\comment}[1]{\textbf{ X {#1} X }}
\title{Weighing Wimps with Kinks at Colliders:\\Invisible Particle
  Mass Measurements from Endpoints}
\author{Alan J Barr}
\email{a.barr@physics.ox.ac.uk}
\affiliation{Department of Physics, Denys Wilkinson Building, Keble
Road, Oxford OX1 3RH, United Kingdom}
\author{Ben Gripaios}  
\email{ben.gripaios@epfl.ch} 
\affiliation{EPFL, BSP 218, 1015 Lausanne, Switzerland}
\affiliation{CERN, PH-TH, 1211 Geneva, Switzerland} 
\author{Christopher G Lester}  
\email{lester@hep.phy.cam.ac.uk} 
\affiliation{Department of Physics, Cavendish Laboratory, JJ Thomson
Avenue, Cambridge, CB3 0HE, United Kingdom}
\begin{abstract}
We consider the application of endpoint techniques to the problem of mass determination for new particles produced at a hadron collider,
where these particles decay to an invisible particle of unknown mass and one or more visible particles of known mass. 
We also consider decays of these types for pair-produced particles and in each case consider situations 
both with and without initial state radiation. 
We prove that, in most (but not all) cases, the endpoint of an appropriate transverse mass observable, considered as a
function of the unknown mass of the invisible particle, has a kink at the true value of the invisible particle mass. 
The co-ordinates of the kink yield the masses of the decaying particle and the invisible particle. 
We discuss the prospects for implementing this method at the LHC.
\end{abstract}   
\preprint{CAVENDISH-HEP-2007/13}
\maketitle 

\section{Introduction}
With the imminent start-up of the Large Hadron Collider (LHC), 
and the hope that this machine could produce the first man-made dark matter,
the question of how to measure the masses of heavy, invisible particles at hadron
colliders has never held more importance.

Cho {\em et al.}\ \cite{Cho:2007qv} have recently made
a remarkable claim regarding a method for measuring 
invisible particle masses. 
They studied identical pairs of three-body superpartner decays ($\tilde{g}\rightarrow q \bar{q} \tilde{\chi}_1^0$), 
with one of the daughter particles being invisible, in the special case 
where the sum of the transverse momenta of the parents vanishes.
They identified ``kinks'' (see, for example, \figref{fig:six-v}) in the graphs of suitably defined transverse-mass variables, 
considered as a function of the ({\em a priori} unknown) mass of the invisible daughter.
The importance of their claim is that it provides, in
principle, a method to determine both of the unknown masses in the
problem -- namely the mass, $m_0$, of the decaying parent particle and the mass,
$m_1$, of the invisible daughter. 
These masses are given simply by the co-ordinates,  $(m_0,m_1)$,
of the kink.

In \cite{Gripaios:2007is}, one of us gave a proof that a similar kink
occurs even in events of a much simpler type, 
that is, in events in which a single parent with a non-zero transverse momentum
distribution undergoes a two-daughter decay.
It was claimed in \cite{Gripaios:2007is} that such
kinks should be a generic phenomenon.

We need hardly stress how useful an  ``transverse mass kink method'' for
determining sparticle masses would be, if experimentally viable. In
particular, if the invisible particle were indeed the dark matter
particle, then knowledge of its mass would be a boon for the
astrophysics and cosmology communities. What is more, knowledge of the
spectrum of, say, superpartner masses, would be of much help to those
attempting to discover the method by which supersymmetry breaking is
mediated to the Standard Model.

When the results of \cite{Cho:2007qv} and \cite{Gripaios:2007is} first
became public, it was not yet clear that 
there was a difference in the 
manner in which the kinks arise.  At first sight, both
papers reported kinks in related transverse variables.
But, as we go on to show in this paper, 
the sources of the kinks in  \cite{Cho:2007qv} and \cite{Gripaios:2007is} are not the same.
The results have no {\em direct} relevance to each other, although each paper
separately provided very important results.

One of the key purposes of this paper is to explain the connections
between the results of \cite{Cho:2007qv} and \cite{Gripaios:2007is}
and the relevance for the LHC.
While Cho {\em et al.}\ dealt with the class of events where the sum of the two parents' transverse momenta is strictly vanishing, 
\cite{Gripaios:2007is} dealt with the class of events in which the transverse momenta could be arbitrarily large.
Neither of these situations is a particularly good approximation at a hadron collider, 
since a realistic sample of events will be somewhere between the two 
-- they will have a distribution of possibly small, but certainly non-zero, transverse momenta. 
We address this more physically relevant case and find that a reasonably sized sample of realistic events will still produce a kink at $(m_0,m_1)$, 
from which, in most cases, we can hope to extract the masses.

We discuss the implications these results have for experimental observability
and explain how the events containing (pairs of) three-body decays considered by
\cite{Cho:2007qv} were able to generate kinks even without approaching the 
asymptotic extrema examined in \cite{Gripaios:2007is}.
We comment on how the kinks seen in \cite{Cho:2007qv} would be modified by a (physically reasonable)
admixture of events in which the parents have significant transverse recoil. 
We also point out that if nature were to provide us with
enough events far above threshold, then pairs of two-daughter decays
would generate a kink, even though the mechanism by which the kink is generated is
independent of the arguments in \cite{Cho:2007qv}.

In doing so we extend the analyses of both \cite{Cho:2007qv} and \cite{Gripaios:2007is}.
We analyze the general case of $N+1$-daughter decays,
including cascade decays. We also examine cases with 
identical pairs of such decays/cascades.
We consider events both with and without recoil
of the parent particle(s) against a secondary system of significant
transverse momentum (respectively SPT and ZPT). 
We also perform Monte Carlo simulations of a variety of different examples.
Since the list of cases studied is not short, the reader may find
it useful to refer to \figref{fig:general_idea} and \figref{fig:cases}, 
which show and label various scenarios diagrammatically.

The layout of the paper is as follows. 
In Section \ref{end-point} we (re)introduce endpoint analyses as methods of measuring masses. 
We discuss the extensions of the transverse mass variable, $m_T$, to cases where
parent particles are pair-produced, and where the mass of the invisible daughters is unknown.
In Sections \ref{single} and \ref{boost}, we consider single parent decays (both SPT and ZPT) and determine
the complete set of extrema of an appropriately-defined $m_T$. We analyze the nature of the various extrema, identify the global maximum, and examine the fraction of events which might be expected to lie near that maximum in each case.
In Section \ref{pair}, we consider events containing pairs of
identical decays of the type discussed in Section \ref{single}. 
In Section \ref{mc}, we present the results of some Monte Carlo simulations which illustrate these results. 
We conclude in Section \ref{sec:conclusions} with a summary of our findings, 
a comment on the relevance of SPT versus ZPT for the LHC,
and a discussion on the measurability of such edges in LHC-like scenarios. 

\section{Endpoint techniques}\label{end-point}
The determination of mass in the complex environment of hadron collisions is not usually
easy. For example, backgrounds from known physics are rather large,
misidentification of particles and jets is a problem, and we have very
little direct information concerning the longitudinal velocity of the
centre-of-mass frame of the two primary interacting partons.

These difficulties are compounded if the new particles are such that they decay with significant branching fraction to particles that interact sufficiently weakly as to be invisible as far as the detector is concerned. In any one event, these invisible particles carry off kinematic information, namely their momenta, and also their energies, if their masses are unknown.

The loss of these particles means that we cannot always measure particle masses on an event-by-event basis. But since we are able, in principle, to observe multiple events, the situation is not beyond hope. The historical example of the discovery, and subsequent mass determination, of the $W$-boson, is rather instructive in this respect.

The $W$-boson was first discovered in the UA1 and UA2 experiments
\cite{Arnison:1983rp,Arnison:1983zy,Banner:1983jy}, through its
sizable leptonic decay to an electron and a neutrino. The latter is, of course, invisible in the detector, and kinematic information thus is lost in each decay. Nevertheless, the transverse mass, defined by
\begin{gather}
m_T^2 = m_e^2 + m_{\nu}^2 + 2(e_e e_{\nu} - {\mathbf p}_e.{\mathbf p}_{\nu}),
\end{gather}
where $e_e = \sqrt{m_e^2 + p_e^2}$, is observable. The electron mass, $m_e$ is known {\em a priori}, and  the neutrino mass, $m_{\nu}$, is negligible.
The transverse momentum of the electron, ${\mathbf p}_e$ (and its magnitude, $p_e$),
can be measured directly, and the transverse momentum of the neutrino,
${\mathbf p}_{\nu}$, can be inferred from the missing transverse momentum
in the event. Now, for a single event, $m_T$ is not of great interest;
its importance lies in the fact that, in the limit of a narrow width for the $W$-boson, 
$m_T$ is bounded above by the
mass of the $W$, $m_W$, and is equal to
$m_W$ in events where the electron and the neutrino have the same
rapidity.  To see this, note that the four-momenta, $(E,\mathbf{p},q)$, obey the constraint
\begin{gather}
m_W^2 = m_e^2  + m_{\nu}^2 + 2(E_e E_{\nu} - {\mathbf p}_e.{\mathbf p}_{\nu}-q_e q_{\nu}).
\end{gather}
Now, the rapidity is given by
\begin{gather}
\eta = \frac{1}{2} \log \frac{E+q}{E-q},
\end{gather}
and the vanishing of the relative rapidity $\eta_e - \eta_{\nu}$ implies 
\begin{gather} \label{w}
E_e q_{\nu} = E_{\nu} q_{e},
\end{gather}
whence
\begin{gather}
E_e E_{\nu} - q_e q_{\nu}= e_e e_{\nu},
\end{gather}
and
\begin{gather}
m_W = m_T.
\end{gather}
Thus, by computing $m_T$ for many events, one can obtain a
distribution of $m_T$ values whose upper endpoint is close to the true
value $m_W$.
Indeed, fitting the $m_T$ distribution using data from CDF provides the most precise single direct measurement of the $W$ mass \cite{Aaltonen:2007ps}.

Similar and more general endpoint techniques are expected to be
useful at the LHC. Extremal values of the kinematic observables
typically contain information about the particle masses, as these are
the parameters which determine shape of the boundary of multi-particle
phase space. 

Although endpoint techniques suffer from the disadvantage that,
ultimately, only a subset of events (those that are near extremal) are
used to generate the mass data, they have a number of obvious
advantages. To begin with, the procedure is straightforward in its
conception and implementation. Secondly, and perhaps most importantly,
the procedure has little or no model-dependence. Were it to be
established that a relatively pure sample of events from a particular
decay channel could be isolated, then the subsequent analysis is pure
kinematics. No assumption about spins, couplings or matrix elements is
required. Finally, we remark that another supposed benefit of
endpoint techniques is that they place less stringent demands on the
degree to which the detector acceptance and backgrounds must be
understood: the location of the step within a Heaviside-like function
may be determined even if the height of the step function has been
modulated by a smoothly varying unknown acceptance and added to a
smoothly varying background of unknown shape.  In a typical
experiment, provided one ends up with a reasonable number of events
that are extremal or nearly so, one should expect to be able to
measure endpoints.

\begin{figure}[htp]
\begin{center}
\includegraphics[width=0.7\linewidth]{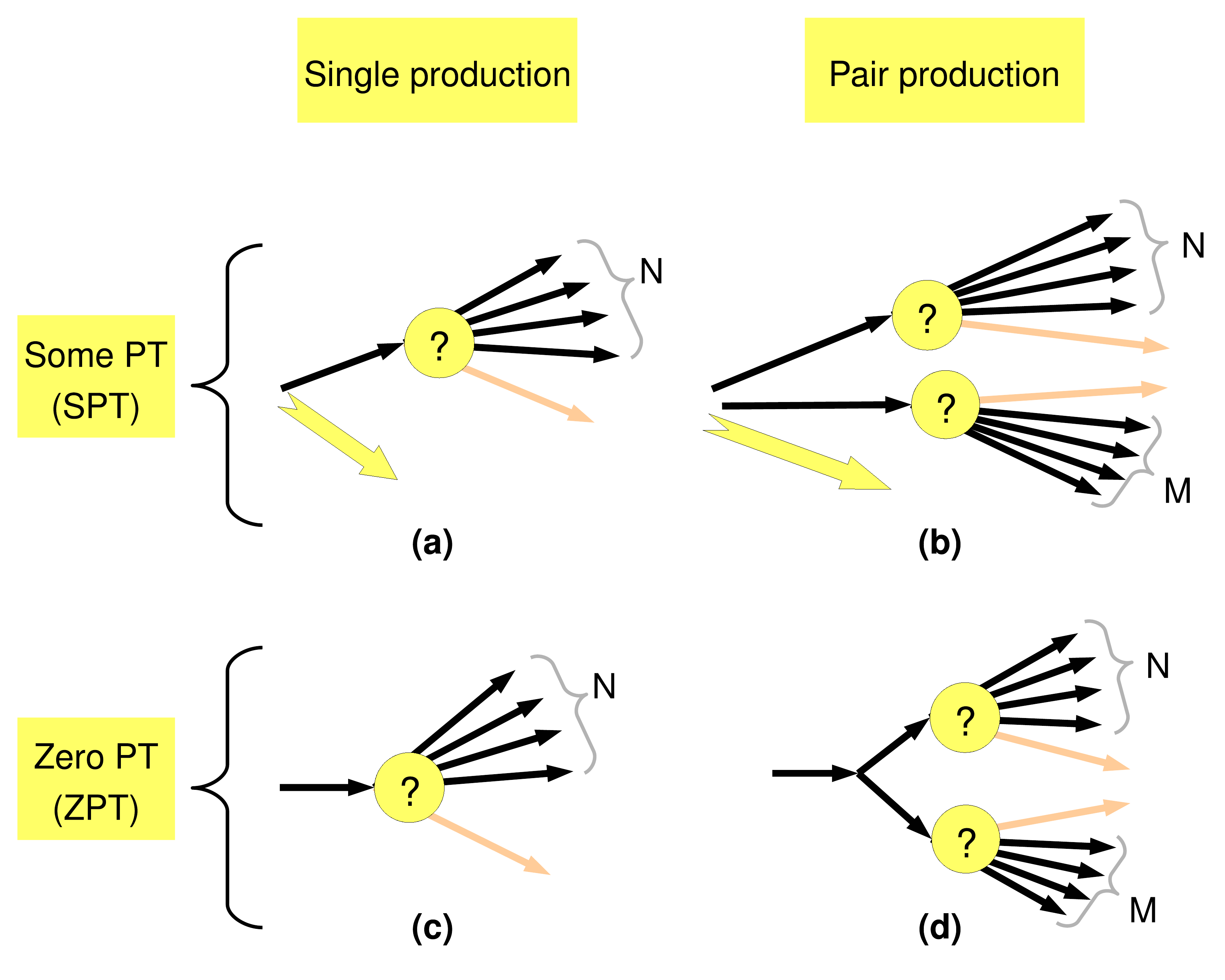}\\
\end{center}
\caption{An illustration of the types of decay considered in this
paper. We consider both single- and pair-production of a parent particle
which can decay into one invisible daughter of unknown mass (represented by the
lightly shaded arrows) and an arbitrary number of visible daughters.
In the analysis and conclusions, we will find it necessary to draw a
distinction between (a,b) events in which the parent(s)
recoils against a system (large yellow arrow) with {\em significant} transverse
momentum (SPT), and (c,d) events in which the parent(s)
recoils against nothing and has {\em zero} transverse momentum (ZPT).
The circular blobs serve to disguise the nature of the mechanism(s) by
which each decay takes place, as our analysis is general in this
regard.  See \figref{fig:cases} for some special cases that are
examined in more detail in the text.
\label{fig:general_idea}
}
\end{figure}
We note that this application of endpoint techniques to mass
determination at the LHC is not new.  For example, the endpoint of the
dilepton invariant mass distribution has long been suggested as a good
observable for constraining the mass-squared differences in
supersymmetric cascade decays \cite{Paige:1996nx}. Here, by contrast, 
we shall
attempt to use the procedure to provide a direct measurement of the
absolute mass scale, rather than differences in masses
or differences in squared-masses. We do this by generalizing the method described above for the $W$-boson.

There are two obvious ways in which we would like to generalize the method used to determine the $W$ mass. Both are motivated, in part, by our theoretical prejudice
as to what we expect, or hope, to discover at the LHC, and both
encounter an immediate obstruction. The types of events we will be
considering are summarized in \figref{fig:general_idea}.

Firstly, we should like to generalize to cases where the invisible
daughter has unknown, but non-negligible mass (\figref{fig:general_idea}(a,c)). One theoretical prejudice for this is that
any viable dark matter candidate ought to be electrically-neutral, colour-singlet and long-lived, hence invisible in the detector. Decays to this particle would be exactly of the type discussed above, except that the dark matter candidate, by definition, has non-negligible mass (perhaps of the order of the Fermi scale or greater, if the dark matter is both thermal and weakly-interacting). In such a case, even more kinematic information is lost, because both the momentum and the mass (and, {\em ergo}, the energy) of the invisible particle are unknown. Hence, $m_T$ is no longer an observable.

Secondly, we should like to generalize to cases where the parent
particles are pair-produced (\figref{fig:general_idea}(b,d)). This is the expectation in, for example,
supersymmetric theories with conserved $R$-parity, or in
little Higgs models with $T$-parity. In such
cases, each of the pair-produced parents of mass $m_0$ can produce
an invisible daughter of unknown mass $m_1$ among its decay
products.  Should this happen, the transverse momenta of the invisible
particles can no longer be individually inferred from the
missing transverse momentum; only the sum of the transverse momenta is now
constrained.  It is, consequently, not possible to observe the
transverse mass of either pair-produced particle.

The generalization of $m_T$ to events with pair-production was first
made in \cite{Lester:1999tx,Barr:2003rg}, where an algorithm was given
for constructing an observable, called $\MTTWO(\chi)$, which is a
function of the transverse momenta of the observed particles and the
missing transverse momentum, and whose definition is based upon the
individual (and unobservable) transverse masses.  (An
entirely different way of considering events of the same type was
suggested in \cite{Cheng:2007xv}.)  Note that, in the
construction of the $\MTTWO(\chi)$ variable, it was necessary to introduce a
parameter $\chi$ which represents a hypothesis for the true mass $m_1$
of the invisible particles generated in the decay of each primary
parton.  In that sense, $\MTTWO(\chi)$ should strictly not be called
an event variable, but rather an ``event function''.  A given event
generates not just one value of $\MTTWO$.  Instead it generates a
distribution of $\MTTWO$ values, one for each value $\chi$ taken as a
hypothesis for the mass of the invisible final-state particles.  When
the hypothesis $\chi$ is chosen to be equal to the true mass $m_1$ of
the invisible particles, the observable $\MTTWO(\chi\mid\chi=m_1)$
shares two properties enjoyed by $m_T$, namely (1) that it is bounded
above by the mass $m_0$ of the decaying particle, and (2) that this
bound may be reached in events in which certain constraints are met,
such as the rapidity of each invisible
particle matching the rapidity of the sum of its sister decay products.
The upper endpoint over events (indicated by a circumflex) of $\MTTWO$, ${\hat
m}_{T2}(\chi\mid{\chi=m_1})$, may thus be used as an estimator for $m_0$.

However, without knowledge of $m_1$, what may be said about the
dependence of $\MTTWO(\chi)$ on the mass hypothesis $\chi$?  Without
knowledge of $m_1$, is one forced to treat ${\hat m}_{T2}(\chi)$ as
providing a one-dimensional constraint in the $(m_0, m_1)$-plane
parameterized by $\chi$.  Is it the case that the ${\hat m}_{T2}$
variable constrains not $m_0$ and $m_1$ but only a relationship
between them?  

Cho {\em et al.} \cite{Cho:2007qv} considered ``\casesixv\ ZPT'' and ``\casesixs\ ZPT'', 
as defined in \figref{fig:cases}. 
Their claim, based on simulations and examples, but made without proof, is
that the ${\hat m}_{T2}(\chi)$ observable, considered as a function of
the hypothetical mass of the invisible particle $\chi$, has a kink (that
is, is continuous, but not differentiable) precisely at the point
where the hypothetical mass equals the true mass. 
In \cite{Gripaios:2007is}, one of us examined ``\casetwo\ SPT'' analytically
and proved the existence of a kink at $\chi=m_1$.
In Section \ref{analysis}, we analyze all the cases of \figref{fig:general_idea} and \figref{fig:cases}, 
including, but not restricted to, those of \cite{Cho:2007qv} and \cite{Gripaios:2007is}.
We go on to produce example Monte Carlo distributions in Section \ref{mc},
including examples of  \casesixv\ ZPT and \casesixs\ ZPT 
(\figref{fig:six-v}, \ref{fig:six-s})
and \casetwo\ SPT (\figref{fig:two}b) which illustrate the sorts of kinks
found in \cite{Cho:2007qv} and  \cite{Gripaios:2007is} respectively.

\section{Analysis}\label{analysis}

\begin{figure}[tbhp]
\begin{center}
\includegraphics[width=0.6\linewidth]{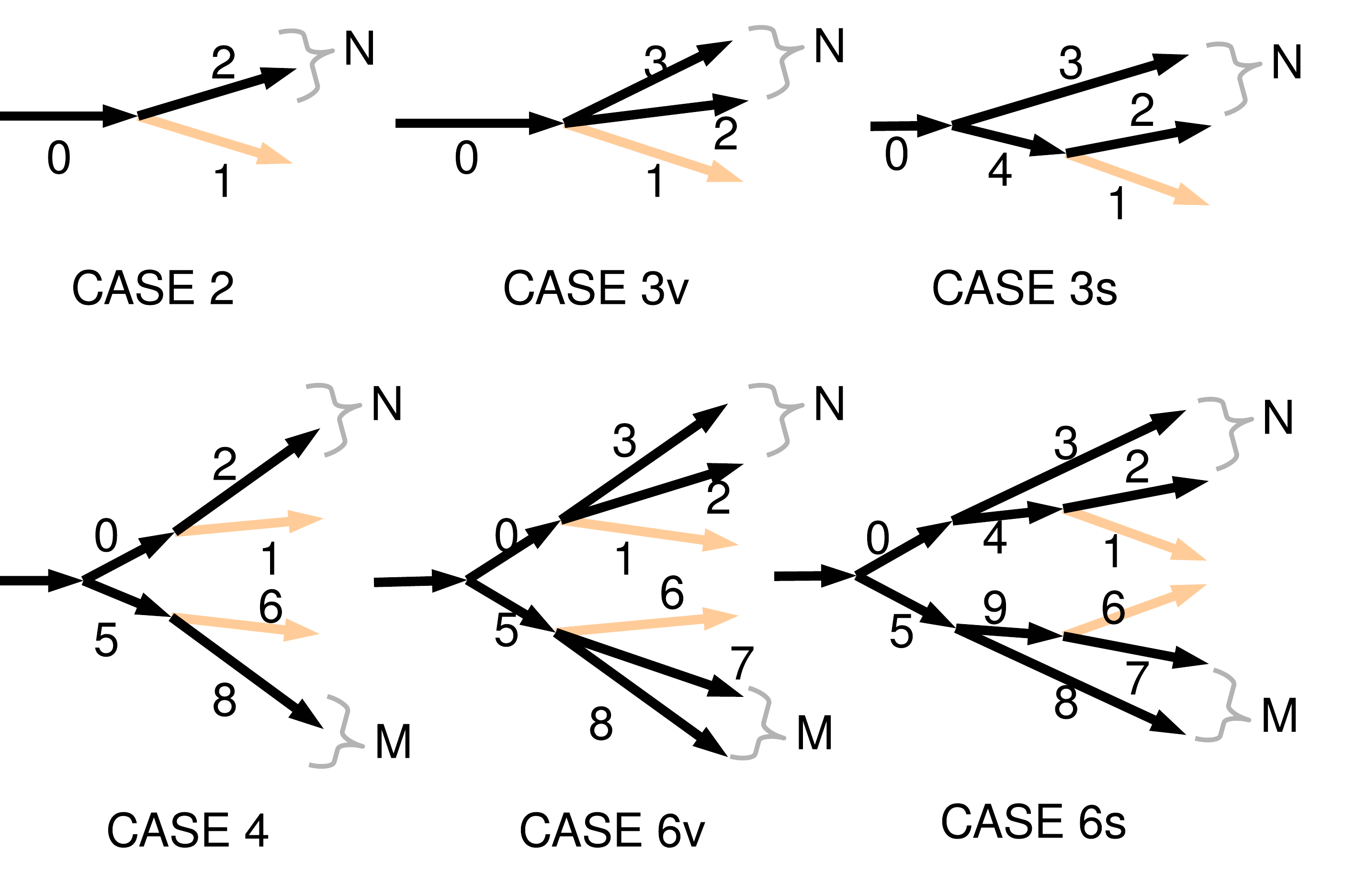}\\
\end{center}
\caption{This figure shows six particular forms of event which are
  used to illustrate points made throughout the analysis and in the
  conclusions. They are all special cases of the generic event-types
  shown in \figref{fig:general_idea}.  Each case will come in both SPT and
  ZPT forms (i.e.\ with and without net transverse momentum as defined
  in \figref{fig:general_idea}) although
  for clarity we omit the recoiling system from the diagram above.
  This figure also defines the numbering system used to denote
  particles in these special cases.  The invisible final state
  particles are represented by a lighter colour, and are numbered
  either 1 or 6.
\label{fig:cases}
}
\end{figure}

Our notation for a single particle decay is as shown in \figref{fig:cases}. The decaying parent is particle 0, the invisible daughter is particle 1,
and the system of visible daughters is labelled $N$.

We denote the energy of particle $i$ by $E_i$, its transverse momentum
by ${\mathbf p}_i$, the magnitude of its transverse momentum by $p_i$, its longitudinal momentum by $q_i$, and its transverse energy by $e_i$. The mass of particle $i$ is $m_i$
and the transverse energy is defined in terms of the transverse momentum by $e_i = \sqrt{p_i^2 + m_i^2}$.
For the case of the invisible particle 1, we will sometimes need to
use the hypothesized mass $\chi$ and the corresponding
transverse energy, $\tilde{e}_1 = \sqrt{p_1^2 + \chi^2}$. The antisymmetrized product is denoted by $E_{[1}q_{2]} \equiv E_1 q_2 - E_2 q_1$.
Finally, we put a circumflex over a function which is evaluated at an extremum. 

\subsection{Single Particle Production (\casetwo, \casethreev,
  \casethrees, etc.)} \label{single}
Consider the decay of particle, $0$, to an invisible particle, $1$, and an $N$-particle system of visible particles, $i$ with individual mass $m_i$ and invariant mass $m_N$. The number of transverse dimensions is arbitrary.
We first assert that $m_N \in [m_<,m_>]$, where, for a true ($N+1$)-daughter point decay,
\begin{align}
m_< &= \Sigma_i m_i, \\
m_> &= m_0 - m_1.
\end{align}
For decays of cascade type, which proceed through intermediate on-shell states, $m_<$ and $m_>$ will be larger and smaller, respectively. For a two-daughter decay, $m_< = m_> = m_2$, whereas
for a three-body cascade decay to massless visible daughters, occurring via an on-shell state of mass $M$, 
\begin{align}
m_< &= 0, \\
m_> &= \sqrt{\frac{(m_0^2 -M^2)(M^2-m_1^2)}{M^2}}.
\end{align}
For further details, see {\em e.g.}\ \cite{Paige:1996nx,Lester:2001zx}.

The problem is to extremize the function
\begin{gather} \label{f2}
f (\chi^2)= \chi^2 + m_N^2 +2 (\tilde{e}_1 e_N -  {\mathbf{p}}_1 \cdot {\mathbf{p}}_N) ,
\end{gather}
which is the square of the transverse mass one would calculate when assuming  
a hypothetical mass, $\chi$, for the invisible particle.\footnote{We could also choose to extremize the same function, but with the invariant mass $m_N$ replaced by the {\em transverse} invariant mass of the visible system. But the transverse invariant mass is also valued in $[m_<,m_>]$, and the extremal values of $f$ defined in this way are the same. Moreover, there are two obvious disadvantages in using the transverse invariant mass. The first is that it only attains its boundary values, $m_{\lessgtr}$, when the relative rapidities of visible particles vanish. There are thus fewer events near the extrema for a given sample size. The second is that it does not use available information, namely the longitudinal momenta of the visible particles.}  
The extremization is
subject to various constraints, namely conservation of energy-momentum, and the mass-shell conditions. We consider two distinct situations. In the former, the decaying particle can have arbitrary transverse momentum with respect to the laboratory frame (SPT). In the latter, the decaying particle has zero transverse momentum in the lab frame (ZPT).

\subsubsection{Single-particle SPT production (\casetwo\
  SPT, \casethreev\ SPT,  \casethrees\ SPT, etc.)} \label{spt}
In the SPT case, a minimal set of constraints can be written as
\begin{align} \label{con}
m_0^2 &= m_1^2 + m_N^2 + 2(E_1 E_N -  {\mathbf{p}}_1 \cdot {\mathbf{p}}_N - q_1 q_N), \\
m_1^2 &= E_1^2 - p_1^2 - q_1^2, \\
m_N^2 &= E_N^2 - p_N^2 - q_N^2.
\end{align}
We introduce Lagrange multipliers $\lambda_{0,1,N}$ and vary with respect to $E_{1,N},\mathbf{p}_{1,N},q_{1,N}$ and $m_N$ to obtain the extremization equations
\begin{align} \label{ext}
0 &= \lambda_0 E_N - \lambda_1 E_1,\nonumber \\
0 &= \lambda_0 E_1 - \lambda_N E_N,\nonumber \\
0 &= \lambda_0 q_N - \lambda_1 q_1,\nonumber \\
0 &= \lambda_0 q_1 - \lambda_N q_N,\nonumber \\
\mathbf{0} &= -\mathbf{p}_N (1 + \lambda_0) +(\lambda_1  +\frac{e_N}{\tilde{e}_1}) \mathbf{p}_1 ,\nonumber \\
\mathbf{0} &= \mathbf{p}_1 (1 + \lambda_0) +(\lambda_N  +\frac{\tilde{e}_1}{e_N}) \mathbf{p}_N,\nonumber \\
0 &= m_N  \Big( 1 + \lambda_0 + \lambda_N + \frac{\tilde{e}_1}{e_N} \Big).
\end{align}
Note that we assume that none of $\tilde{e}_1, e_{1,N}$ vanish so that we can freely multiply or divide by any energy or transverse energy.
Now, the first four of these equations together imply that
\begin{gather} \label{a}
\lambda_0^2 =\lambda_1 \lambda_N,
\end{gather}
and that either
\begin{gather}
0 = \lambda_0  =\lambda_1 = \lambda_N,
\end{gather}
or that
\begin{gather} \label{b}
0 =E_{[1}q_{N]},
\end{gather}
which, just as in (\ref{w}), implies that the relative rapidity of the visible system and the invisible particle should vanish. 
In the former case, equations (\ref{ext}) imply
\begin{gather} \label{c}
\mathbf{0} = \tilde{e}_1 \mathbf{p}_N -e_N \mathbf{p}_1,
\end{gather}
or equivalently that
\begin{gather} \label{d}
m_N \mathbf{p}_1 = \chi \mathbf{p}_N.
\end{gather}
But using (\ref{c}) in (\ref{f2}) then yields
\begin{gather}  \label{e}
f = (\chi + m_N)^2.
\end{gather}
In this case also, the last of equations (\ref{ext}) has no solution for $m_N \in [m_<,m_>]$, which implies that there is no stationary point in $m_N$. Thus, the extrema of $f$
must arise at the boundary values of $m_N$. The extremal values are
\begin{align} \label{hat1} 
\hat f_I &= (\chi + m_{<})^2, \nonumber \\
\hat f_{II}&= (\chi + m_{>})^2.
\end{align}
For a true ($N$+1)-daughter point decay,
\begin{align} 
\sqrt{\hat f_I} = \chi +  \Sigma_i m_i ,\nonumber \\
 \sqrt{\hat f_{II}} =\chi - m_1 + m_0.
\end{align}
It is straightforward to check that all of equations (\ref{ext}), bar the last one, have been satisfied. One should also check that the constraints, 
(\ref{con}), can be satisfied. They can, for some given value of $p_1$. In the case $m_> = m_0 - m_1$, for example, one finds that
\begin{gather}
p_1^2 = \frac{\chi^2 m_1^2}{(\chi - m_1)^2}.
\end{gather}
The singular behaviour at $\chi=m_1$ arises, because there are extrema for all values of $p_1$ at this point.

The second possibility (\ref{b}) is that $0 =E_{[1}q_{N]}$, in which case the first constraint in (\ref{con}) simplifies to
\begin{gather}
m_0^2 = m_1^2 + m_N^2 + 2(e_1 e_N - \mathbf{p}_1 \cdot \mathbf{p}_N).
\end{gather}
Now the penultimate two equations of (\ref{ext}), together with (\ref{a}), imply that
\begin{gather}
\mathbf{0} = (\tilde{e}_1 \mathbf{p}_N -e_N \mathbf{p}_1) (e_N \lambda_0 - \tilde{e}_1 \lambda_1).
\end{gather}
One of these two factors must therefore vanish; if it is the first, then we are led back to (\ref{c}) and the extremum (\ref{e}). If it is the second,
we find, from the first equation of (\ref{ext}), \footnote{We assume all Lagrange multipliers non-vanishing, otherwise we end up with the extremum obtained previously.} that
\begin{gather} \label{f}
E_1 e_N = E_N \tilde{e}_1.
\end{gather}
Now since $0 =E_{[1}q_{N]}$ in this case, we can square and add $(E_1 q_N)^2$ to the left-hand side and $(E_N q_1)^2$ to the right-hand side to obtain
\begin{gather}
e_1 = \tilde{e}_1,
\end{gather}
or rather
\begin{gather}
\chi = m_1,
\end{gather}
for which $f = m_0^2$.
So this possibility generates the well-known maxima of $f$ at $\chi = m_1$, with $0 =E_{[1}q_{N]}$.

We should also consider whether $f$ takes extremal values elsewhere on the boundary. Indeed, we already saw in (\ref{hat1}) the extrema occurring at the boundary values of the invariant mass $m_N$. There are two other boundaries that are consistent with the constraints, namely the boundary at large $q_{1,N}$ and the boundary at large $p_{1,N}$ (both of which imply large $E_{1,N}$ via the constraints).  It is easy to see that the boundary at large $q_{1,N}$ does not give rise to new extrema for $f$, since $f$ has no explicit dependence on $q_{1,N}$.
Let us look instead at large values of the transverse momenta $p_1$ and $p_N$. We should still satisfy the extremization equations obtained by variation with respect to $E_{1,N}$ and $q_{1,N}$, and so we still find that either the Lagrange multipliers or the relative rapidity should vanish. In the former case, we reproduce the extremum (\ref{hat1}). In the latter case,
the first constraint in (\ref{con}) becomes 
\begin{gather} \label{g}
m_0^2 = m_1^2 + m_N^2 + 2(e_1 e_N - \mathbf{p}_1 \cdot \mathbf{p}_N).
\end{gather}
and $f$ can be written as
\begin{gather} \label{asymf}
f (\chi^2)= m_0^2 + \chi^2 - m_1^2 +2 e_N (\tilde{e}_1 - e_1) .
\end{gather}
At large $p_1$ and $p_N$, equation (\ref{g}) becomes, at leading order,
\begin{gather}
0 = p_1 p_N - \mathbf{p}_1 \cdot \mathbf{p}_N,
\end{gather}
which is solved by choosing $\mathbf{p}_1$ and  $\mathbf{p}_N$ to be parallel.
At next-to-leading order, (\ref{g}) becomes
\begin{gather}
m_1^2\frac{p_N}{ p_1} + m_N^2\frac{ p_1}{ p_N} = m_0^2 - m_1^2 -m_N^2,
\end{gather} 
whence
\begin{gather}
\frac{p_N}{p_1} = \frac{m_0^2 - m_1^2 -m_N^2 \pm \sqrt{(m_0^2 - m_1^2 -m_N^2)^2 - 4m_1^2 m_N^2}}{2 m_1^2}.
\end{gather} 
To understand how the two values arise, note that in the rest frame of particle 0,
the magnitudes of the momenta, $p_{1,N}$, are completely fixed (for a given value of the invariant mass $m_N$). The two values above are obtained by performing an infinite boost either parallel or anti-parallel to
$\mathbf{p}_1$ (in the rest frame of 0).

Now let us expand $f$ in (\ref{asymf}) to next-to-leading order. We find
\begin{align} \label{as}
f &= m_0^2 + (\chi^2 - m_1^2)\Big(1+ \frac{p_N}{p_1}\Big), \\
&=m_0^2 + (\chi^2 - m_1^2)\Big(1+  \frac{m_0^2 - m_1^2 -m_N^2 \pm \sqrt{(m_0^2 - m_1^2 -m_N^2)^2 - 4m_1^2 m_N^2}}{2 m_1^2}\Big).
\label{h}
\end{align} 
Again, the extrema with respect to $m_N$ are obtained at the boundary, such that the extrema are
\begin{align} \label{hat2}
\hat{f}_{III} &= m_0^2 + (\chi^2 - m_1^2)\Big(1+  \frac{m_0^2 - m_1^2 -m_{<}^2 + \sqrt{(m_0^2 - m_1^2 -m_<^2)^2 - 4m_1^2 m_<^2}}{2 m_1^2}\Big), \nonumber \\
\hat{f}_{IV}&= m_0^2 + (\chi^2 - m_1^2)\Big(1+  \frac{m_0^2 - m_1^2 -m_>^2 + \sqrt{(m_0^2 - m_1^2 -m_>^2)^2 - 4m_1^2 m_>^2}}{2 m_1^2}\Big), \nonumber \\
\hat{f}_{V}&= m_0^2 + (\chi^2 - m_1^2)\Big(1+  \frac{m_0^2 - m_1^2 -m_>^2 - \sqrt{(m_0^2 - m_1^2 -m_>^2)^2 - 4m_1^2 m_>^2}}{2 m_1^2}\Big), \nonumber \\
\hat{f}_{VI}&= m_0^2 + (\chi^2 - m_1^2)\Big(1+  \frac{m_0^2 - m_1^2 -m_<^2 - \sqrt{(m_0^2 - m_1^2 -m_<^2)^2 - 4m_1^2 m_<^2}}{2 m_1^2}\Big).
\end{align}
There are thus four possible extremal values for $\hat{f}$ which occur at asymptotically large momenta.\footnote{Two in the case of a two-daughter decay with $m_< =m_>$.} 
They obey the following order relations above the kink (the order relations are simply reversed below the kink)
\begin{gather}\label{order}
\hat{f}_{III} \ge \hat{f}_{IV} \ge \hat{f}_{V} \ge \hat{f}_{VI}.
\end{gather}
The central relation is trivial; we prove the other two by establishing that 
\begin{gather}\label{mono}
m_0^2 - m_1^2 -m_N^2 \pm \sqrt{(m_0^2 - m_1^2 -m_N^2)^2 - 4m_1^2 m_N^2}
\end{gather} 
is monotonically decreasing on $m_N \in [0,\infty]$ for the $+$ branch, and monotonically increasing for the $-$ branch.
Indeed, differentiating with respect to $m_N^2$ yields
\begin{gather}
-1 \mp \frac{m_0^2 + m_1^2 -m_N^2}{ \sqrt{(m_0^2 - m_1^2 -m_N^2)^2 - 4m_1^2 m_N^2}}.
\end{gather} 
Furthermore, the magnitude of the fraction is greater than one for all values of $m_N^2$. Hence (\ref{mono}), has the claimed monotonicity properties, and the order relations in (\ref{order}) follow.

For a true point decay with massless visible daughters, the gradients are given by $1$, $m_0/m_1$, $m_0/m_1$ and $(m_0/m_1)^2$.

In summary, the possible extrema for $f$ are given by $\hat{f}_{I-VI}$ in equations (\ref{hat1}) and (\ref{hat2}).

Up until now, we have said nothing about the nature of the extrema of $f$, that is whether they correspond to maxima, minima, or saddle points. 
Now, it is clear from (\ref{f2}) that the extrema of the type
\begin{gather}
\hat{f}_{I,II} = (\chi + m_{\lessgtr})^2
\end{gather}
correspond to minima in phase space. That is, if we shift the momenta slightly, subject to the constraints, then $f$ will increase. Moreover, if we increase $m_N$ from $m_<$, then $f$ will also increase. Thus
\begin{gather}
\hat{f}_{I} = (\chi + m_{<})^2
\end{gather}
is a minimum. On the other hand, if we decrease $m_N$ from $m_>$, then $f$ will decrease, implying that
\begin{gather}
\hat{f}_{II}= (\chi + m_{>})^2
\end{gather}
is a saddle point. What about the asymptotic extrema? Well, we know that the $-$ branch in (\ref{as}) corresponds to a minimum value for the coefficient of 
$(\chi^2 - m_1^2)$, with respect to variations in the momenta, whereas the $+$ branch corresponds to a maximum. But we also know that the coefficient is monotonically decreasing as a function of $m_N$ on the $+$ branch and monotonically increasing on the $-$ branch. Putting this together, we see that, above the kink, $\hat{f}_{III}$ is a maximum, $\hat{f}_{VI}$ is a minimum, and $\hat{f}_{IV}$ and $\hat{f}_{V}$ are saddle points. In contrast, below the kink, $\hat{f}_{VI}$ is the maximum and $\hat{f}_{III}$ a minimum.

Since there is only ever a single maximum, for any value of $\chi$, it must be the global maximum. Since there are always two minima, it is not immediately clear which is the global minimum. In the case of a point decay to massless visible particles, it is easy to show that $\hat{f}_{I}$ is always the global one.

In summary, the global maximum for $\chi<m_1$ is given by
\begin{gather} \label{sptmax1}
\hat{f} (\chi | \chi<m_1)= m_0^2 + (\chi^2 - m_1^2)\Big(1+  \frac{m_0^2 - m_1^2 -m_<^2 - \sqrt{(m_0^2 - m_1^2 -m_<^2)^2 - 4m_1^2 m_<^2}}{2 m_1^2}\Big),
\end{gather}
whereas for $\chi>m_1$, it is given by
\begin{gather} \label{sptmax2}
\hat{f} (\chi | \chi>m_1) = m_0^2 + (\chi^2 - m_1^2)\Big(1+  \frac{m_0^2 - m_1^2 -m_{<}^2 + \sqrt{(m_0^2 - m_1^2 -m_<^2)^2 - 4m_1^2 m_<^2}}{2 m_1^2}\Big).
\end{gather}
These functions coincide at $\chi=m_1$, but have different gradients there, giving rise to a kink.
\subsubsection{Single-particle ZPT production (\casetwo\
  ZPT, \casethreev\ ZPT,  \casethrees\ ZPT, etc.)} \label{zpt}
To study decays in which the decaying particle has zero transverse momentum, we simply append the constraint
\begin{gather} \label{newcon}
\mathbf{0} = \mathbf{p}_1 + \mathbf{p}_N,
\end{gather}
to the set (\ref{con}). As before we find that either all Lagrange multipliers vanish or the relative rapidity vanishes. Then, we find from (\ref{con}) and (\ref{newcon}) that
\begin{gather}
p_1^2 = \frac{(m_0^2 - m_1^2 - m_N^2)^2 - 4m_1^2 m_N^2}{4m_0^2},
\end{gather}
whence
\begin{gather}
e_1 = \frac{m_0^2 + m_1^2 - m_N^2}{2m_0}
\end{gather}
and
\begin{gather}
e_N = \frac{m_0^2 - m_1^2 + m_N^2}{2m_0}.
\end{gather}
Thus,
\begin{gather}
f = m_0^2 + (\chi^2 - m_1^2) + \frac{m_0^2 - m_1^2 + m_N^2}{2m_0}\Bigg( \sqrt{(\chi^2-m_1^2) + \Big(\frac{m_0^2 + m_1^2 - m_N^2}{2m_0}\Big)^2} - \frac{m_0^2 + m_1^2 - m_N^2}{2m_0} \Bigg).
\end{gather}
Note that the radicand is always positive-definite for $\chi >  0$, and $0<m_N<m_0-m_1$ (the largest possible endpoints for $m_N$), such that $f$ is real.

We have not yet considered the behaviour of $f$ as we vary $m_N$. Let us now prove that $f$ is a monotonically increasing function of $m_N$ for $\chi > m_1$, and a monotonically
decreasing function for $\chi < m_1$, such that the global maximum is obtained at the upper endpoint of $m_N$ for $\chi > m_1$ and at the lower endpoint for $\chi < m_1$. To do so, it is convenient to define the quantities 
\begin{align}
A &= m_0^2,\nonumber \\
B &=  4m_0^2 (\chi^2 - m_1^2),\nonumber \\
C &=   m_1^2  -m_N^2 ,\nonumber \\
D &= m_0^2 + (\chi^2 - m_1^2),
\end{align}
which are all positive semi-definite apart from $B > -4m_0^2 m_1^2$.
In terms of these quantities, $f$ may be written as
\begin{gather}
f = D + \frac{A-C}{4A}\Big( \sqrt{B+(A+C)^2} - (A+C) \Big),
\end{gather}
such that
\begin{gather}
\frac{\partial f}{\partial m_N^2} =  \frac{1}{4 A \sqrt{B+(A+C)^2}}\Big( B + 2C(A+C) - 2C \sqrt{B+(A+C)^2} \Big).
\end{gather}

To exhibit the claimed monotonicity properties of $f$, we wish to show that, for $B \lessgtr 0$, 
$$\Big( B + 2C(A+C) - 2C \sqrt{B+(A+C)^2} \Big) \lessgtr 0.$$
Now, since $  AC$ and $(A+C)^2$ are both greater than $4m_0^2m_1^2$, we have that $B+4AC > 0$ and that $B +(A+C)^2>0$. From the first of these, $B  \lessgtr 0$ implies 
\begin{gather}
B(B+4AC) > 0,
\end{gather}
which itself implies, after adding an identical term to both sides, that
\begin{gather}
(B + 2C(A+C))^2 \lessgtr 4C^2 (B +(A+C)^2 ).
\end{gather}
Since $C >0$ and $B +(A+C)^2>0$, we can safely take the square root on the right-hand side to obtain
\begin{gather} \label{woo}
|B + 2C(A+C)| \lessgtr 2 C \sqrt{B+(A+C)^2}.
\end{gather}
Now $B + 2C(A+C)$ can be either positive or negative. If the latter, then $B$ is necessarily negative, and, moreover, $B + 2C(A+C) - 2C \sqrt{B+(A+C)^2}$ is the sum of two negative-definite terms and is negative. If the former, then (\ref{woo}) becomes
\begin{gather}
B + 2C(A+C) -2 C \sqrt{B+(A+C)^2} \lessgtr 0.
\end{gather}
In either case, we have the result claimed, namely that $B\lessgtr 0$ implies $\Big( B + 2C(A+C) - 2C \sqrt{B+(A+C)^2} \Big) \lessgtr 0$, which in turn implies that $f$ is monotonic, either 
increasing or decreasing as specified above. 
Thus, the global maximum of $f$ for  $\chi<m_1$ is given by
\begin{multline}\label{zptmax1}
\hat{f} (\chi | \chi<m_1) = \\ m_0^2 + (\chi^2 - m_1^2) + \frac{m_0^2 - m_1^2 + m_<^2}{2m_0}\Bigg( \sqrt{(\chi^2-m_1^2) +\Big(\frac{m_0^2 + m_1^2 - m_<^2}{2m_0}\Big)^2} - \frac{m_0^2 + m_1^2 - m_<^2}{2m_0} \Bigg),
\end{multline}
whereas for $\chi>m_1$, the global maximum is given by
\begin{multline} \label{zptmax2}
\hat{f} (\chi | \chi>m_1) =\\ m_0^2 + (\chi^2 - m_1^2) + \frac{m_0^2 - m_1^2 + m_>^2}{2m_0}\Bigg( \sqrt{(\chi^2-m_1^2) + \Big(\frac{m_0^2 + m_1^2 - m_>^2)}{2m_0}\Big)^2} - \frac{m_0^2 + m_1^2 - m_>^2}{2m_0} \Bigg).
\end{multline}
These two values coincide at $\chi = m_1$ but have different gradients there, iff.\ $m_> \neq m_<$. This condition is trivially broken in the case of decay to a single visible daughter, in which case $m_> = m_< = m_2$. In all other cases, we will get a kink.

It is worth observing that the maxima of the ZPT case never coincide with maxima of the SPT case, not even as local (rather than global) maxima.
Indeed, the only maximum of the ZPT case which coincides with an extremum of the SPT case occurs for a true point decay, for which $m_N=m_0 -m_1$ and $\mathbf{p}_1=\mathbf{0}$. But, as we pointed out in the previous subsubsection (\ref{spt}), this extremum is a saddle point in the SPT case.
\subsection{Further analysis of single particle decays (\casetwo, \casethreev,  \casethrees, etc.)}\label{boost}
Section \ref{single} provides a comprehensive description of the extremal values of $f$ for single particle decays, but is not terribly intuitive.

In order to gain some intuition as to the physical origin of the kinks and extrema, in this subsection we investigate the
dynamics by first looking at the decay kinematics in the rest frame of the parent particle,
applying the physical constraints there, and then boosting the system of daughters to the lab frame.
As before, the parent particle, `0', decays to an invisible particle, `1', and a system of visible particles, `$N$'.
$(E_N,\mathbf{p}_N,q_N)$ can be either the four-momentum of a single particle in the two-daughter decay case, 
or the sum of four-momenta of all the visible daughters in the $N$+1-daughter decay case, with $N+1> 2$.

We have shown already that maxima of $f$ occur
when the visible and invisible systems are produced with vanishing relative rapidity \eqref{b}. 
Since it is the maxima that we are interested in,
for the rest of this section we restrict ourselves to vanishing relative rapidity, in which case
the transverse mass-squared is given by
\begin{gather}
f = m_0^2 + (\chi^2 - m_1^2) + 2 e_N(\tilde{e}_1-e_1) .
\end{gather}
Moreover, since we are interested in the existence of kinks
near $\chi=m_1$, we simplify subsequent expressions by expanding $f$ 
as a Taylor series in $\chi^2$, about $\chi^2 = m_1^2$. Thus
\begin{align}
f & = m_0^2 + (\chi^2-m_1^2) \left( 1 + \frac{e_N}{e_1} \right) + \dots
\label{eq:taylor}
\end{align}

We remind ourselves that since we are interested in the maximal value of $f$, 
for values of $\chi$ greater than $m_1$, we must take the maximal value of 
\begin{gather}
\alpha \equiv 1 + \frac{e_N}{e_1} , \label{alpha}
\end{gather}
whereas for $\chi$ less than $m_1$, we must take the minimal value of $\alpha$.
There will be a kink in $\hat{f}$ if the maximal and minimal values of $\alpha$ are different.

We first consider the decay in the rest frame of the parent.
In that frame (denoted by primed quantities), for fixed values of $m_0, m_1$ and $m_N$, the momenta and transverse energies of the daughters
are fixed:
\begin{align}
(p^\prime)^2  &\equiv |{\mathbf{p}}^\prime_{1,2}|^2 
              = \frac{(m_0^2-(m_1+m_N)^2)(m_0^2-(m_1-m_N)^2) }{4 m_0^2}, \label{pprime}\\
(e^\prime_i)^2 &= (p^\prime)^2 + m_i^2. \label{eprime}
\end{align} 

We now boost back to the lab frame. We define $\beta = |\mathbf{p_0}/e_0|$ to be the transverse velocity 
of the parent particle in the lab frame. We define $\theta$
to be the angle, as measured in the rest frame of the parent, 
between the velocity vector of the lab frame and the transverse momentum vector of 
the visible particle system, $\mathbf{p}_N$. 
This means that, to an observer in the lab frame, the condition $\theta=0$ corresponds to the 
invisible particle being thrown forwards in the decay, {\em i.\ e.\ }in the same direction as the parent's velocity, 
while $\theta=\pi$ corresponds to the invisible particle being thrown backwards,{\em  i.\ e.\  }against the parent's velocity.

A simple Lorentz transformation gives the lab-frame transverse energies and momenta of the daughters,
\begin{align}
e_N & = \gamma ( e_N^\prime - p^\prime\beta\cos\theta ), \\
e_1 & =  \gamma (e_1^\prime + p^\prime\beta\cos\theta ) ,
\end{align}
where $\gamma \equiv (1-\beta^2)^{-\frac{1}{2}}$ is the Lorentz factor associated with the velocity $\beta$.
This means that $\alpha$ is given by
\begin{align}
\alpha
       & = 1+ \frac{e_N^\prime -  p^\prime\beta\cos\theta}{e_1^\prime + p^\prime\beta\cos\theta}, \nonumber \\
       & = \frac{2 m_0^2}{m_0^2+m_1^2-m_N^2 + 2 m_0 p^\prime \beta\cos\theta} ,  \nonumber \\
       & = \frac{2 m_0^2}{m_0^2+m_1^2-m_N^2 + \beta\cos\theta \sqrt{\left(m_0^2-(m_1-m_N)^2\right) \left(m_0^2-(m_1+m_N)^2\right)}} .\label{eq:gengrad}
\end{align}

Equation \ref{eq:gengrad} tells us everything about the gradient around $\chi^2=m_1^2$ for any two-daughter, 
three-daughter, or indeed $N$+1-daughter decay.
We recall that $N$ represents the system of visible daughters and has a mass equal to the 
invariant mass of the visible daughters.
For the $N$+1-daughter decay, with $N+1>2$, the visible-daughter invariant mass, $m_N$, 
can take any value within the range
$m_N \in [m_<,m_>]$. We recall that, for a true point decay, $m_<$  
is equal to the sum of the masses of the visible daughters produced in the decay,
and $m_>$ is $m_0-m_1$.

We can see that, if some finite experimental sample contains events in which  {\em either} the parent is highly boosted, such that $\beta\cos\theta$ has values significantly different from zero, {\em or} $m_N$ 
spans some reasonable range, then the maximal and minimal values of the gradient obtained from the sample will be significantly different, and an observable kink 
in $\hat{f}$ can be expected at $\chi=m_1$. Indeed, in principle, we need just two events with vanishing relative rapidity to generate a kink at $\chi=m_1$.

Thus, the presence or absence of the kink at $(m_0,m_1)$ is not, in fact, contingent upon whether our experimental sample contains events that correspond to the global maxima, whether SPT or ZPT.
Rather, we simply need a reasonable number of events with vanishing (or near-vanishing) relative rapidity and significantly differing values of either $m_N$ or $\beta \cos \theta$. Either of these will suffice to generate a kink, and indeed our later simulations will illustrate this rather well.

That said, it is important to remark that, if our sample does not contain events at or close to the global maximum, then it is possible that further kinks will arise at $\chi \neq m_1$. The reasoning for this is as follows.
Suppose our sample contains events with vanishing relative rapidity, giving rise to a kink at $\chi = m_1$. If an event corresponding to the global maximum is not contained in the sample, then it is possible that other events in the sample, with non-vanishing relative rapidity, can exceed the maximum value of $f$ generated by the events with vanishing relative rapidity. Now, this cannot occur at $\chi = m_1$, because we have saturated the global maximum at this point. So, the kink at $(m_0,m_1)$ cannot be erased in this way. But the events with non-vanishing rapidity can exceed the maximum value of $f$ generated by the events with vanishing relative rapidity at values of $\chi \neq m_1$. If this does indeed occur, then a second kink will be generated at the value of $\chi \neq m_1$ where the values of $f$ coming from the two different types of event coincide.

So it is possible, albeit unlikely, that a finite experimental sample of events will give rise to spurious kinks, leading to a discrete ambiguity in the extracted values of the masses $m_0$ and $m_1$. We note, however, that this phenomenon is not observed in any of our simulations.

Let us put this issue to one side, and return to our discussion of the kink at $(m_0,m_1)$.
In the case where the range of values of $m_N$ is small --- and in particular {\casetwo} where $m_N$ is single-valued --- we must rely on different values of $\beta\cos\theta$ to produce a kink.
The global-maximum gradient on either side of the kink is obtained asymptotically as the transverse momenta become both large in magnitude and collinear. 

Some other important properties of the extremal gradients are as follows.
\begin{itemize}
\item{
The maximum and minimum gradients will occur when $\beta\cos\theta\rightarrow \mp 1 $.  
The values of these extremal gradients are ({\em cf.\ }Eqs. \ref{sptmax1} and \ref{sptmax2})
\begin{gather}
\alpha^{\max}_{\min} = 1+ \frac{m_0^2 - m_1^2 - m_<^2 \pm \sqrt{(m_0^2 - m_1^2 - m_<^2)^2 - 4 m_1^2 m_<^2}}{2m_1^2}.
\label{extremal-gradients}
\end{gather}
In the limit where all the visible particles are massless, $m_<=0$, these
become:
\begin{align}
\alpha^\mathrm{max} & = {\left(\frac{m_0}{m_1}\right)}^2,\\
\alpha^\mathrm{min} & = 1 .
\end{align}
}
\item{
When $m_> = m_0-m_1$, then $p^\prime=0$, and the invisible daughter is at rest in the parent's rest frame. The gradient is then
independent of the motion of the parent (parameterized by $\beta$ and $\theta$) and 
takes value $m_0/m_1$.
In \casetwo\ this configuration is only obtained if masses are precisely at the threshold.
In the $N+1$-daughter point-decay case with $N\ge 2$ (e.g. \casethreev), it represents the kinematic configuration in which the 
visible daughters have their maximum possible invariant mass. This configuration cannot be realised in cascade decays, such as \casethrees.
}
\end{itemize}
\begin{figure}
\begin{center}
\includegraphics[width=0.5\linewidth]{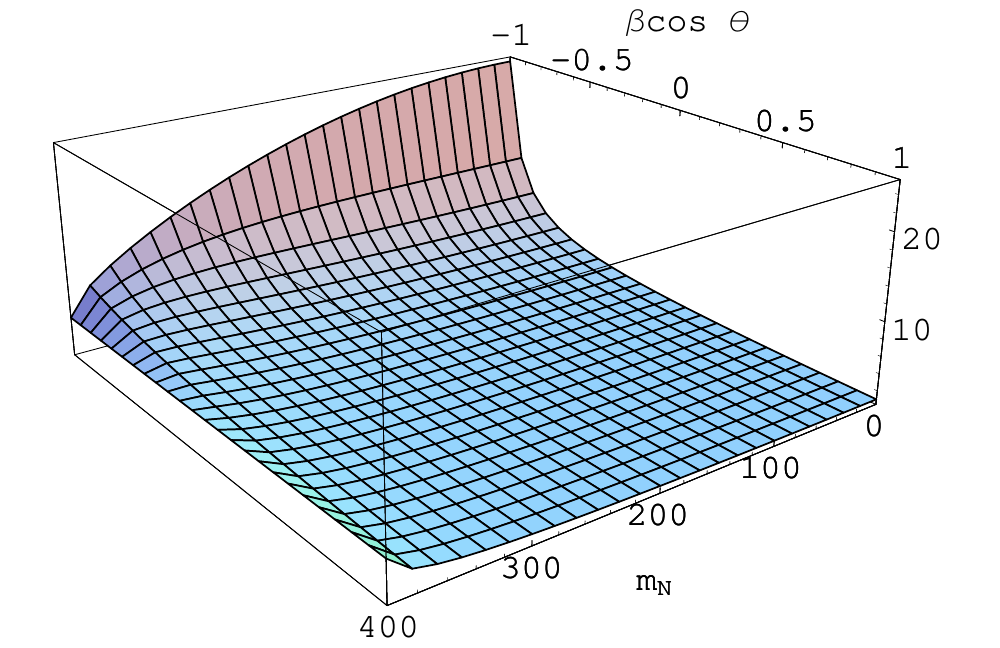}\\
\caption{The surface shows the gradient, $\alpha = \frac{df}{d\chi^2}$, near $\chi^2=m_1^2$ 
as a function of $\beta\cos\theta$, and of $m_N$, the invariant mass of the visible particles produced in the decay.
The values of the other masses are $m_0=500$ and
$m_1=100$.
\label{fig:sharkfin}
}
\end{center}
\end{figure} 
In \figref{fig:sharkfin}, we show the gradient as a function of  $\beta\cos\theta$ and $m_N$, for 
the particular values of $m_0=500$ and $m_1=100$.
As expected, the maximal value of the gradient occurs at $\beta\cos\theta\rightarrow - 1$, $m_N=0$, 
and is $m_0^2/m_1^2 = 25$.
The minimal value occurs at $\beta\cos\theta\rightarrow + 1$, $m_N=0$,  and is 1.
When $m_N = m_0-m_1 = 400$ we obtain gradient $m_0/m_1 = 5 $ for any value of $\beta\cos\theta$.

Note that in a real collider experiment there will usually be very few events near $\beta\cos\theta=\pm 1 $
since these events require $\beta\rightarrow1$ and hence asymptotically high energy. 
Indeed, for the common case of heavy particles being produced near threshold, 
most events will have $\beta\approx 0$.
These threshold-production (ZPT) events have gradients
\begin{align}
\alpha_{\beta=0} & = 1 + \frac{e_N^\prime}{e_1^\prime}, \\
                 & = \frac{2m_0^2}{m_0^2+m_1^2-m_N^2}.
\end{align}
For the $N$+1-daughter point decay with $m_< = \Sigma_i m_i$ and $m_>=m_0-m_1$, the ZPT extremal gradients are
\begin{align}
\alpha_{\beta=0}^{\max} & = \frac{m_0}{m_1},\\
\alpha_{\beta=0}^{\min} &= \frac{2m_0^2}{m_0^2+m_1^2-m_<^2}.
\end{align}

Returning to the SPT case (in which the global maximum is obtained at asymptotically large momenta) we should like to understand how close to the global maximum we can get in events
with large, but finite, momenta, such as we might hope to achieve in an experiment.
To this end, let us ask what the relative decrease is in the gradient of $f$ that results from a small shift in $\beta \cos \theta$ away from the maximal values of $\pm 1$. For $\beta \cos \theta = 1 - \delta_+$, we find that the relative change in the gradient is given by
\begin{gather}
\frac{p^\prime \delta_+}{e^\prime_1+p^\prime},
\end{gather}
whereas for $\beta \cos \theta = -1 + \delta_-$, the relative change in the gradient is given by
\begin{gather}
\frac{p^\prime \delta_-}{e^\prime_1-p^\prime}.
\end{gather}
Thus, the relative change for a given shift in the $\delta$s is always greater above the kink than it is below, irrespective of the mass values. So in a finite sample of events, in which the distribution of $\cos\theta$ is roughly uniform, we always expect to get closer to the maximum below the kink than we do above it.

\begin{figure}
  \begin{center}
    \begin{minipage}[b]{.49\linewidth}
      \begin{center}
        \includegraphics[width=0.75\linewidth]{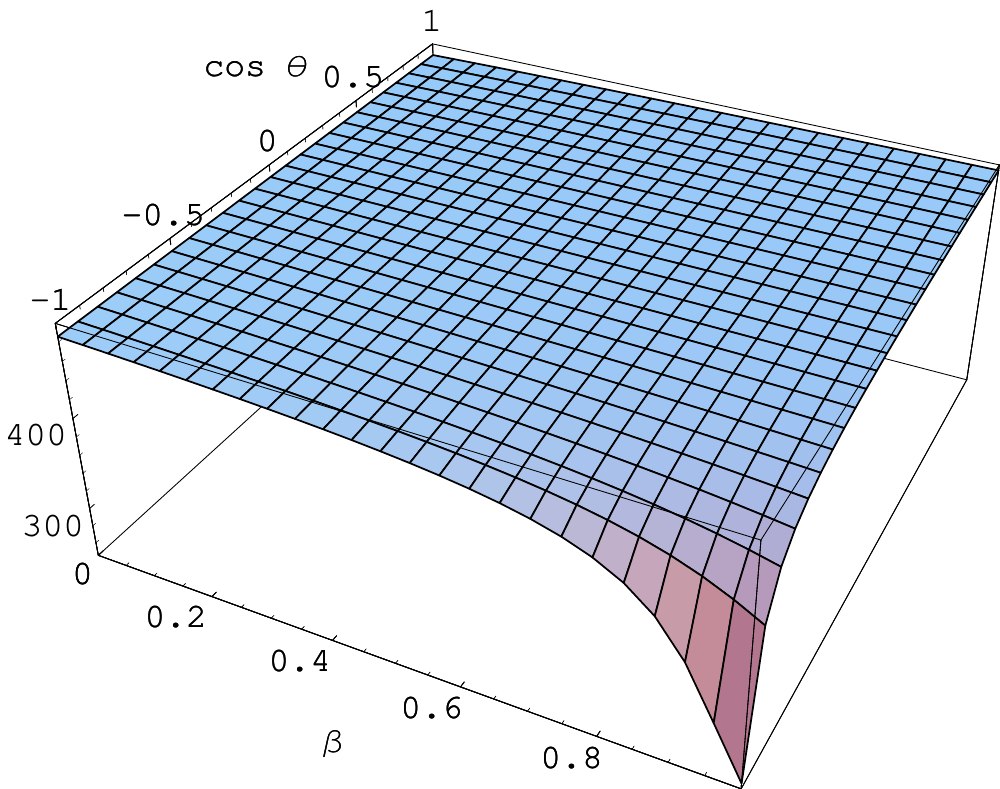}\\
        {\bf (a)}
      \end{center}
     \end{minipage}
   \begin{minipage}[b]{.49\linewidth}
     \begin{center}
       \includegraphics[width=0.75\linewidth]{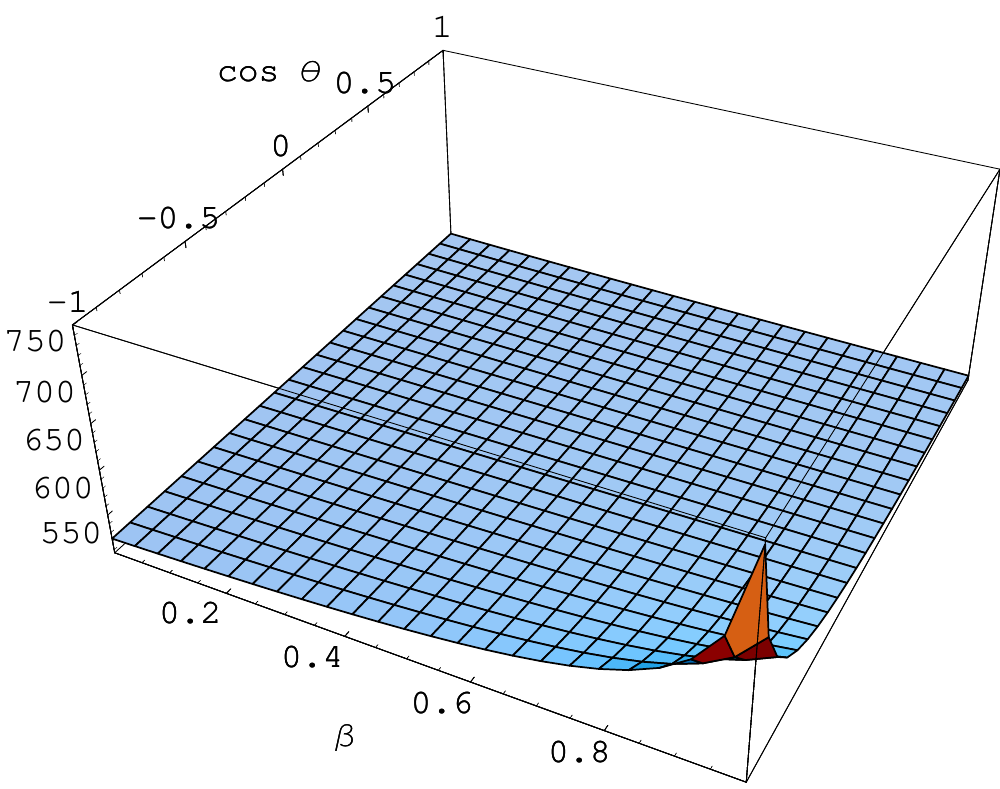}\\
       {\bf (b)}
     \end{center}
   \end{minipage}
\caption{
The surface shows the maximum transverse mass 
that would be found for a two-daughter decay when the hypothesized mass, $\chi$,
of the invisible particle is {\bf (a)} less than and {\bf (b)} greater than its true mass.
The invariant mass is plotted as a function of the transverse velocity, 
$\beta = p_0/e_0$, of the  parent particle in the lab frame, 
and of the cosine of the angle, $\theta$, as described in the text.
Decays in which the visible and invisible daughters are produced with vanishing relative rapidity will lie
on the surface, while all other events will lie below it. 
In both cases the masses are: parent $m_0=500$; visible daughter $m_2=0$; invisible daughter $m_1=100$.
The hypothesized mass of the invisible daughter ($\chi$) is 50 in {\bf (a)} and 150 in {\bf (b)}.
\label{fig:nose_and_wall}
}
  \end{center}
\end{figure}

A graphical representation of this result for particular mass values 
can be seen in \figref{fig:nose_and_wall}, where 
$m_T = \surd f$ is plotted as a function of $\cos\theta$ and $\beta$ for two particular values of $\chi$.
Firstly, we see that the maxima are obtained at the expected values of $\theta$ and $\beta$. We also see
that, for angles close to the respective maxima, the fall-off is much greater above the kink ($\chi>m_1$, \figref{fig:nose_and_wall}b) than it is below
($\chi<m_1$, \figref{fig:nose_and_wall}a). Indeed, a large fraction of the surface lies near the upper limit for 
$\chi<m_1$, 
while only a very small fraction of events, with $\beta\cos\theta\approx -1$, lie near the maximum value for
$\chi > m_1$.

We might therefore expect that in a finite sample of events, we are likely measure a maximal gradient close to the predicted one when $\chi<m_1$. However we are likely to underestimate the maximum gradient when $\chi>m_1$ because of the scarcity of events in this kinematic region. Our Monte Carlo simulations (described in Section \ref{mc}) confirm these expectations.

\subsection{Pair production at SPT and ZPT (\casefour, \casesixv,  \casesixs, etc.)}\label{pair}
Now let us consider the case, relevant to supersymmetric theories with conserved R-parity, in which parent particles are pair-produced, and each parent
decays into an invisible daughter particle and one or more visible daughter particles. This case is complicated by the fact that there are now
two invisible particles, and thus two transverse momenta go unmeasured. Only their sum can be inferred from  measurement of the total missing transverse momentum. Thus, the individual transverse masses for each decay are no longer observables, and we need to construct a new observable.

A suitable observable was put forward in
\cite{Lester:1999tx,Barr:2003rg}, and is defined as follows. We label
the pair of parent particles as $0$ and $5$ and their respective
invisible daughters by $1$ and $6$ (see figure~\ref{fig:cases}). There are $N$ and $M$ visible daughters, respectively, with invariant masses $m_N$ and $m_M$. For
simplicity, we assume that the produced pair of particles have identical mass $m_0$, 
that the invisible particles have identical mass $m_1$.\footnote{This is, for example, the case for supersymmetric decays to the lightest stable superpartner.}

Following
\cite{Lester:1999tx,Barr:2003rg}, we construct two transverse mass
functions, namely
\begin{align}
f &= \chi^2 + m_N^2 + 2 (\tilde{e}_1 e_N - \mathbf{p}_1 \cdot \mathbf{p}_N),\\
g &= \chi^2 + m_M^2 + 2 (\tilde{e}_ 6 e_M - \mathbf{p}_6 \cdot \mathbf{p}_M).
\end{align}
Even if the mass $\chi$ were known, these would not be observables, since one cannot separately determine the transverse momenta 
$\mathbf{p}_1$ and $\mathbf{p}_6$ of the invisible particles in a collider experiment. One can only determine their sum, which equals the total missing transverse momentum, $\mathbf{p}$, say.
An observable can then be constructed as follows. Consider all possible partitions of the measured missing transverse momentum $\mathbf{p}$ into the unmeasured invisible transverse momenta. Only one of these partitions will, of course, be the correct one.
For each partition, select the largest value of $f$ and $g$, $\mathrm{max} (f,g)$. Finally, minimize $\mathrm{max} (f,g)$ over all possible partitions.

In this way, one obtains an observable function of the hypothesized invisible mass $\chi$. This observable, called $m^2_{T2}(\chi)$ has the same property as the usual transverse mass observable: at $\chi=m_1$, it is bounded above by $m_0^2$ \cite{Lester:1999tx,Barr:2003rg}.

\begin{figure}
\begin{center}
\includegraphics[width=14cm]{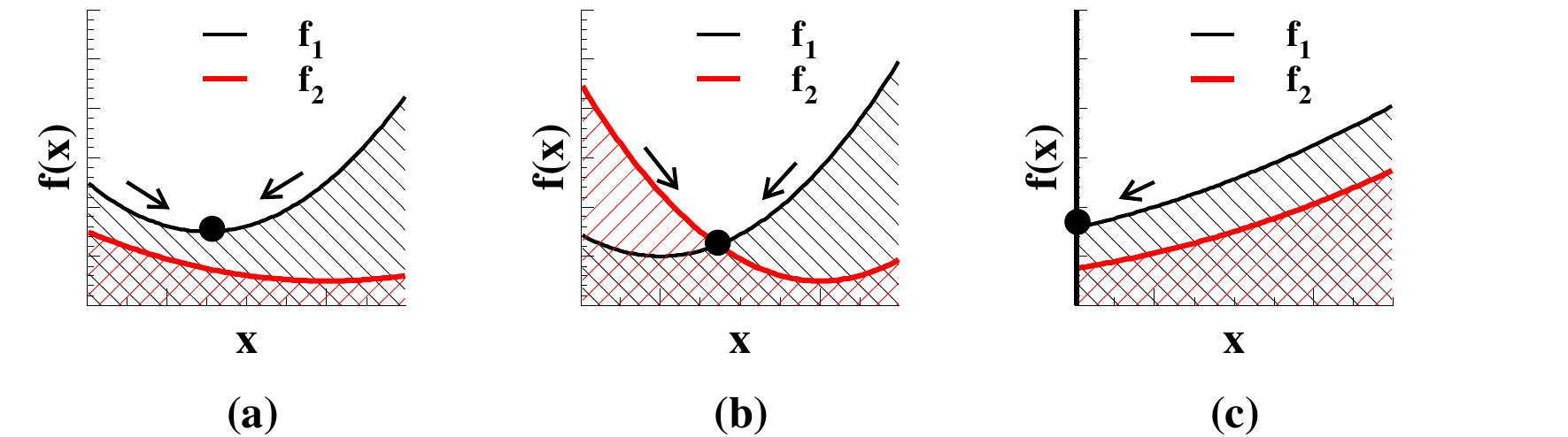}
\end{center}
\caption{\label{mtx}
A diagram demonstrating that the minimisation over some parameter of the maximum of two differentiable functions
may occur either at {\bf (a)} a minimum value of one of them, or {\bf (b)} when they
are equal, or {\bf (c)} at the boundary of the domain.
}
\end{figure}

To study $m^2_{T2}$ away from the point $\chi=m_1$, let us define the hypothesized invisible momenta of particles $1$ and $6$ in the trial partition to be $\mathbf{p}_1^\star$ and $\mathbf{p}_6^\star$.
Since $\mathbf{p}_1^\star + \mathbf{p}_6^\star = \mathbf{p}$ we can write 
\begin{align} \label{hyp}
f &= \chi^2 + m_N^2 + 2 \sqrt{(p_1^\star)^2 + \chi^2}   \sqrt{\mathbf{p}_N^2 + m_N^2}-2 \mathbf{p}_1^\star \cdot \mathbf{p}_N,\\
g &= \chi^2 + m_M^2 + 2  \sqrt{(\mathbf{p}-\mathbf{p}_1^\star)^2 + \chi^2} \sqrt{\mathbf{p}_M^2 + m_M^2} - 2(\mathbf{p}-\mathbf{p}_1^\star)\cdot  \mathbf{p}_M.
\end{align}
The prescription for constructing $m^2_{T2}$ given above instructs us to find the minimum with respect to variations of $\mathbf{p}_1^\star$, of the maximum of $f,g$. There are three ways in which this minimum can arise (see \figref{mtx}) . Either {\bf (a)} it is a minimum of one of the $f,g$ that lies above the other one of $g,f$, or {\bf (b)}  it is a point at which $f=g$, or {\bf (c)} it occurs
at a boundary. In the case at hand, it is not
difficult to show that $f,g$, considered as functions of $\mathbf{p}_1^\star$,
are unbounded above as $\mathbf{p}_1^\star \rightarrow \pm \infty$ and each has a
unique minimum with $f,g = (\chi+m_{N,M})^2$. For $f$, this minimum occurs at $\chi \mathbf{p}_N = m_N \mathbf{p}_1^\star$, and for $g$, it occurs at 
$\chi \mathbf{p}_M = m_M (\mathbf{p} - \mathbf{p}_1^\star)$.  
For minima of type {\bf {(a)}} the value of $\MTTWO$ that results is given by $\chi+m_{N,M}$, and thus the largest value of $\MTTWO$ that can arise is given by the largest
upper endpoint of the two $m_{N,M}$ intervals. Such extrema always lie below $m_0$ at $\chi=m_1$ (except in the special case of a point decay, in which the upper endpoint, $m_0-m_1$, gives $\MTTWO = m_0$). Thus, they do not, in general affect the behaviour at $(m_0,m_1)$ in which we are principally interested.

We are thus left to consider minima of the type {\bf (b)}, for which $f=g$. Now, to find the partition of $\mathbf{p}$ that minimizes $f=g$ for a given event requires us to solve
 $f=g$, which is a quartic equation in one of the assigned invisible momenta. Rather than do this, we note that the entire procedure of computing $\MTTWO$ by minimizing over partitions and then maximizing $\MTTWO$ over all events, can be phrased as a single extremization problem. Indeed, we wish to find the extremum which is a minimum of $f$ (or $g$, it does not matter which) with respect to variations of the hypothesized transverse momenta and a maximum with respect to the true event variables, subject to the usual energy-momentum constraints, supplemented with the constraint $f=g$. 
 
Explicitly, we wish to extremize
\begin{gather}
f = m_N^2 + \chi^2 + 2(\tilde{e}_1^\star e_N - \mathbf{p}_1^\star \cdot \mathbf{p}_N),
\end{gather}
where $\tilde{e}_1^\star = \sqrt{p_1^{\star 2}+\chi^2}$, subject to 
\begin{align} \label{paircon}
0&= m_N^2 + \chi^2 + 2(\tilde{e}_1^\star e_N - \mathbf{p}_1^\star \cdot \mathbf{p}_N) - m_M^2 + \chi^2 + 2(\tilde{e}_6^\star e_M - \mathbf{p}_6^\star \cdot \mathbf{p}_M),\nonumber \\
0&=  -m_0^2 + m_1^2 +m_N^2 + 2 (E_1 E_N - \mathbf{p}_1 \cdot \mathbf{p}_N-q_1 q_N),\nonumber \\
0&=  -m_0^2 + m_1^2 +m_M^2 + 2 (E_6 E_M - \mathbf{p}_6 \cdot \mathbf{p}_M - q_6 q_M),\nonumber \\
\mathbf{0}&= -\mathbf{p} +\mathbf{p}_1 +\mathbf{p}_6,\nonumber \\
\mathbf{0}&= -\mathbf{p} +\mathbf{p}_1^\star +\mathbf{p}_6^\star.
\end{align}
In the case of ZPT decays, we must further supplement these with the constraint
\begin{gather}\label{pairz}
\mathbf{0}= \mathbf{p} +\mathbf{p}_N +\mathbf{p}_M.
\end{gather}
As always, variation with respect to $E$ and $q$ will force either the relevant Lagrange multipliers to vanish, or the relative rapidities to vanish. That is
\begin{align}
0&= E_{[1}q_{N]},\nonumber \\
0&= E_{[6}q_{M]}.
\end{align}
We are interested in the latter case, since this will give rise to values of $\MTTWO (\chi)$ which go through the point $(m_0,m_1)$. In this case, the two energy-momentum constraints simplify to
\begin{align}
0&=  -m_0^2 + m_1^2 +m_N^2 + 2 (e_1 e_N - \mathbf{p}_1 \cdot \mathbf{p}_N),\nonumber \\
0&=  -m_0^2 + m_1^2 +m_M^2 + 2 (e_6 e_M - \mathbf{p}_6 \cdot \mathbf{p}_M ).
\end{align}
Let us now explicitly introduce Lagrange multipliers $\mu_{1,2,3}$ for the first three constraints in (\ref{paircon}). The final two constraints we use to eliminate 
$\mathbf{p}_6$ and $\mathbf{p}_6^\star$. We also add a vector Lagrange
multiplier $\mathbf{\mu}_4$ for constraint equation (\ref{pairz}) in the case of ZPT.  The extremization equations, obtained by varying with respect to $\mathbf{p}_1^\star,\mathbf{p}_1,\mathbf{p}_N,\mathbf{p}_M,\mathbf{p},m_N$ and $m_M$ are
\begin{align} \label{mess}
\mathbf{0}&=(1+\mu_1)\Big( \frac{\mathbf{p}_1^\star}{\tilde{e}_1^\star}e_N - \mathbf{p}_N \Big) -\mu_1 \Big( \frac{\mathbf{p}_1^\star-\mathbf{p}}{\tilde{e}_6^\star}e_M+\mathbf{p}_M\Big) ,\nonumber \\
\mathbf{0}&= \mu_2 \Big( \frac{\mathbf{p}_1}{e_1}e_N - \mathbf{p}_N \Big)  + \mu_3 \Big(\frac{\mathbf{p}_1-\mathbf{p}}{e_6}e_M+\mathbf{p}_M \Big) ,\nonumber \\
\mathbf{0}&= (1+\mu_1)\Big( \frac{\mathbf{p}_N}{e_N}\tilde{e}_1^\star-\mathbf{p}_1^\star\Big)  +\mu_2 \Big( \frac{\mathbf{p}_N}{e_N}e_1 - \mathbf{p}_1\Big) +\mathbf{\mu}_4 ,\nonumber \\
\mathbf{0}&= -\mu_1 \Big(\frac{\mathbf{p}_M}{e_M}\tilde{e}_6^\star-(\mathbf{p} - \mathbf{p}_1^\star) \Big) +\mu_3 \Big( \frac{\mathbf{p}_M}{e_M}e_6 - (\mathbf{p}-\mathbf{p}_1) \Big) +\mathbf{\mu}_4,\nonumber \\
\mathbf{0}&= -\mu_1 \Big( \frac{\mathbf{p}-\mathbf{p}_1^\star}{\tilde{e}_6^\star}e_M - \mathbf{p}_M \Big)  + \mu_3 \Big( \frac{\mathbf{p}-\mathbf{p}_1}{e_6}e_M - \mathbf{p}_M\Big) + \mathbf{\mu}_4 ,\nonumber \\
0&= (1+\mu_1)\Big( \tilde{e}_1^\star + e_N\Big)  + \mu_2 \Big( e_1 + e_N \Big) ,\nonumber \\
0&= -\mu_1 \Big( \tilde{e}_6^\star + e_M \Big) + \mu_3 \Big( e_6 + e_M \Big)  ,
\end{align}
where it is to be understood that $e_6$ and $e_6^\star$ depend implicitly upon $\mathbf{p}_1$, $\mathbf{p}_1^\star$ and $\mathbf{p}$.

We have not attempted to find the general set of solutions to (\ref{mess}) and (\ref{paircon}). However, it is possible to show that, in the ZPT case, a solution of 
(\ref{paircon}), (\ref{pairz}), and all but the last two equations of (\ref{mess}) is given by
\begin{align} \label{soln}
\mathbf{p}_1^\star &= \mathbf{p}_1, \nonumber \\
\mathbf{p}_6^\star&= \mathbf{p}_1,\nonumber \\
\mathbf{p}_6 &= \mathbf{p}_1,\nonumber \\
\mathbf{p}_N&= -\mathbf{p}_1,\nonumber \\
\mathbf{p}_M &= - \mathbf{p}_1,\nonumber \\
\mu_1 &= -\frac{1}{2} ,\nonumber \\
\mu_2 &= \mu_3 = \frac{e_1}{\tilde{e}_1}\frac{e_N^2 - \tilde{e}_1^2}{2(e_N + e_1)^2},\nonumber \\
\mathbf{\mu}_4 &= \mathbf{p}_1\frac{(\tilde{e}_1+e_N)(\tilde{e}_1+e_1)}{2\tilde{e}_1 (e_1+e_N)}, \nonumber \\
m_N&=m_M.
\end{align}
In the above, $p_1$ is then fixed by the second of equations (\ref{mess}). Now, these values do not represent a stationary point of $\MTTWO$, because they do not solve the last two equations of (\ref{mess}). If however, the two decay channels are identical, such that the endpoints of the $m_{N,M}$ intervals coincide, then we see that we obtain extrema at the upper and lower endpoints of the common invariant mass interval $[m_<,m_>]$. 
These extrema correspond to events in which both of the decaying parents are at rest in the lab frame, and in which each of the visible daughter systems obtains either its maximum or its minimum value. They are, thus, simply pairwise copies of the extrema we found for the single particle ZPT decays in subsubsection \ref{zpt}. They are, moreover, the extrema exhibited by Cho {\em et al.} in \cite{Cho:2007qv}. Since they are just pairwise copies of the extrema we found in subsubsection \ref{zpt}, it is trivial to see that the resulting values of $\MTTWO$ generate a kink at $(m_0,m_1)$ with $\MTTWO^2$ given by
\begin{multline}
\hat{\MTTWO^2} (\chi<m_1) = \\m_0^2 + (\chi^2 - m_1^2) + \frac{m_0^2 - m_1^2 + m_<^2}{2m_0}\Bigg( \sqrt{(\chi^2-m_1^2) +\Big(\frac{m_0^2 + m_1^2 - m_<^2}{2m_0}\Big)^2} - \frac{m_0^2 + m_1^2 - m_<^2}{2m_0} \Bigg),
\end{multline}
and
\begin{multline}
\hat{\MTTWO^2} (\chi>m_1) =\\ m_0^2 + (\chi^2 - m_1^2) + \frac{m_0^2 - m_1^2 + m_>^2}{2m_0}\Bigg( \sqrt{(\chi^2-m_1^2) + \Big(\frac{m_0^2 + m_1^2 - m_>^2)}{2m_0}\Big)^2} - \frac{m_0^2 + m_1^2 - m_>^2}{2m_0} \Bigg).
\end{multline}

Note that we have not shown that the extrema that we found correspond to the global maximum. To do so would require a complete solution of equations (\ref{mess}) and (\ref{paircon}). But we can show that a kink is necessarily present nevertheless. Indeed, suppose there exists a maximum whose value of $\MTTWO$ is greater or equal to the values found above for some values of $\chi$. At $\chi = m_1$, the $\MTTWO$ value of such a new maximum is necessarily $m_0$, since we know that $\MTTWO (\chi = m_1)$ is bounded above by $m_0$. Now, considering values of $\chi$ in the vicinity of $\chi = m_1$, we see that any new maximum can only accentuate the kink at $(m_0,m_1)$. This proves that $\MTTWO$ has a kink at $(m_0,m_1)$ in the ZPT case, and indeed in any class of events that includes the ZPT such as a subset. Since the SPT case is such a class, we have also proven, as a corollary, that there is necessarily a kink in the SPT case as well.

Note that the sole exception to this result is \casefour, for either ZPT or SPT. In this case $m_<=m_>=m_2$ and the events described above do not generate a kink. This is just as was the case for the single particle ZPT case, \casetwo. Of course, this does not prove that there is no kink in this case, since we have not excluded the possibility of a higher maximum. Nevertheless, our later simulations will suggest that there is no kink in the ZPT case.

Another immediate result that follows is that, just as for single parent decays, the extrema of the ZPT case we have identified are not extrema of the SPT case. Indeed, in the SPT case, we should solve equations (\ref{mess}) and (\ref{paircon}), but without (\ref{pairz}) and with $\mathbf{\mu_4} = \mathbf{0}$. But $\mathbf{\mu_4} = \mathbf{0}$ is not consistent with the solution we found in (\ref{soln}). Thus, configurations of this type are not extrema of the SPT case.

What then, is the maximum of $\MTTWO$ in the SPT case?
To answer this, it is first useful to observe
 that the value of $\MTTWO^2$ measured in an event cannot exceed the maximum of the individual transverse masses squared $f,g$ that would be measured in the same event, if the invisible momenta could be determined. This follows from the definition of $\MTTWO$: the value of $\MTTWO^2$ that one computes for a given event, {\em viz.} the minimum with respect to partitions of the invisible momenta of $\mathrm{max} (f,g)$, cannot exceed the value of $\mathrm{max} (f,g)$ that would be obtained with the true values of the invisible momenta, because one of the partitions must correspond to the true values of the invisible momenta. So $\MTTWO^2$ cannot exceed $\mathrm{max} (f,g)$ in a given event. This, of course, was the original motivation for the definition of $\MTTWO$ at $\chi = m_1$, but it is true for all values of $\chi$. 

It follows that the maximal value of $\MTTWO^2$, obtained by considering some set of events, is bounded above by the maximal value of $\mathrm{max} (f,g)$
for the same set of events. If, furthermore, we can show that the bound is saturated, then we can infer the maximal value of $\MTTWO^2$ for the set of events from the maximal
values of $f$ or $g$ determined in the previous section.

This argument allows us to determine the maximal value of \MTTWO\ in the SPT case, where the initial transverse momentum of the pair is allowed to be arbitrary. 
Indeed, consider the event configuration in which each of the two decays in the pair corresponds to the global maximum of the single particle SPT decay  found in Section~\ref{spt}, with arbitrarily 
large transverse momentum. Thus, the pair of decaying particles have equal and arbitrarily large transverse momentum, as measured in the lab frame. Furthermore, they both decay in the same way in their rest frames: for $\chi < m_1$ the invisible daughters are emitted in the forwards direction, and the visible daughter systems of invariant mass $m_N=m_M=m_<$ are emitted in the backwards direction. For $\chi > m_1$, the invisible particles are emitted in the backwards direction in the rest frame.

It is straightforward to show that the process of minimizing over the partitions of the invisible momenta in this configuration singles out the partition with the true values for
the invisible momenta. This shows that this configuration saturates the bound on \MTTWO.

In summary then, we have shown that in the case of identical pair decays, either SPT or ZPT, there is always a kink, except perhaps in \casefour\ ZPT.
In the case of SPT, the extrema correspond to pairwise copies of the global maxima in the single-decay cases, and are the global maximum.
In the case of ZPT, the extrema correspond to pairwise copies of the global maxima in the single-decay cases, but it is not clear whether or not they are the global maximum.\footnote{It is worth remarking that the ZPT configuration in which the parents have arbitrarily large momenta, but are produced back-to-back, does not give a higher maximum. Indeed, the minimization over partitions in this case picks out $\MTTWO =  \surd f = \chi + m_{\lessgtr}$.}

\section{Monte Carlo Simulations} \label{mc}

As a concrete illustration of this method,
we perform simulations of events for particular scenarios,
analysing both single-parent cases, 
and also examples of pair-production of parent particles, as shown in \figref{fig:cases}.

We generate Monte Carlo events by two different methods:
\begin{itemize}
\item{ Firstly, we use the {\tt
HERWIG}\cite{Corcella:2002jc,Moretti:2002eu,Marchesini:1991ch} Monte
Carlo generator, with LHC beam conditions, to produce unweighted
supersymmetric particle pair-production events.  These events have the
advantage that the transverse boost of the particles should bear some
resemblance to that expected for real events, and so we use this
generator when investigating SPT events with a ``physically
reasonable'' $p_T$ distribution.\footnote{We recognise that it will
somewhat underestimate the high-$p_T$ tail, as extra radiation ought
to have matrix element corrections which will not be well modelled by
{\tt HERWIG}'s parton shower.}  We always generate the SUSY particles
in pairs, but when we are considering the kinematics of single
particle decays, we treat each decay independently, ignoring any
missing transverse momentum with its origin in the other `side' of the
event.  }
\item{ Secondly, to investigate the behaviour of the transverse
kinematic endpoints under some unphysical but nevertheless highly
instructive limiting cases, we use a home-grown toy Monte Carlo (MC)
event generator.  This toy MC allows us to either (a) generate events
with the primary particle(s) having {\em exactly} zero net transverse
momentum, thereby allowing us to examine the ZPT limiting case, or (b)
generate events where the transverse momenta are non-zero (SPT) and
entirely under our control.  In the latter case, except where stated
otherwise, we choose the $p_T$ distribution to have an extremely long
unphysical tail (denoted ``SPT large pt'' in figures) in order to
highlight the effects of large transverse momenta on the kinematic
endpoint distributions.  The toy MC generates all
decays according to phase space alone, implements no showering and
keeps all particles on mass shell.  It defines an arbitrary mass scale
$m_S=m_0+m_1$ in terms of which its generation scheme is fixed as
follows.  In the single-particle production cases with ZPT, the parent
particle 0 starts its life at rest in the lab frame.  In the ZPT
pair-production cases, particles 0 and 5 are generated back to back
from the isotropic decay of an initial state at rest in the laboratory
frame with centre-of-mass energy $m_0+m_5+T$, where the kinetic energy
$T$ differs from event-to-event, being an exponentially distributed
random variable with mean $m_S$.  In the single-particle production
cases with SPT, the parent particle and an associated state with mass
$m_\mathrm{ISR}$, against which it recoils, are produced back-to-back
from an initial state at rest in the lab-frame which has energy $m_0 +
T$.  Both $m_\mathrm{ISR}$ and the kinetic energy $T$ differ from
event to event, being random variables: $m_\mathrm{ISR}$ is
distributed exponentially with mean $m_S$, while the kinetic energy
$T$ is distributed as the absolute value of a
Cauchy-distributed\footnote{There are two isolated cases in the paper
where we replace this Cauchy distribution with an Exponential
distribution with mean $m_S$, thus significantly reducing the high
$p_T$ tail to a more physically reasonable size.  We do this on the
two occasions in the paper in which we wish to use the toy MC to
produce a less extreme and more physically relevant SPT distribution
which is intermediate between its usual ZPT and usual SPT output.
These two isolated cases are labelled ``SPT (medium pT)'' where
they occur.} random variable with mean 0 and scale $m_S$.  For
pair-production with SPT, the two parent particles and an associated
state with mass $m_\mathrm{ISR}$ against which both jointly recoil,
are produced in the isotropic three-body decay of an initial state at rest in
the lab frame which has energy $m_0 + m_5 + T$, where $m_\mathrm{ISR}$
and $T$ have the same distributions as they had for single-particle
SPT production.}
\end{itemize}
\subsection{Single production with two-daughter decay (\casetwo)} \label{mc-two}

The predicted gradients at the kink are given  for two-daughter decays as a
function of the kinematic variables of the decay in
\eqref{extremal-gradients}.  We recall that we expect qualitatively
{\em different} behaviour around $\chi=m_1$ depending on whether or
not the parent has transverse momentum in the lab frame.  To be
concrete: we expect to see a kink in the $m_T^\mathrm{max}(\chi)$ curve when the
parents can have transverse momentum but do not expect to see a kink
when parents are only produced at rest.

\begin{figure}
  \begin{center}
    \begin{minipage}[b]{.4\linewidth}
      \begin{center}
        \includegraphics[width=0.99\linewidth]{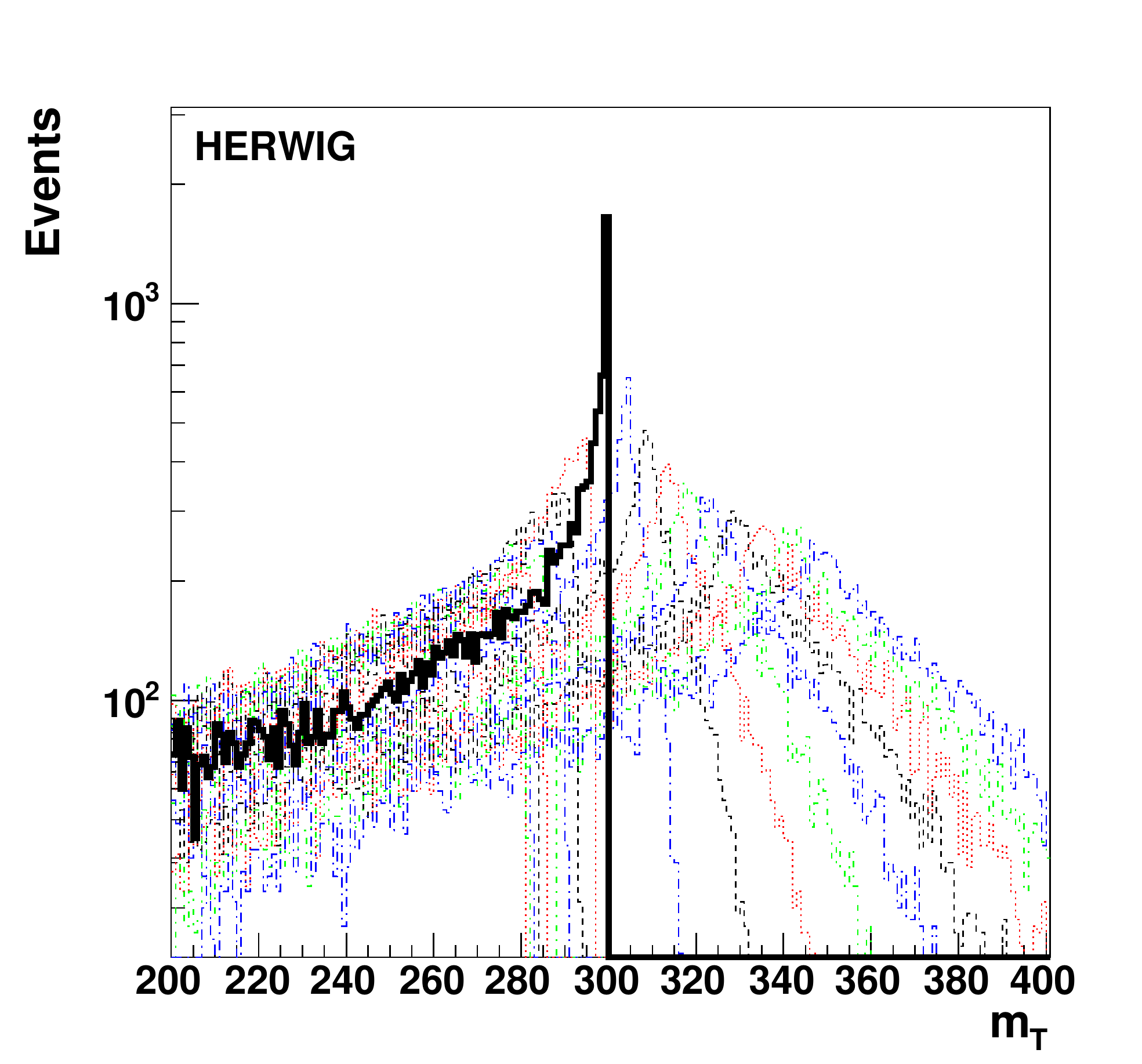}\\
        {\bf (a)}
      \end{center}
     \end{minipage}
   \begin{minipage}[b]{.4\linewidth}
     \begin{center}
       \includegraphics[width=0.99\linewidth]{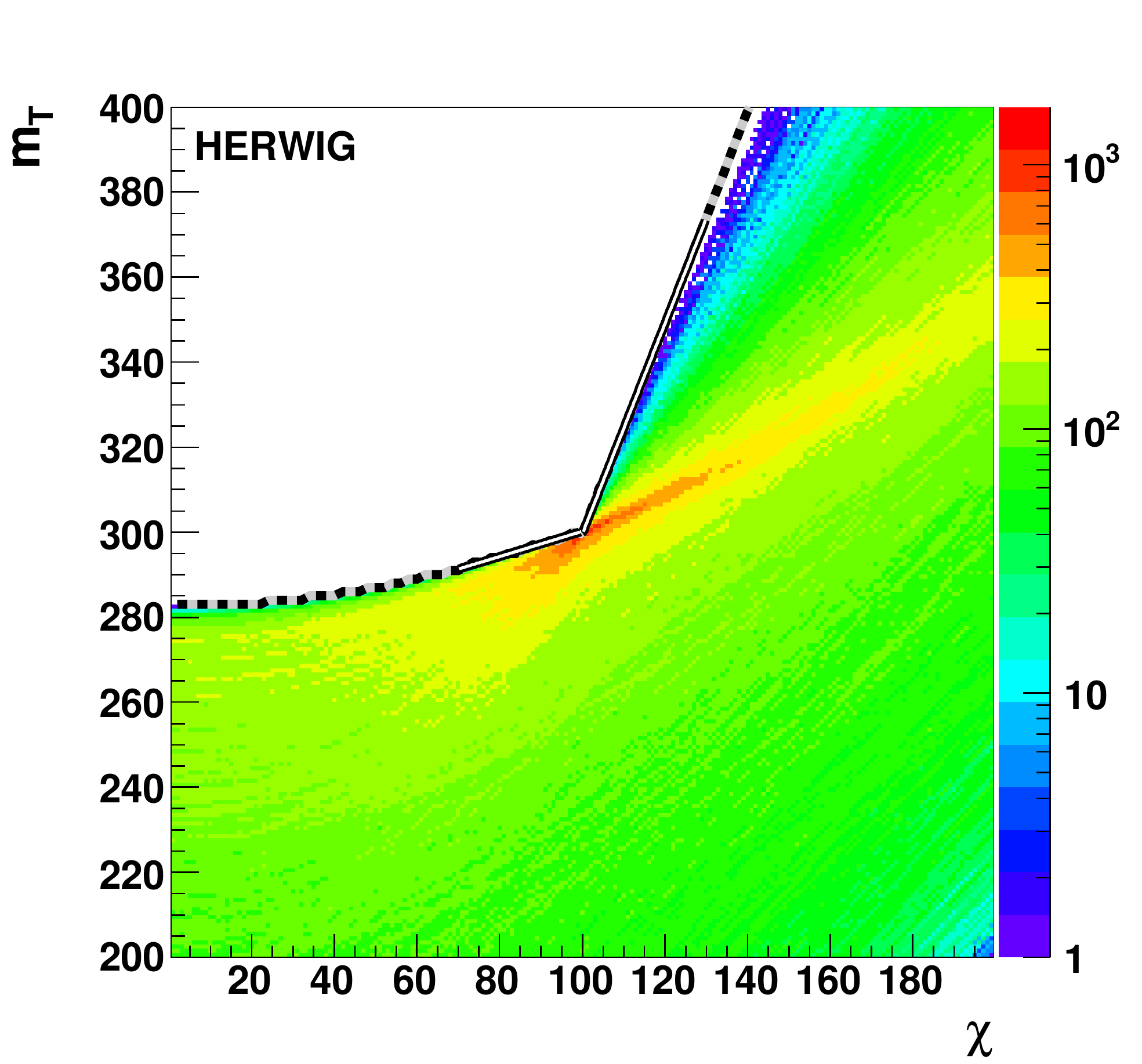}\\
       {\bf (b)}
     \end{center}
   \end{minipage}
\\
\begin{minipage}[b]{.4\linewidth}
     \begin{center}
       \includegraphics[width=0.99\linewidth]{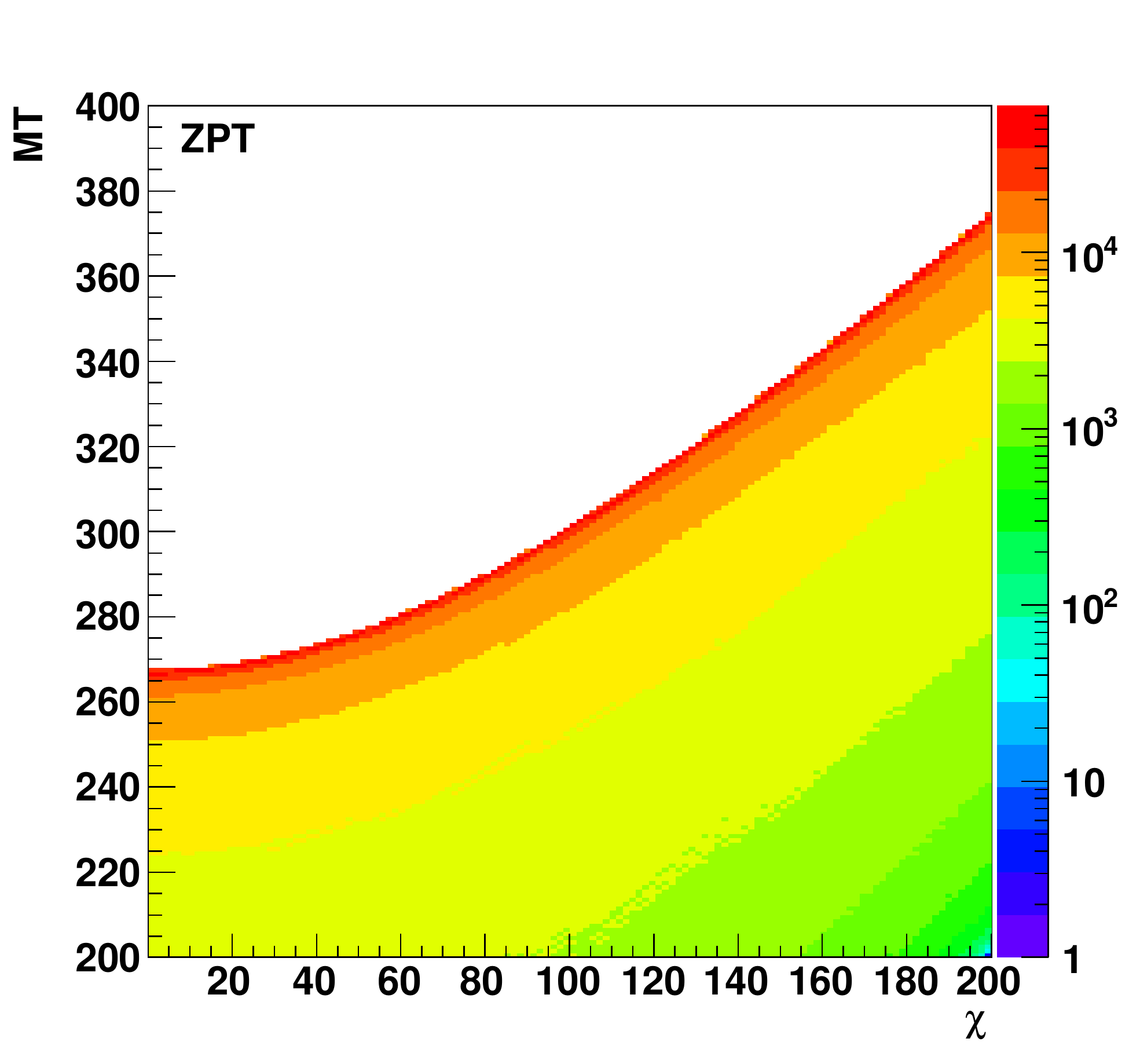}
      {\bf (c)}
     \end{center}
\end{minipage}
\begin{minipage}[b]{.4\linewidth}
     \begin{center}
       \includegraphics[width=0.99\linewidth]{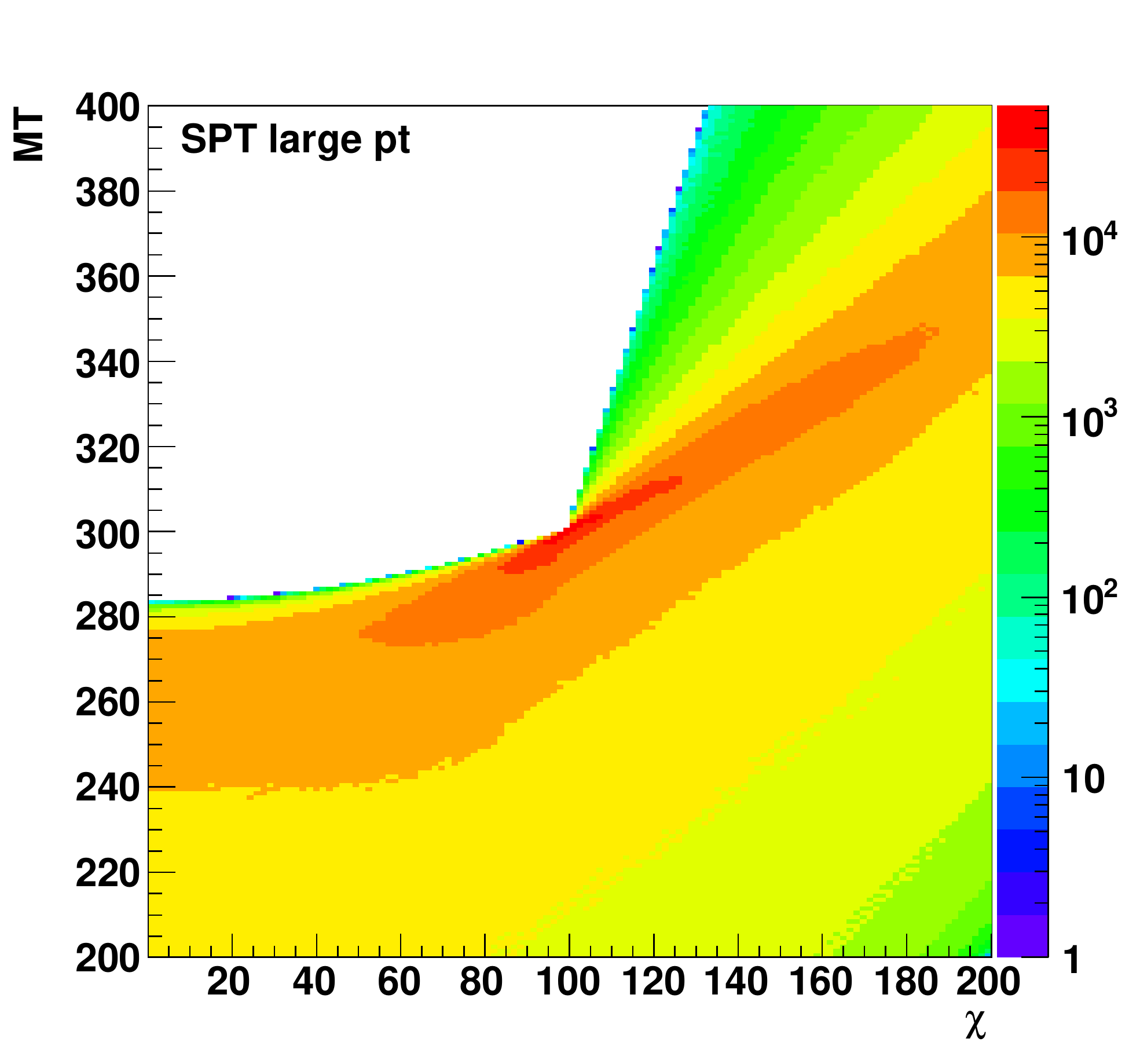}
      {\bf (d)}
     \end{center}
\end{minipage}
\caption{(\casetwo)
{\bf (a)} Example distributions of the transverse mass, $m_T(\chi) = \surd f(\chi)$, for the 
two-daughter decay with $m_0 = 300$, $m_1 = 100$, and $m_2=0$ ({\tt HERWIG}).
The different histograms represent the same events but have different values of $\chi$. 
The thick solid line is for $\chi = m_1$, while the thinner broken lines are for other values of $\chi$ uniformly spanning the range $0<\chi<200$. 
{\bf (b-d)} 
The plots show the distribution of the number of events (colour scale) which produce particular values of $m_T(\chi) = \surd f(\chi)$ ($y$-axis) for different values of $\chi$ ($x$-axis).
{\bf (b)} {\tt HERWIG} events. 
The upper limit, $m_T^\mathrm{max}(\chi)$ is indicated as a dashed line.
The short solid lines are fits to $m_T^\mathrm{max}(\chi)$ above and below $\chi=m_1$.
{\bf (c)} For the phase space Monte Carlo in which the mother has been constrained to be at rest in the lab frame (ZPT).
{\bf (d)} For the phase space Monte Carlo in which the mother can have large transverse momentum (SPT) in the lab frame.
\label{fig:two}
}
  \end{center}
\end{figure}

We generated {\tt HERWIG} events with slepton ($\tilde{\ell}$) pair production, followed by the decay
$\tilde{\ell}\rightarrow\ell\tilde{\chi}_1^0$. 
The slepton mass was 300~GeV, and the $\tilde{\chi}_1^0$ mass was 100~GeV. The slepton width was set to be much less than 1~GeV.
The visible lepton is sufficiently light that it can be assumed massless in this context.
We show in \figref{fig:two}a several distributions of $m_T(\chi) = \surd f(\chi)$ for different values of $\chi$.
We note that the upper kinematic limit is well saturated, making an experimental determination of the endpoint a reasonable expectation
for this case.

In \figref{fig:two}b we show the same information in a different format, now plotting the
number of events (colour scale) with particular values of $m_T(\chi)$ ($y$-axis) as a function of $\chi$ ($x$-axis).
Using these distributions, we estimate the value of the maximum value, $\surd \hat{f}(\chi)$ 
from the mid-point of the uppermost non-zero bin, which we denote $m_T^\mathrm{max}(\chi)$.
A graph of  $m_T^\mathrm{max}(\chi)$ as a function of $\chi$ shows clearly the kink at at $(\chi=m_1$, $m_T=m_0)$; this is shown as a dashed line 
superimposed on \figref{fig:two}b.
The gradients $\left.\frac{d m_T^\mathrm{max}}{d\chi}\right|_{\chi=m_1\pm\varepsilon}$ 
were estimated from short line fits to that graph below and above the kink, 
and yielded gradients of 0.325 below and 2.56 above the kink.
The predicted extremal values calculated for the example masses, using \eqref{extremal-gradients} and converted to
\begin{gather}
\left.\frac{d\surd \hat{f}}{d\chi}\right|_{\chi=m_0\pm\varepsilon} = \left.\frac{\chi}{m_0}\frac{d\hat{f}}{d\chi^2}\right|_{\chi=m_0\pm\varepsilon},
\end{gather}
are $\frac{1}{3}$ for $\chi<m_1$ and 3 for $\chi>m_1$. 
As discussed in Section \ref{boost}, we expect the measured value to be closer to the predicted one below the kink as compared to above it, 
since the high energies required mean that a much smaller proportion of events lie near the global maximum above the kink, as compared to below it.

Note that the value of $m_T^\mathrm{max}$ we have used for the gradient determination is generated from the
{\em single} maximal event at each value of $\chi$. In an experiment which observes significantly fewer similarly-distributed events
than we have generated ($10^4$ in \figref{fig:two}a-b), 
one would be unlikely to find an event at this limit.
A typical experimenter observing, say, a factor of ten fewer events without background
(or for whom e.g. ten events are required to get a statistically significant determination of the position of the end-point 
due to the presence of backgrounds)
might expect to observe a maximum a little below our dotted within the neighbouring cyan regions of the plot.

If we restrict the parent to have no transverse momentum in the lab
frame (\figref{fig:two}c, ZPT) we find that the kink disappears.  In
its place we see a smooth curve passing through the point where the
kink used to be, with the gradient at that point measured to be
$\sim0.60$.  A plot where the parent's transverse momentum, $p_T$, can
take very large values (\figref{fig:two}d, SPT) shows a slightly more
prominent kink than the {\tt HERWIG} plot.  As we increase the
parent's $p_T$ (ZPT$\rightarrow${\tt HERWIG}$\rightarrow$SPT) the
fraction of events contributing to the kink in a significant way
increases.  Both the absence of the kink for \casetwo\ at ZPT and the
increasing number of events found near the kinematic limit $(\surd
\hat{f})$ as $p_T$ (and hence $\beta$) increases confirm the
predictions of \eqref{eq:gengrad}.  The gradients measured on either
side of the kink in \figref{fig:two}d are found to be 0.335 and 2.99.
These gradients are closer to the asymptotically achievable
values of $1/3$ and $3$ than the gradients seen in the {\tt
HERWIG} events, in line with expectation.

\subsection{Single production with point-like three-daughter decay (\casethreev)}\label{mc-three}
For the three-daughter decay, the range of allowed values of $m_N$ means that 
we expect to see a kink regardless of whether the parent has any transverse momentum in the lab frame \eqref{eq:gengrad}.

\begin{figure}
  \begin{center}
    \begin{minipage}[b]{.4\linewidth}
      \begin{center}
        \includegraphics[width=0.99\linewidth]{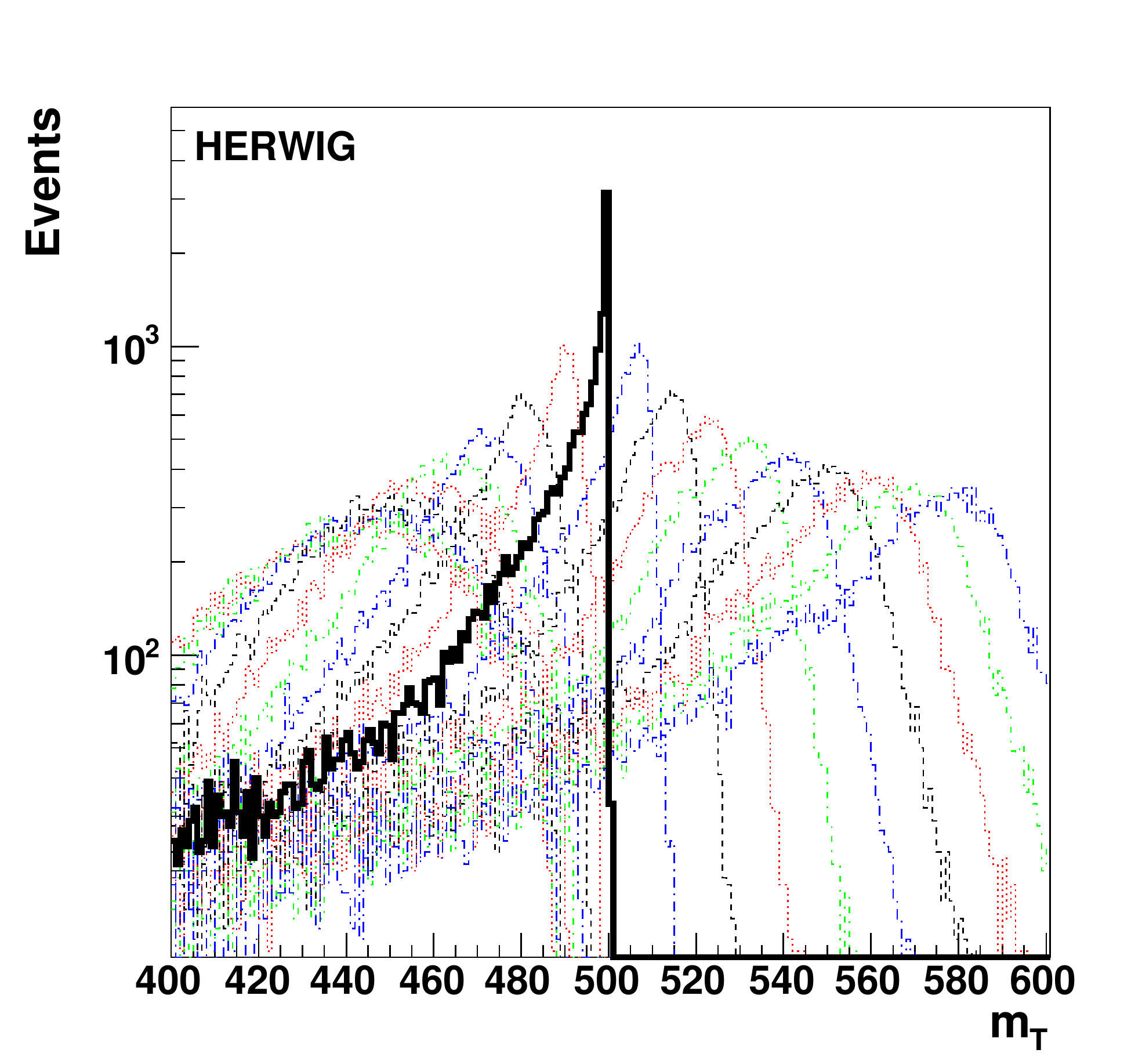}\\
        {\bf (a)}
      \end{center}
     \end{minipage}
   \begin{minipage}[b]{.4\linewidth}
     \begin{center}
       \includegraphics[width=0.99\linewidth]{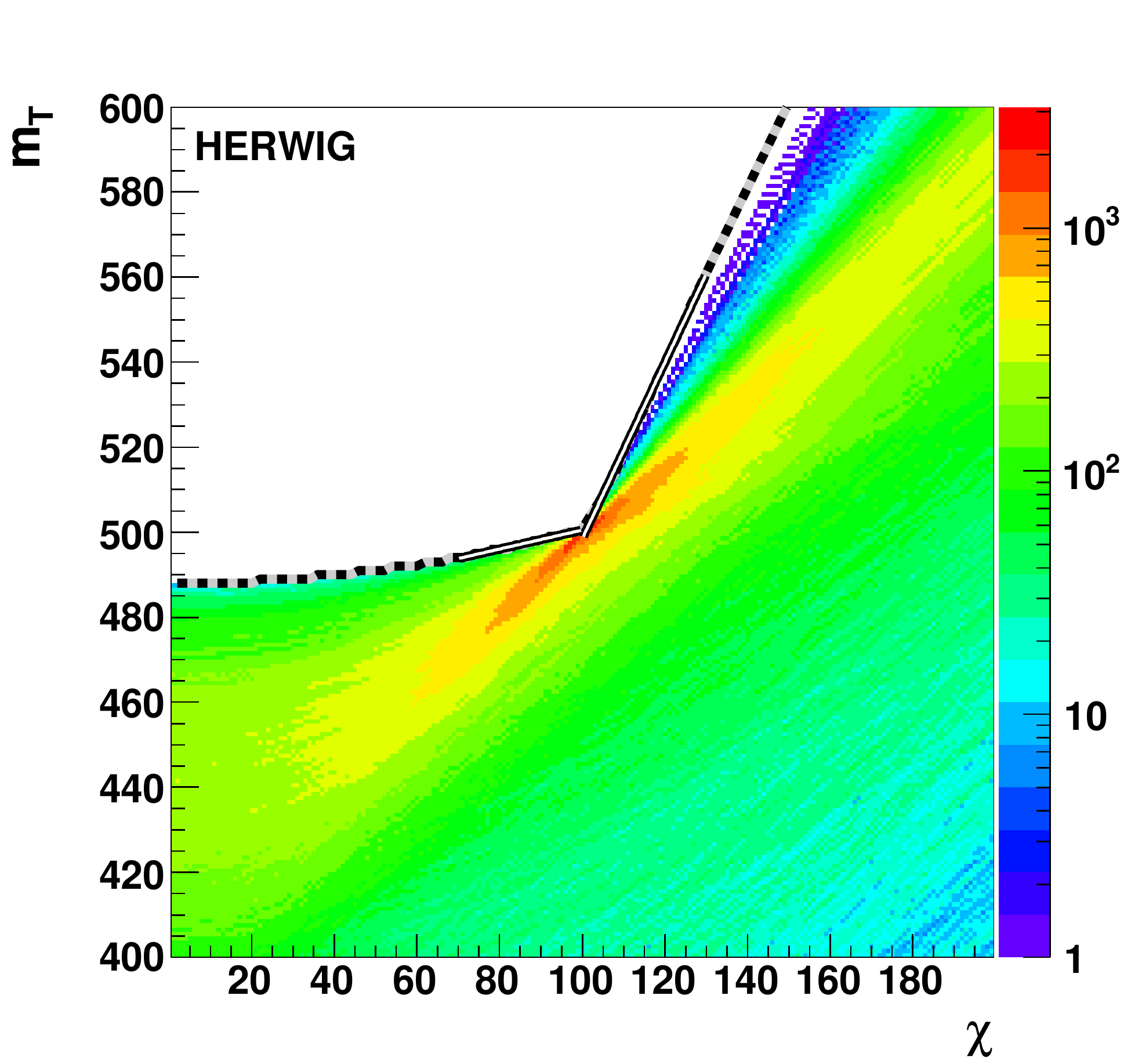}\\
       {\bf (b)}
     \end{center}
   \end{minipage}
\\
   \begin{minipage}[b]{.4\linewidth}
     \begin{center}
        \includegraphics[width=0.99\linewidth]{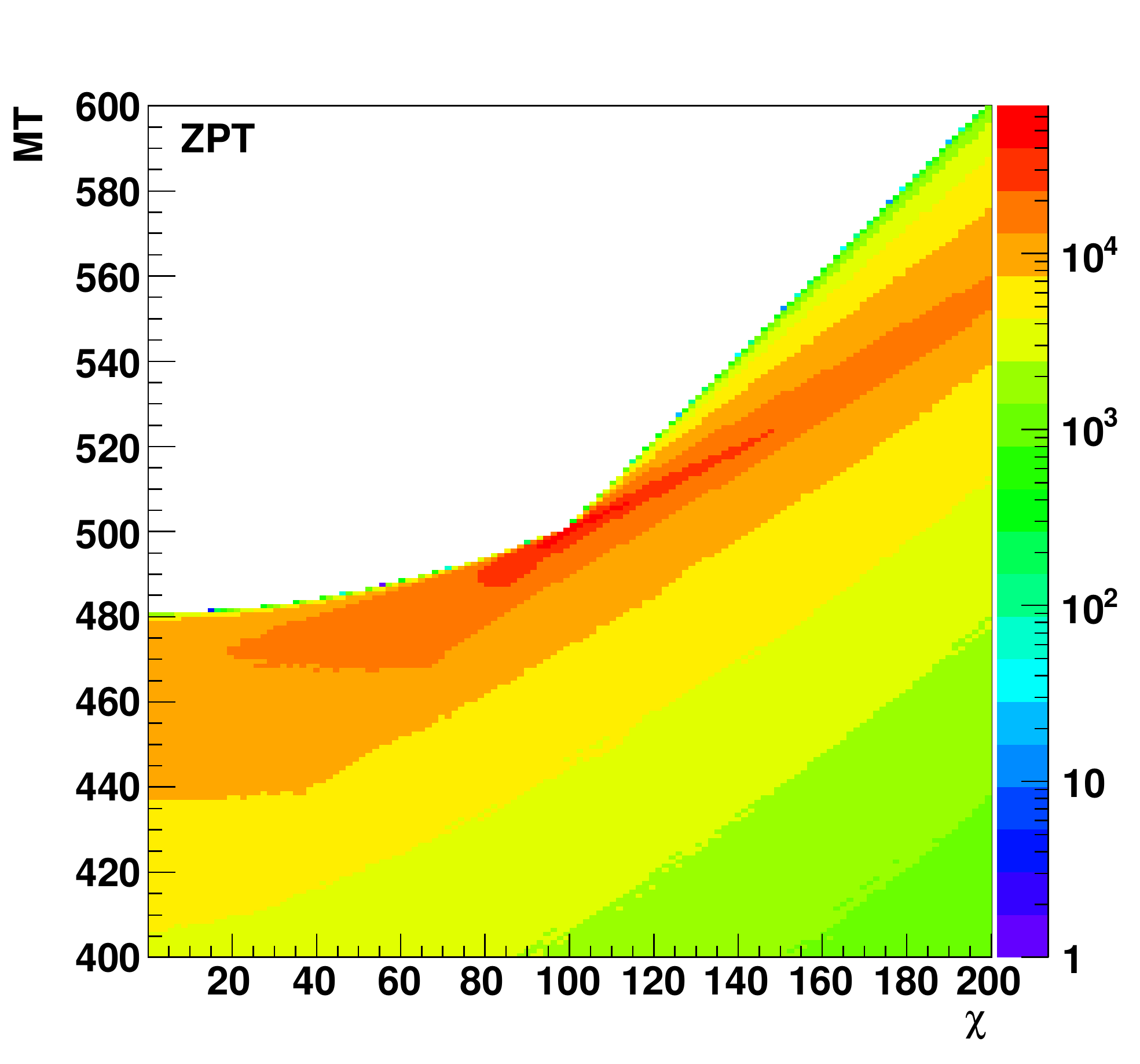}
	{\bf (c)}
     \end{center}
   \end{minipage}
   \begin{minipage}[b]{.4\linewidth}
     \begin{center}
        \includegraphics[width=0.99\linewidth]{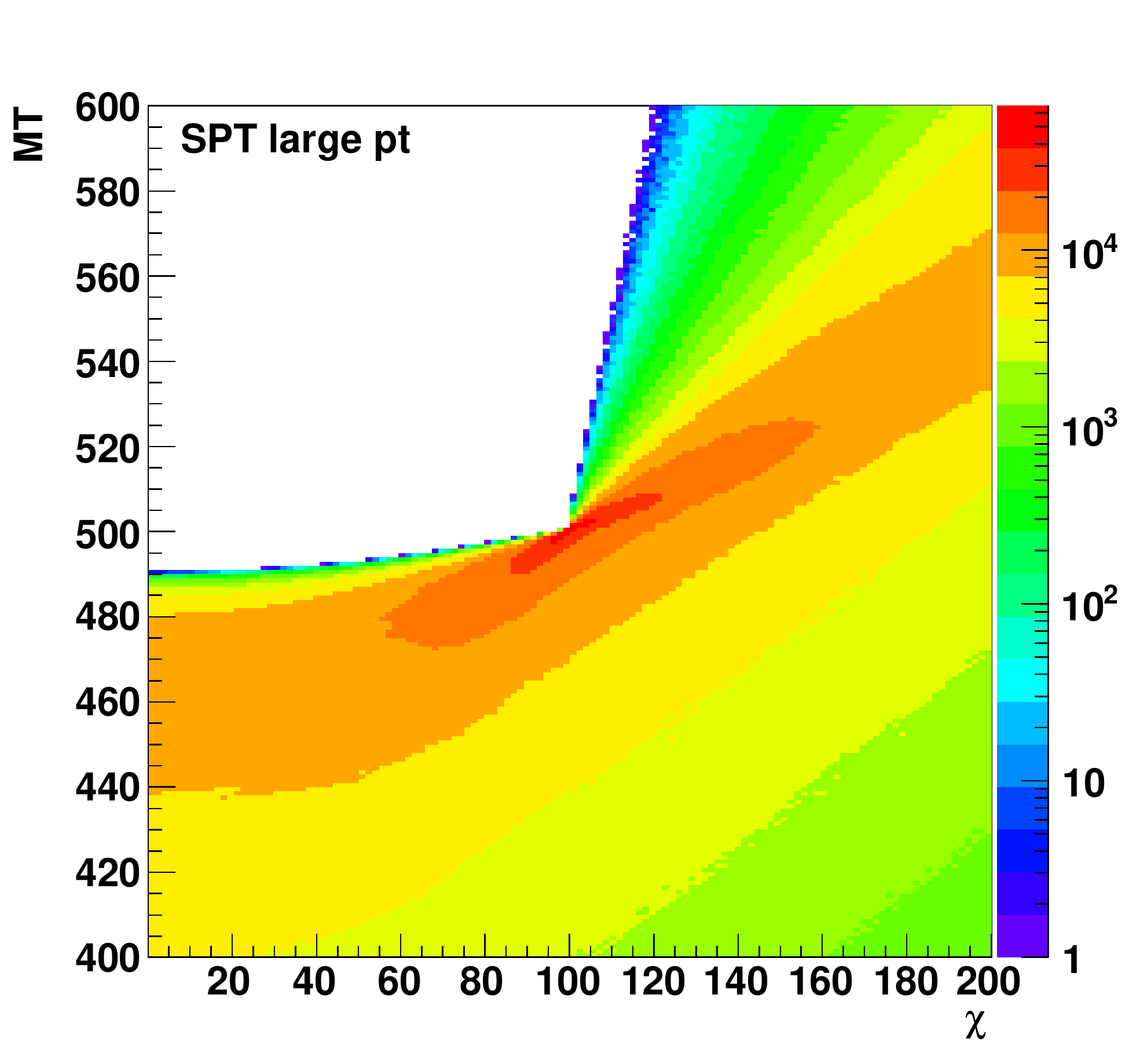}
        {\bf (d)}
     \end{center}
   \end{minipage}
 \end{center}
\caption{(\casethreev)
As for \figref{fig:two} but for the three-daughter 
decay, and  with $m_0 = 500$, $m_1 = 100$, and $m_{2,3}=0$.
\label{fig:three-v}
}
\end{figure}

As for Section \ref{mc-two}, we generate events with {\tt HERWIG}, but this time for the initial process of gluino pair production.
The gluinos then decay according to $\tilde{g}\rightarrow q \bar{q} \tilde{\chi}_1^0$. The mass of the gluino was 500~GeV, 
and that of the $\tilde{\chi}_1^0$, 100~GeV. 
Again, the quarks can be considered massless for our purposes, since for simplicity
we only allow decays to light quarks.

In \figref{fig:three-v}a we plot distributions of $m_T = \surd f$ for
this three-daughter decay for different values of $\chi$.  
The upper limit is well saturated, and the edge of this distribution,
$m_T^\mathrm{max}(\chi)$ also shows clearly the kink at $(\chi=m_1$,
$m_T=m_0)$ (\figref{fig:three-v}b).  The fitted gradients $\frac{d m_T^\mathrm{max}}{d\chi}$
are 0.223 below and 2.1 above the kink.
The predicted gradients are 1/5 below the kink
and 5 above the kink for SPT events, and $5/13 \simeq 0.3846$ and 1, respectively, for ZPT events.  The observed gradient on either side of the kink lies between the SPT global maximum
and the ZPT maximum. This reflects the fact that the gluino has finite, but non-zero, $p_T$
in the lab frame. As expected, we get closer to the SPT maximum below the kink than we do above it.

As predicted by the analysis, for the three-daughter case, the kink is
retained in ZPT events where the parent is constrained to have no
transverse momentum in the lab frame (\figref{fig:three-v}c).  The
gradients $\frac{d m_T^\mathrm{max}}{d\chi}$ measured either side of
the kink in \figref{fig:three-v}c are 0.385 and 0.998 which are both
very close to their predicted values.

Also, as expected, if the parent can have very large transverse
momentum (SPT), the kink is even more pronounced
(\figref{fig:three-v}d).  In this case the measured gradients below
and above the kink are found to be respectively 0.214 and 4.878.  Both
gradients have changed from the values measured in the {\tt HERWIG}
events, coming to within about 5\% of the limiting values
(attainable only with infinite momenta)
predicted earlier.

\subsection{Single production with three-daughter cascade decay (\casethrees)}
\begin{figure}
  \begin{center}
    \begin{minipage}[b]{.32\linewidth}
      \begin{center}
        \includegraphics[width=0.99\linewidth]{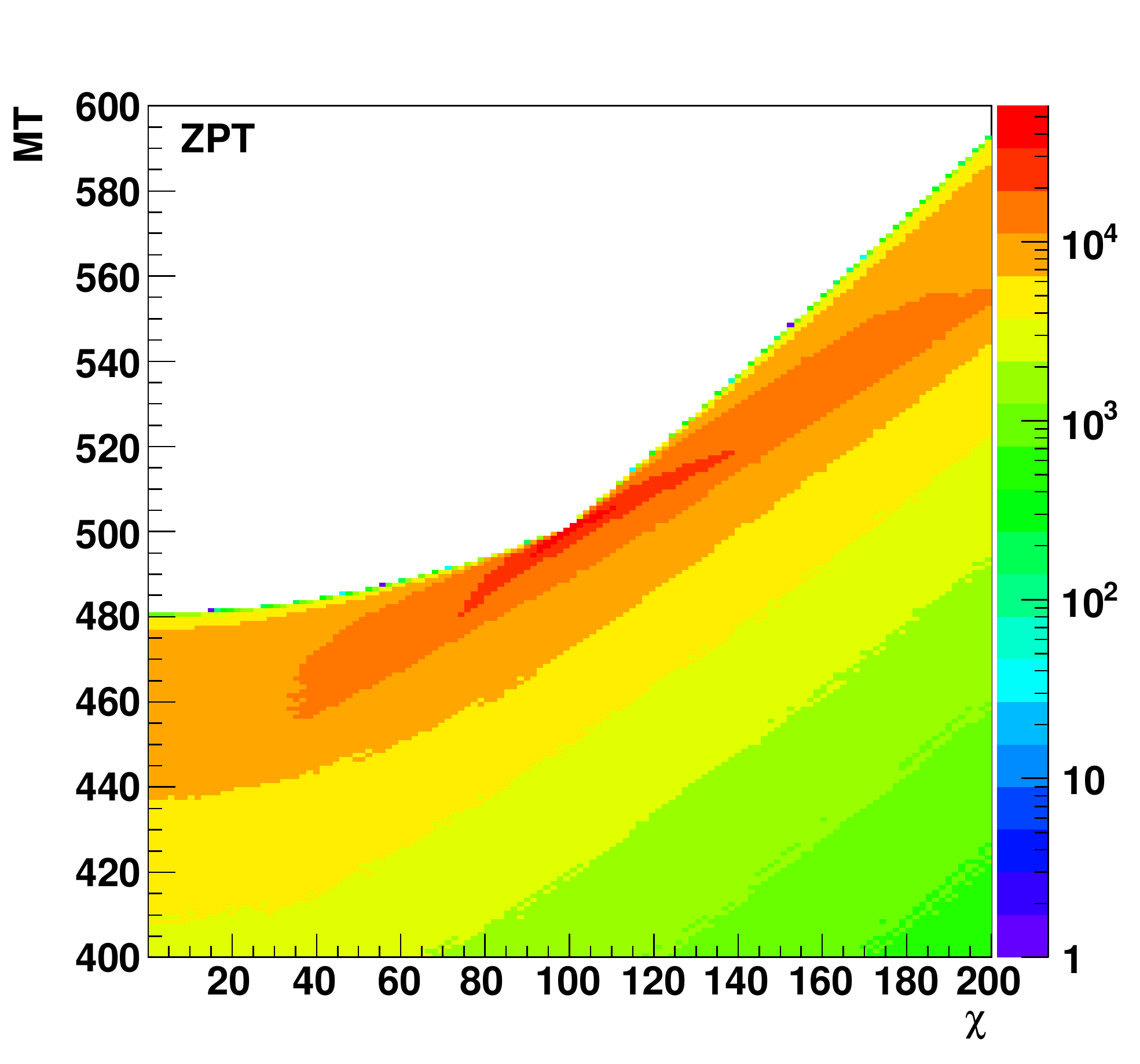}\\
			{\bf (a)}
      \end{center}
    \end{minipage}
    \begin{minipage}[b]{.32\linewidth}
      \begin{center}
	\includegraphics[width=0.99\linewidth]{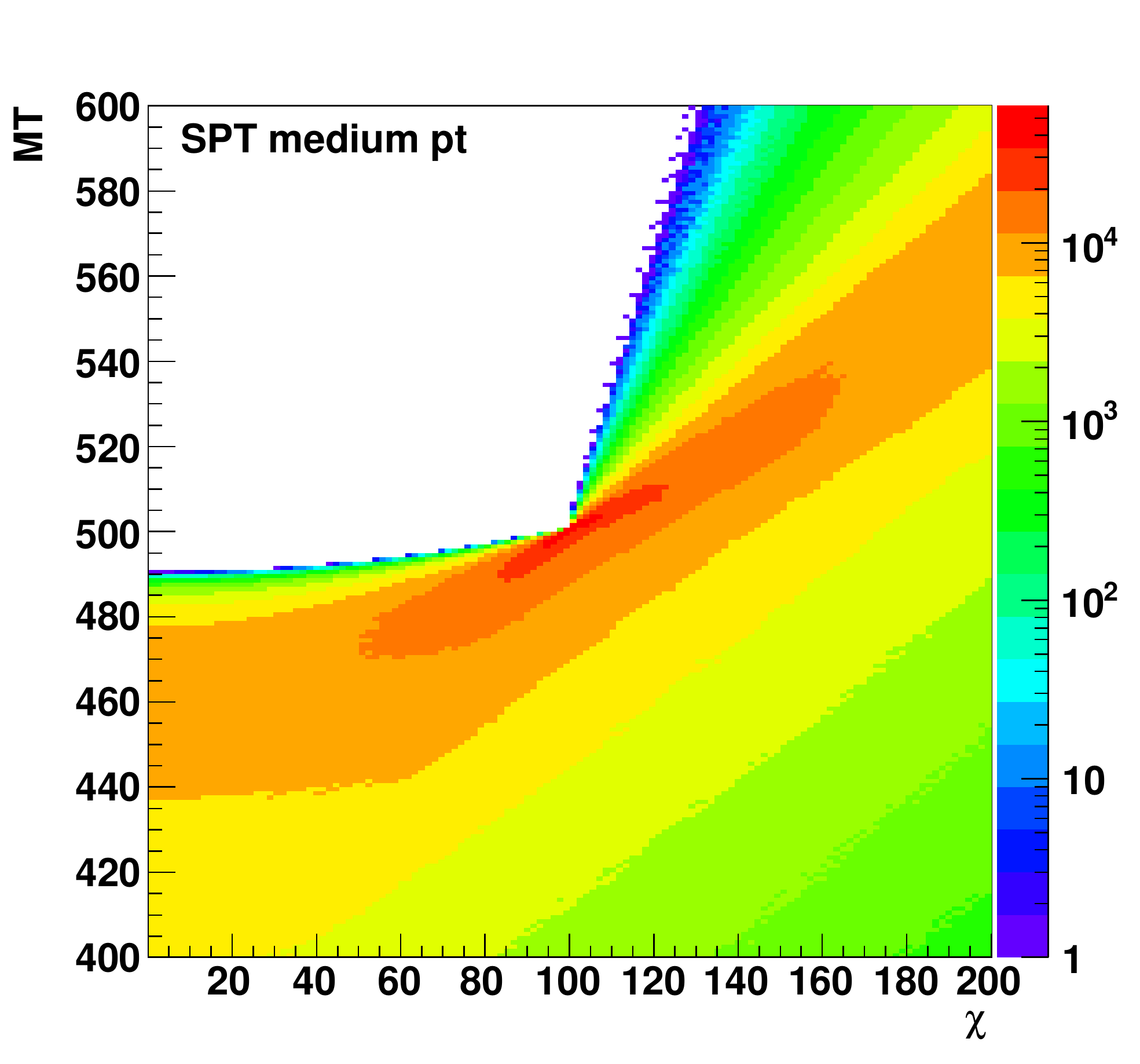}\\
			{\bf (b)}
      \end{center}
    \end{minipage}
    \begin{minipage}[b]{.32\linewidth}
      \begin{center}
	\includegraphics[width=0.99\linewidth]{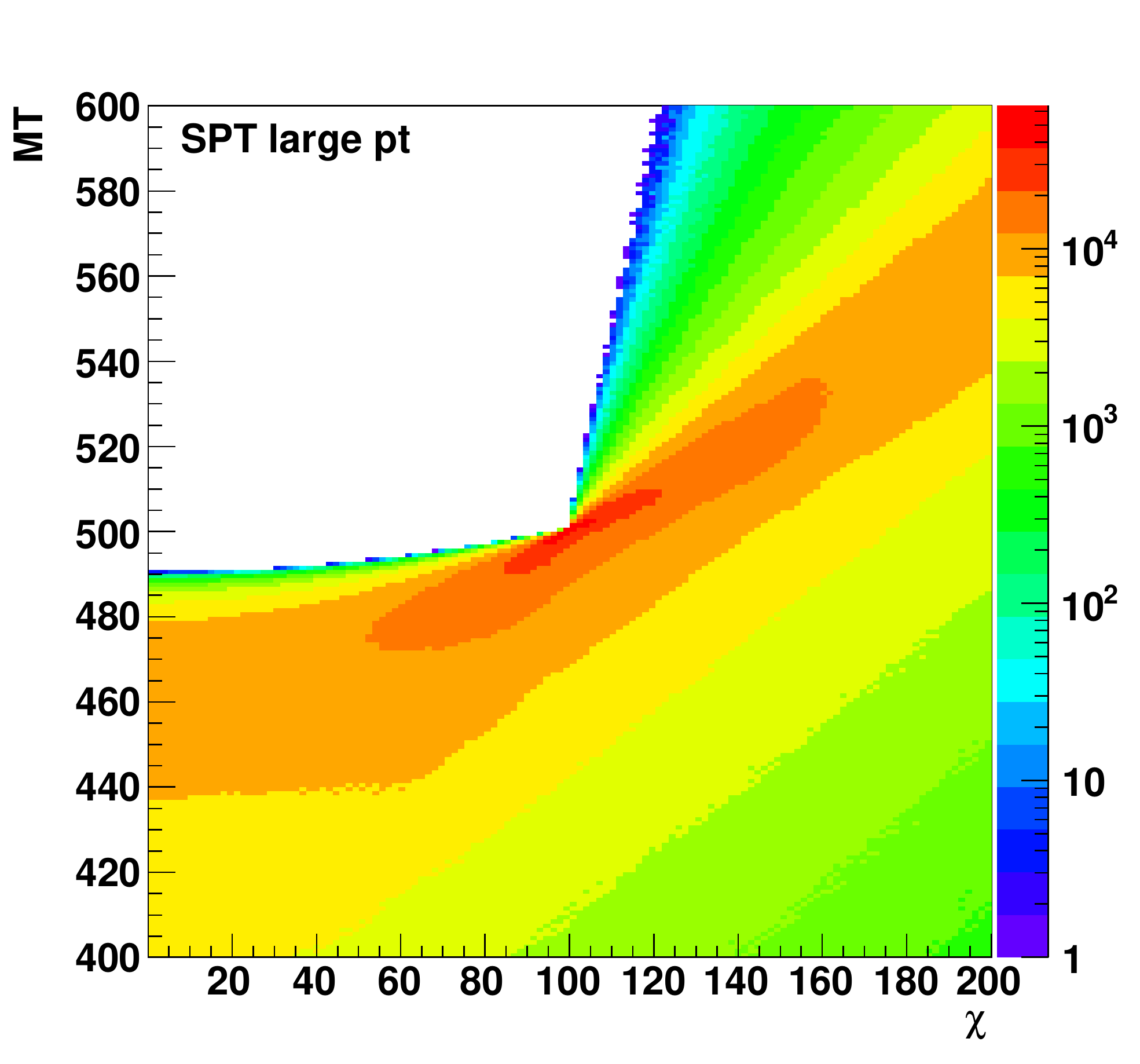}\\
			{\bf (c)}
      \end{center}
    \end{minipage}
  \end{center}
 \caption{(\casethrees) These figure show the $m_T(\chi)$ distributions
   obtained from phase-space Monte-Carlo-generated events containing a
   single parent ($m_0=500$) undergoing a sequence of two-body decays
   via an on-shell intermediate particle ($m_4=300$) resulting in a
   final-state containing two massless visible daughters ($m_2=m_3=0$)
   and an invisible massive daughter ($m_1=100$).
    {\bf (a)} ZPT  (zero pt),
    {\bf (b)} SPT  (medium pt), 
    {\bf (c)} SPT  (large pt), 
    \label{fig:three-s}
  }
\end{figure}

The gradients on either side of $\chi=m_1$ for
three-daughter {\em cascade} decays (\casethrees) are similar to \casethreev\ ({\em direct} decays),
except that the mass-shell requirement of the intermediate particle places an extra
constraint, narrowing the range of allowed values of $m_N$.
Provided that none of the decays are near-threshold,
the range of values of $m_N$ will be significant, and 
we therefore expect to see kinks in \casethrees\ for both SPT and ZPT.
We show the ZPT, SPT (medium pt) and SPT (large pt) cases in \figref{fig:three-s}, 
for the particular values of $m_0=500, m_1=100, m_2=m_3=0, m_4=300$.
Neither of the decays is near threshold and so, as expected, the kink is 
visible even at ZPT.

\subsection{Pair production with two-daughter decays (\casefour)}
\begin{figure}
  \begin{center}
    \begin{minipage}[b]{.4\linewidth}
      \begin{center}
        \includegraphics[width=0.99\linewidth]{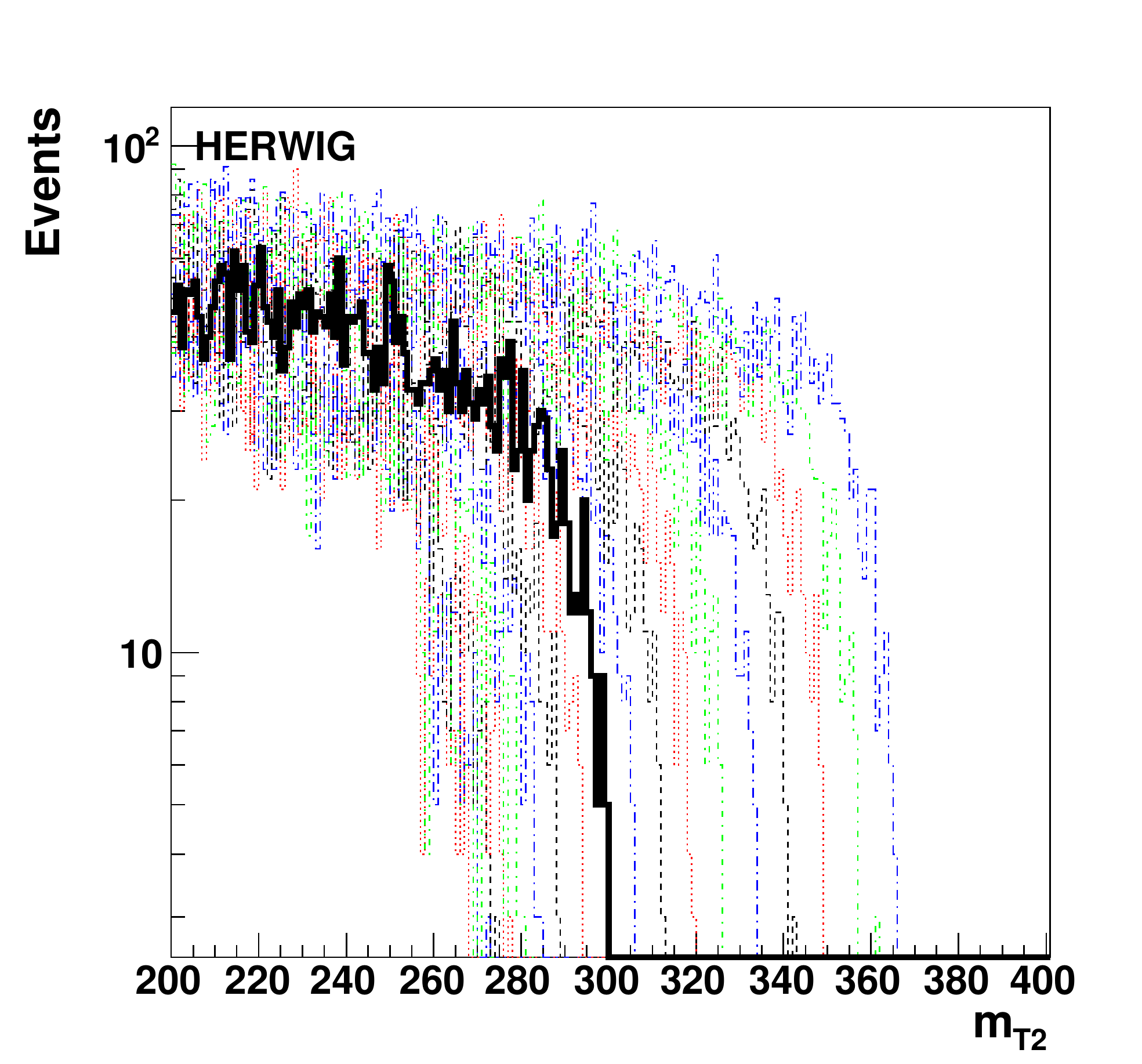}\\
        {\bf (a)}
      \end{center}
     \end{minipage}
   \begin{minipage}[b]{.4\linewidth}
     \begin{center}
       \includegraphics[width=0.99\linewidth]{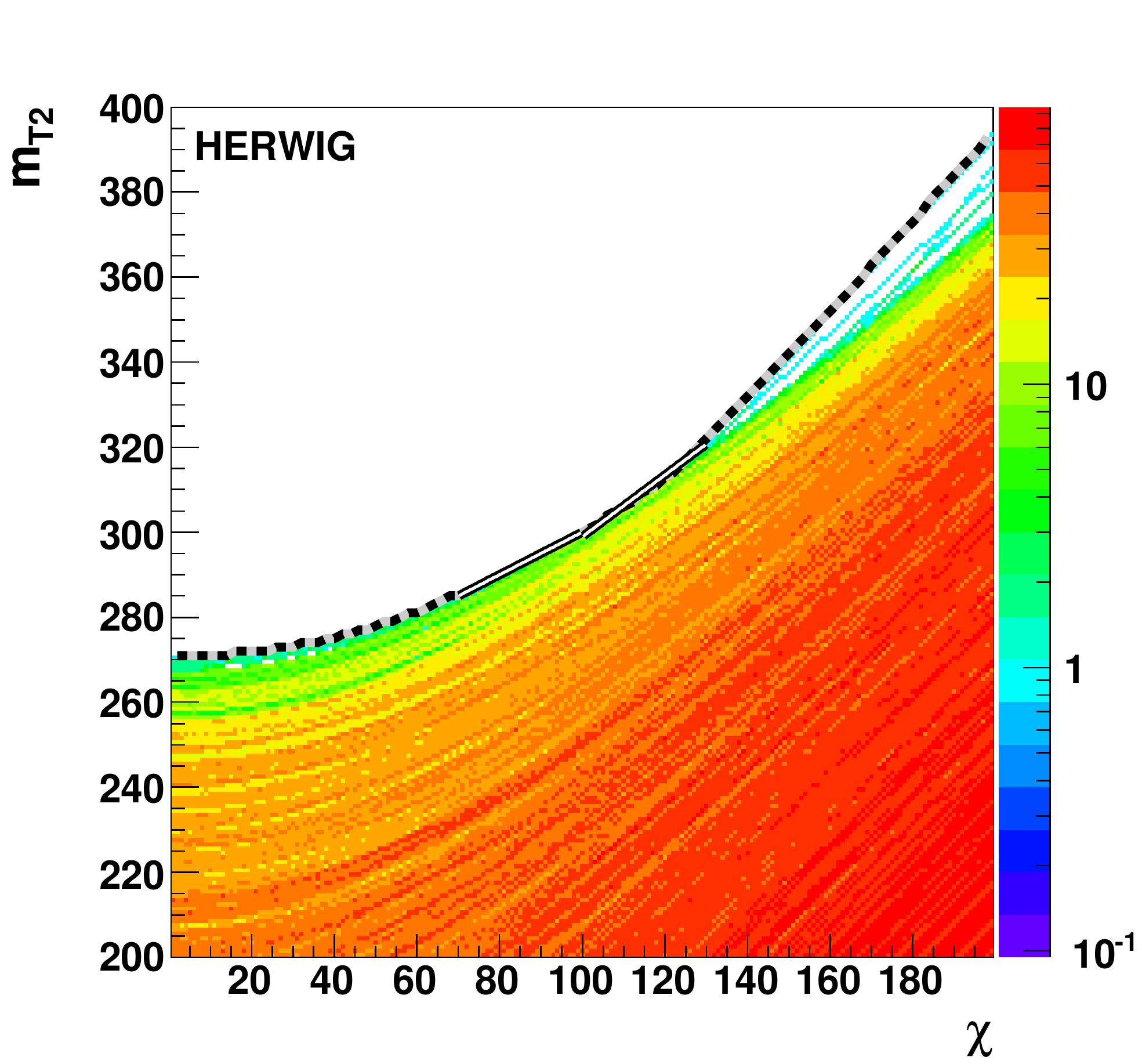}\\
       {\bf (b)}
     \end{center}
   \end{minipage}
\begin{minipage}[b]{.4\linewidth}
     \begin{center}
       \includegraphics[width=0.99\linewidth]{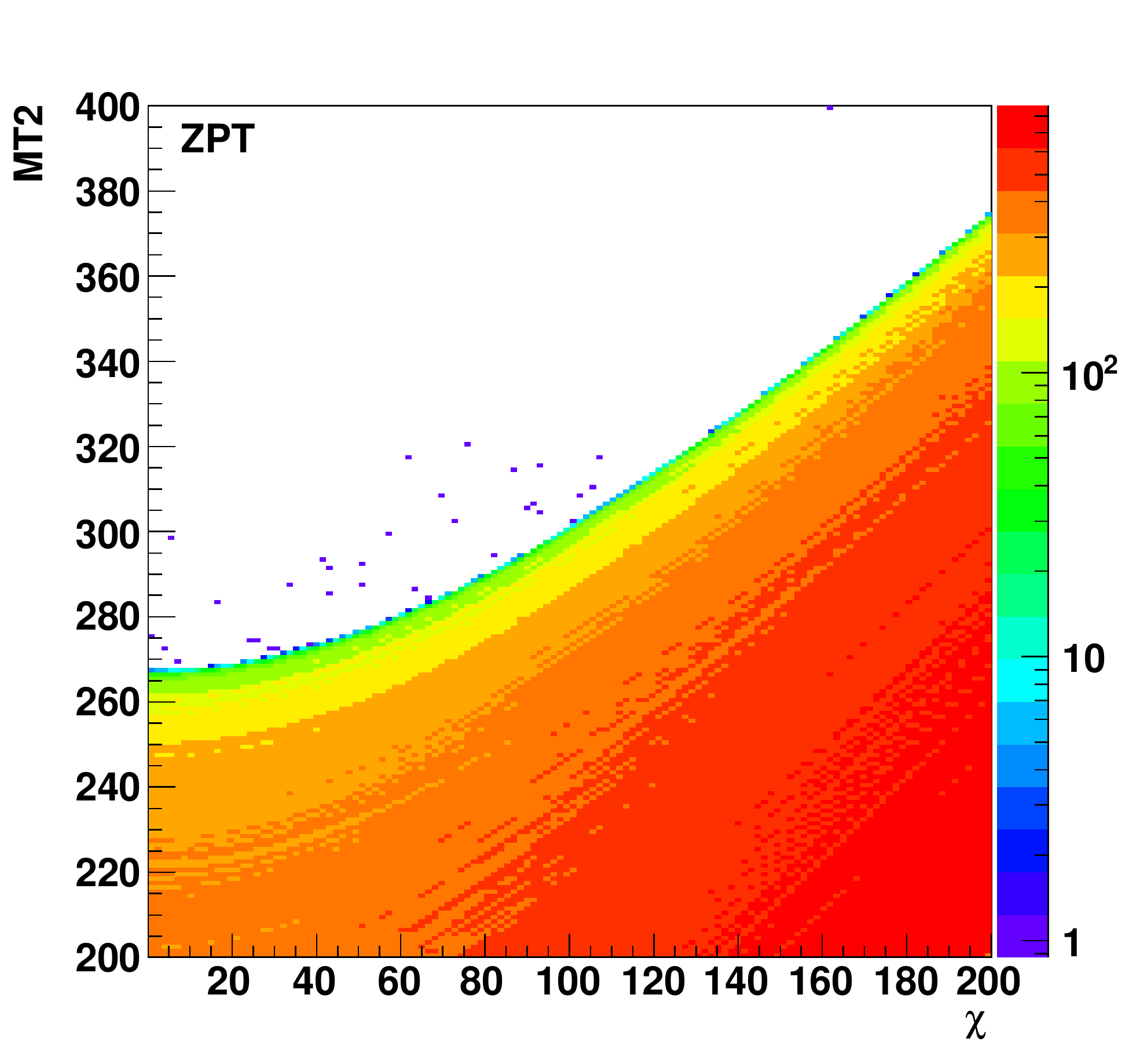} \\
       {\bf (c)}
    \end{center}
   \end{minipage}
\begin{minipage}[b]{.4\linewidth}
     \begin{center}
       \includegraphics[width=0.99\linewidth]{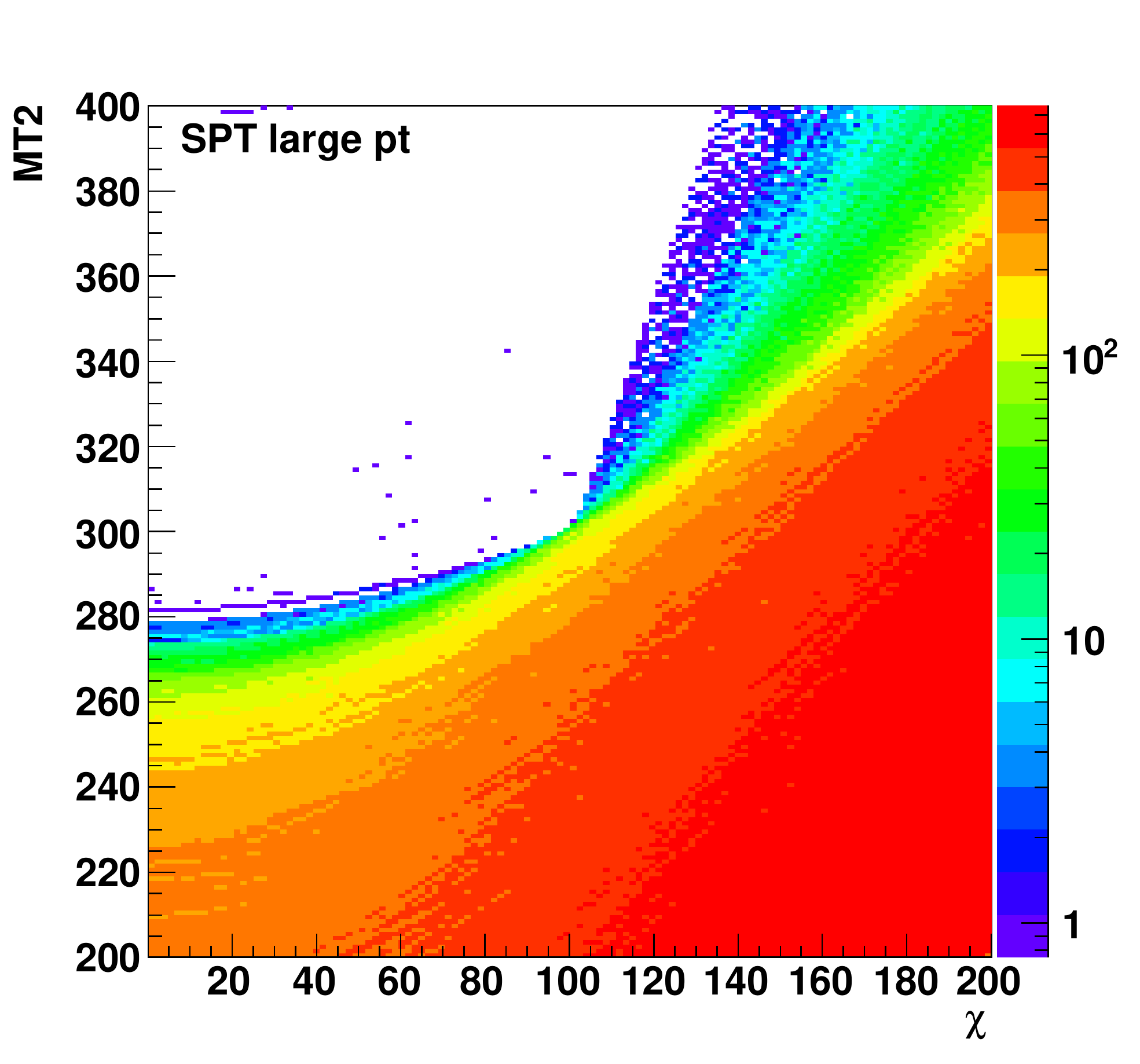} \\
       {\bf (d)}
    \end{center}
   \end{minipage}
\caption{(\casefour)
As for \figref{fig:two} but for pair decays, and so plotting the variable $\MTTWO(\chi)$.
$m_{0,5} = 300$, $m_{1,6} = 100$, $m_{2,7}=0$.
The outlying points in parts (c) and (d) originate from occasional failed numerical minimisations
when calculating \MTTWO\ and have no physical significance.
\label{fig:four}
}
\end{center}
\end{figure}
We conjectured before that the only possible source for a kink in pairs of two-daughter decays (\casefour)
is from large boosts of the sleptons resulting in significant differences in $\beta\cos\theta$. 
In \figref{fig:four}b, we find that any kink in $\MTTWO^\mathrm{max}$ is hardly discernible in pair-decays of
sleptons in our {\tt HERWIG} events. This is to be expected, since to get events close to the SPT maximum for pair decays, we need both parents to be
produced with large energy, and this is unlikely to occur at the LHC.  

When the system of the pair of decaying particles is produced at rest
in the transverse plane (ZPT), there is no visible kink whatsoever
(\figref{fig:four}c).  If finally we construct an unphysically large
$p_T$ distribution for the parents, we confirm that a kink could
``technically'' become visible (\figref{fig:four}d).  However, we note
that even here the kink is not visible at the yellow-green boundary,
and only becomes discernable at the green-cyan boundary, indicating
that the size of the tail in the \MTTWO\ distribution which generates
the kink is very small - about 1\% of the height of the rest of the
distribution.  It seem highly unlikely, therefore, that a kink would
ever be seen in \casefour\ in presence of any background, not even
if the signal cross section was truly gigantic, for detector resolution
would almost almost certainly smear away the details of this tiny tail.

\subsection{Pair production with point-like three-daughter decays (\casesixv) \label{sec:mc6v}}
\begin{figure}
\begin{center}
\begin{minipage}[b]{.4\linewidth}
      \begin{center}
        \includegraphics[width=0.99\linewidth]{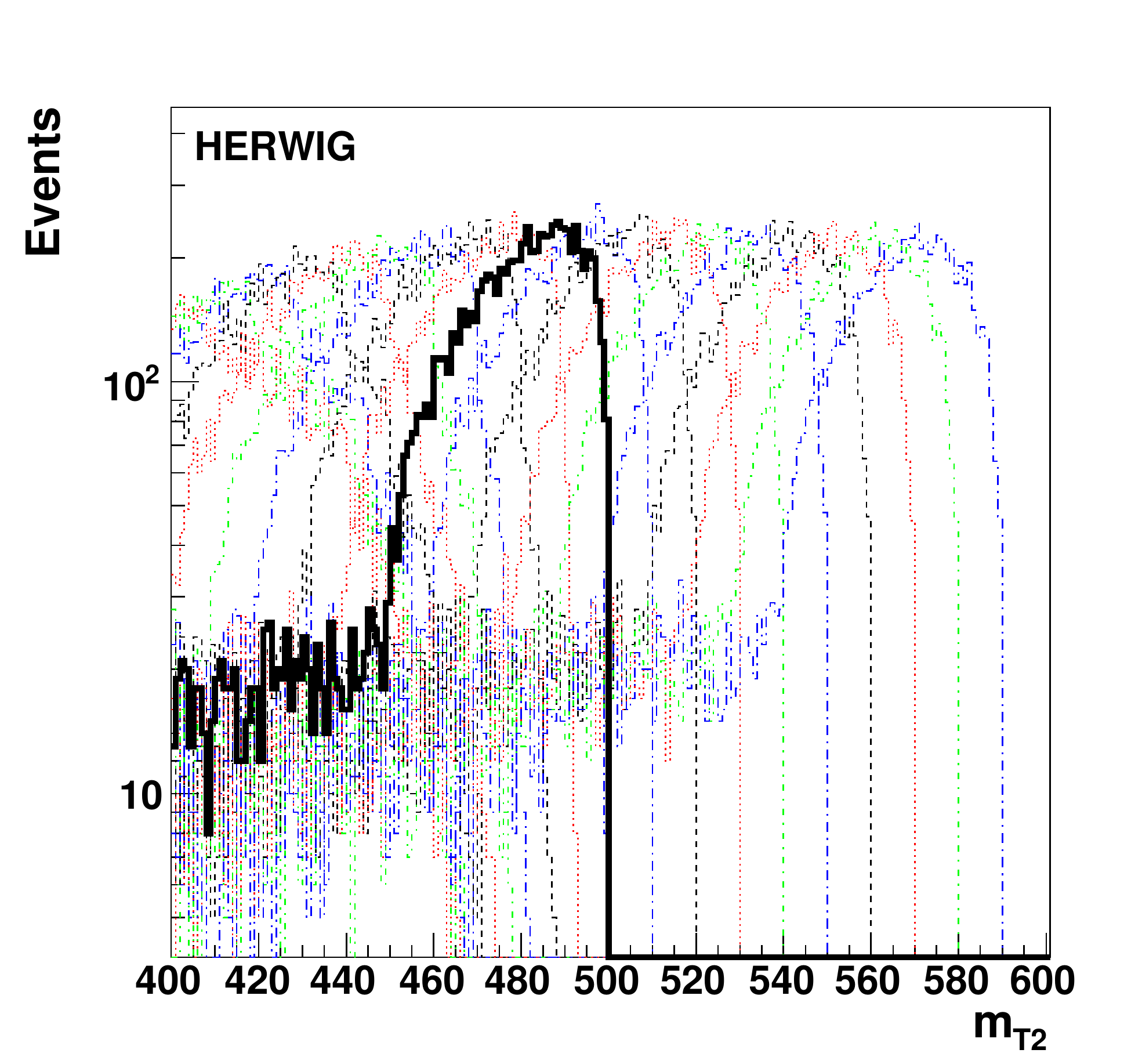}\\
        {\bf (a)}
      \end{center}
     \end{minipage}
   \begin{minipage}[b]{.4\linewidth}
     \begin{center}
       \includegraphics[width=0.99\linewidth]{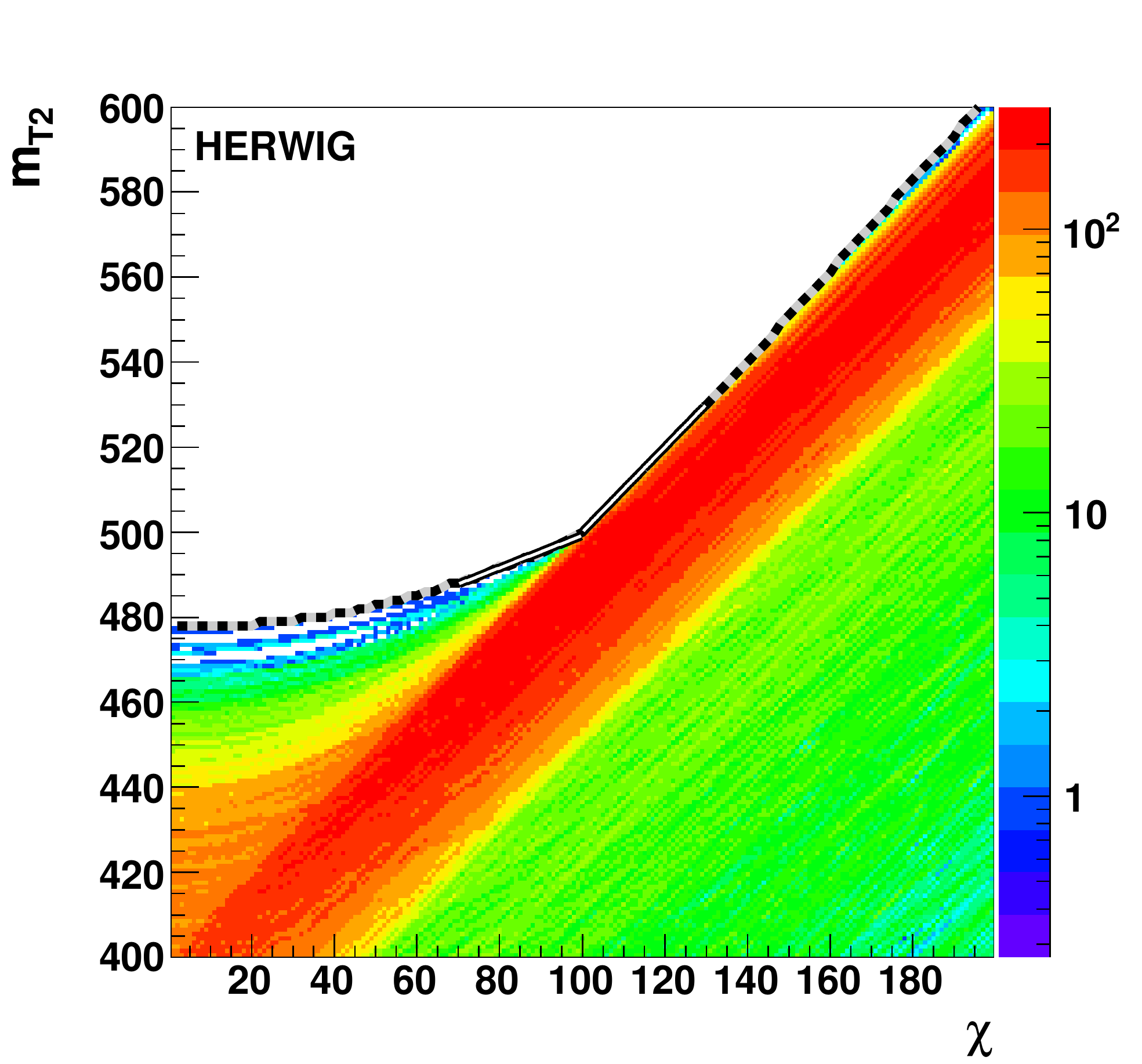}\\
       {\bf (b)}
     \end{center}
   \end{minipage}
\\
   \begin{minipage}[b]{.4\linewidth}
     \begin{center}
        \includegraphics[width=0.99\linewidth]{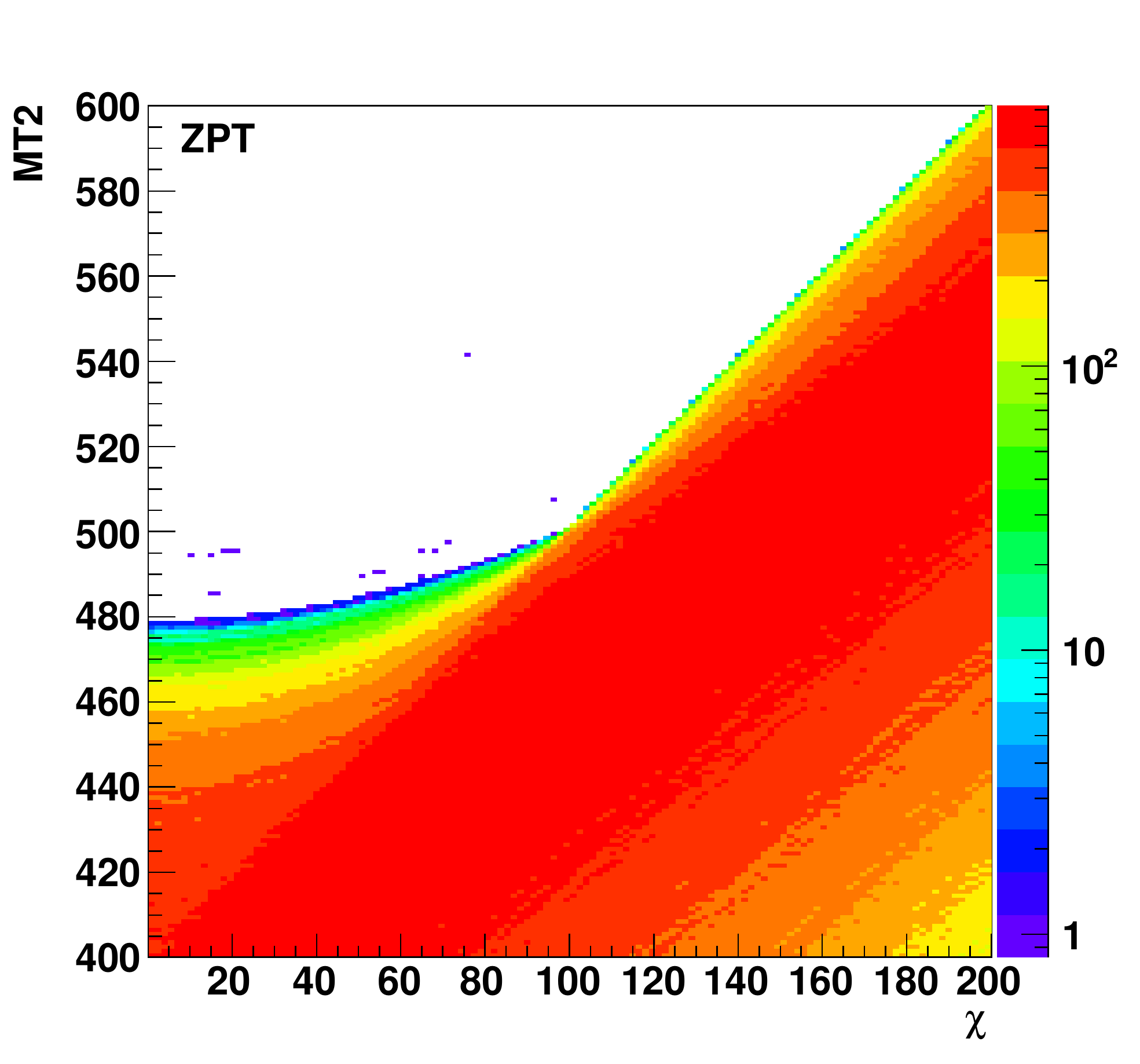}
	{\bf (c)}
     \end{center}
   \end{minipage}
   \begin{minipage}[b]{.4\linewidth}
     \begin{center}
        \includegraphics[width=0.99\linewidth]{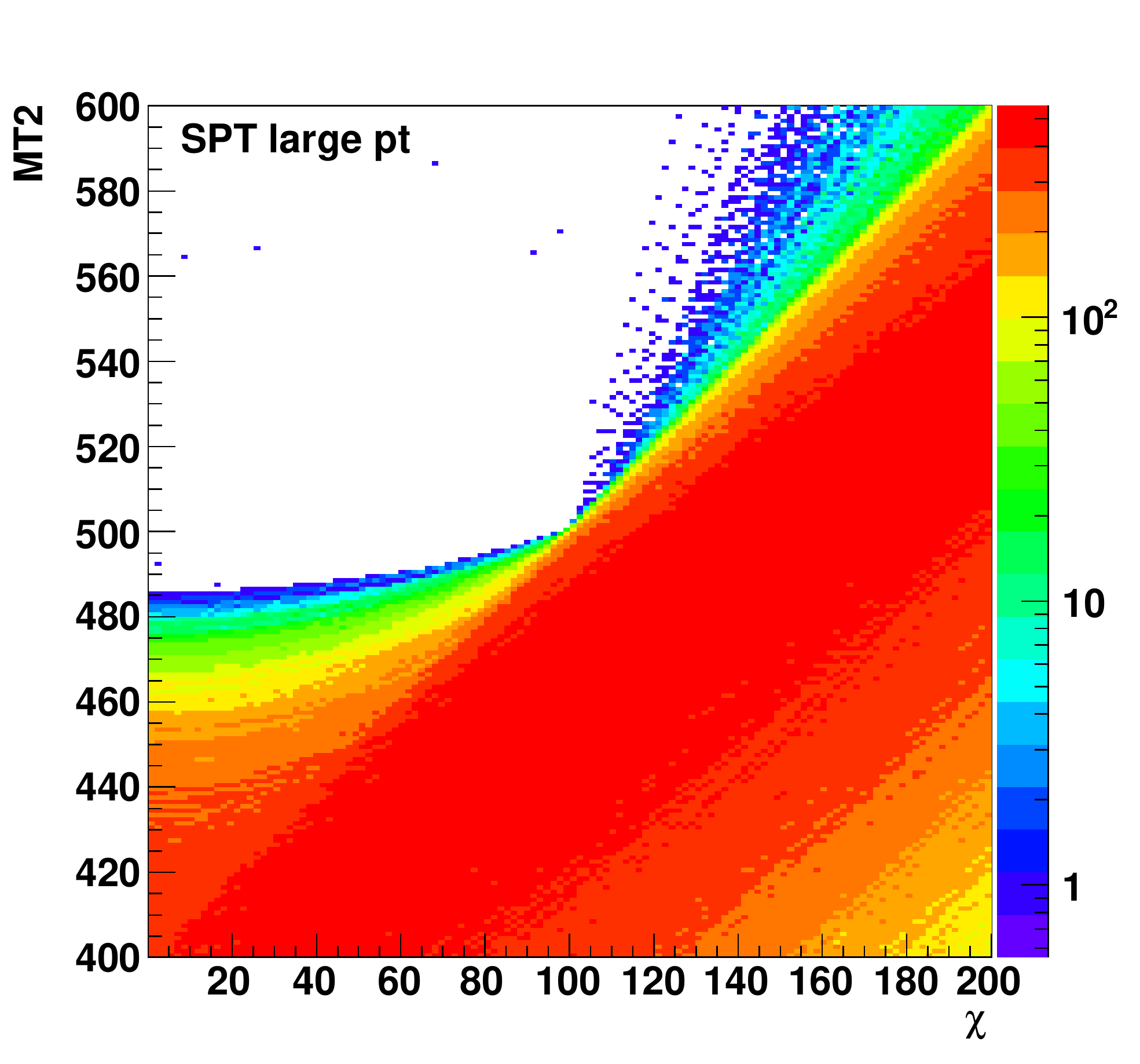}
        {\bf (d)}
     \end{center}
   \end{minipage}
 \end{center}
\caption{(\casesixv)
As for \figref{fig:three-v} but plotting the variable \MTTWO\ for a pair
of three-daughter decays.
$m_{0,5} = 500$, $m_{1,6} = 100$, and $m_{2,3,7,8}=0$.
\label{fig:six-v}
}
\end{figure}


It was in a pair of three-daughter decays that the presence of a kink was first noticed \cite{Cho:2007qv}.
For our simulations of this case we use the same events as in Section \ref{mc-three} but now we treat the pair of decays together,
recognising that if the parents are pair-produced, then the transverse momenta of the invisible 
daughters are not individually known.
For the ({\tt HERWIG}) SPT case, the $\MTTWO$ distribution (\figref{fig:six-v}a) shows a large number of events near the upper edge.
The kink at $\chi=m_1$ is clear (\figref{fig:six-v}b) and the gradients below and above the edge are found
to be 0.38 and 1.00, respectively, to be compared to the SPT global maxima of 0.2 and 5, respectively, and to the ZPT maxima of
$5/13 \sim 0.38$ and 1, respectively. Note that, despite the presence of non-zero $\mathbf{p}_T$, we find that the observed values correspond very closely to the ZPT maxima, suggesting that the relevant events are very close to threshold.

For the ZPT case we also generate events each containing a pair
of virtual three-body decays, but this time we use the toy Monte Carlo
to impose the requirement that the sum of the parent particles has
zero transverse momentum \figref{fig:six-v}c.  Again we see a kink.

Notice that the \casesixv\ results contrast with those from
\casefour.  In \casefour\ a kink was only seen in SPT, whereas in
\casesixv\ kinks are seen in both SPT and ZPT.  This means that events
containing pairs of three-body decays stand a better chance of
generating observable kinks at the LHC than do events containing pairs
of two-body decays, for the former can generate kinks without
needing to gain large transverse boosts from ISR, while the latter need
these boosts.

\begin{figure}
\begin{center}
  \begin{minipage}[b]{.4\linewidth}
    \begin{center}
      \includegraphics[width=0.99\linewidth]{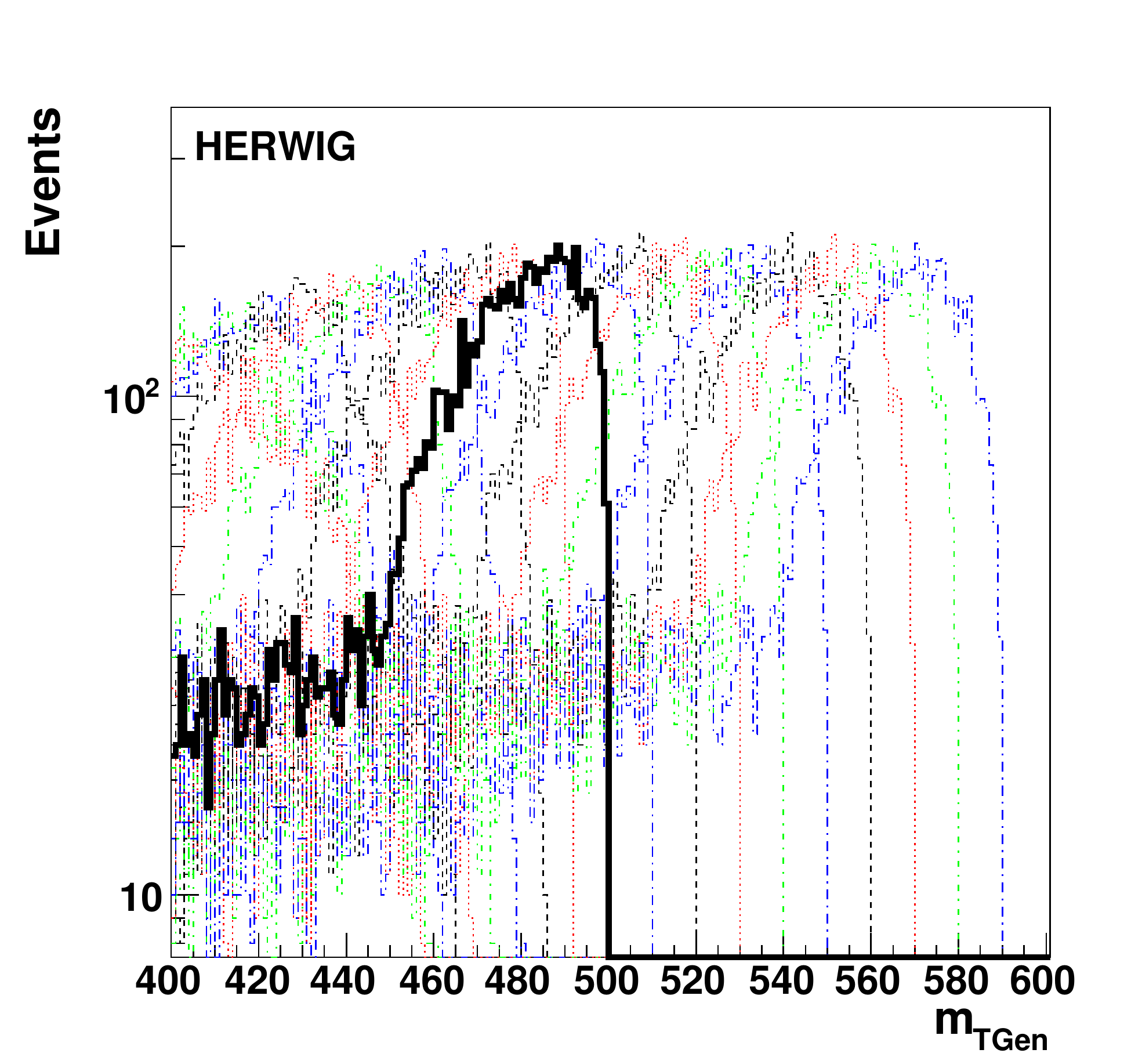}\\
		      {\bf (a)}
    \end{center}
  \end{minipage}
  \begin{minipage}[b]{.4\linewidth}
    \begin{center}
      \includegraphics[width=0.99\linewidth]{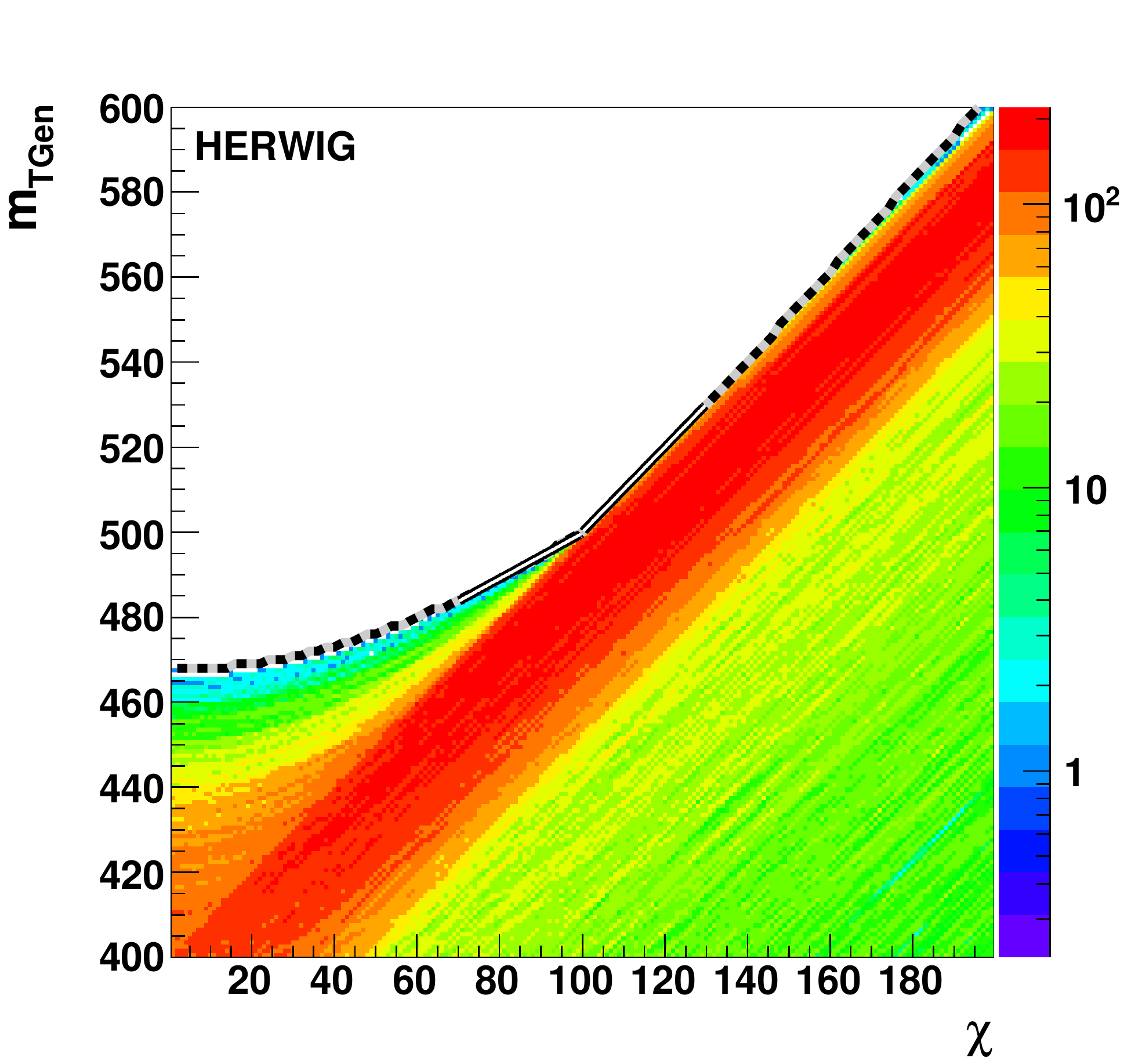}\\
		      {\bf (b)}
    \end{center}
  \end{minipage}
\\
   \begin{minipage}[b]{.4\linewidth}
     \begin{center}
        \includegraphics[width=0.99\linewidth]{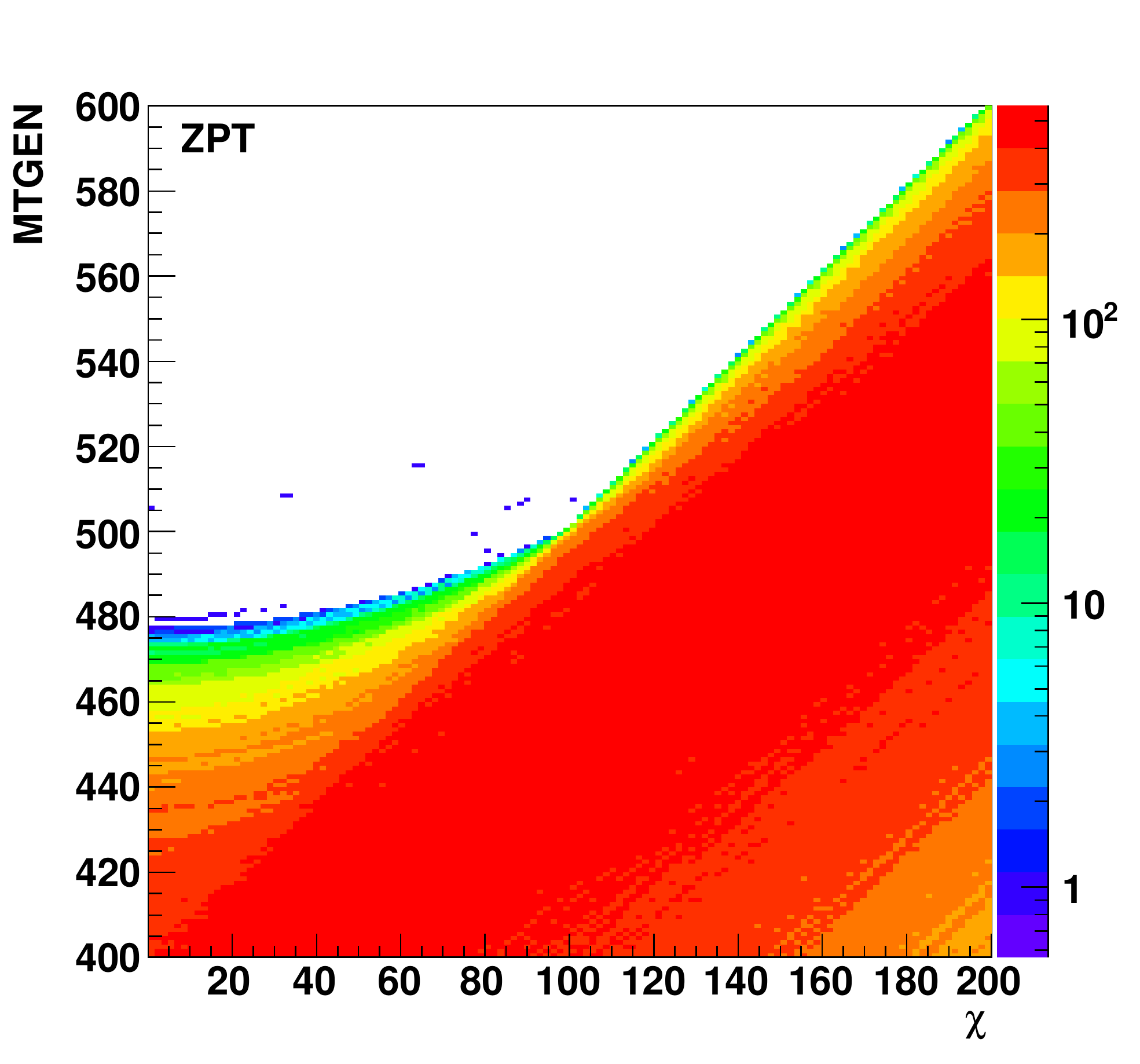}
	{\bf (c)}
     \end{center}
   \end{minipage}
   \begin{minipage}[b]{.4\linewidth}
     \begin{center}
        \includegraphics[width=0.99\linewidth]{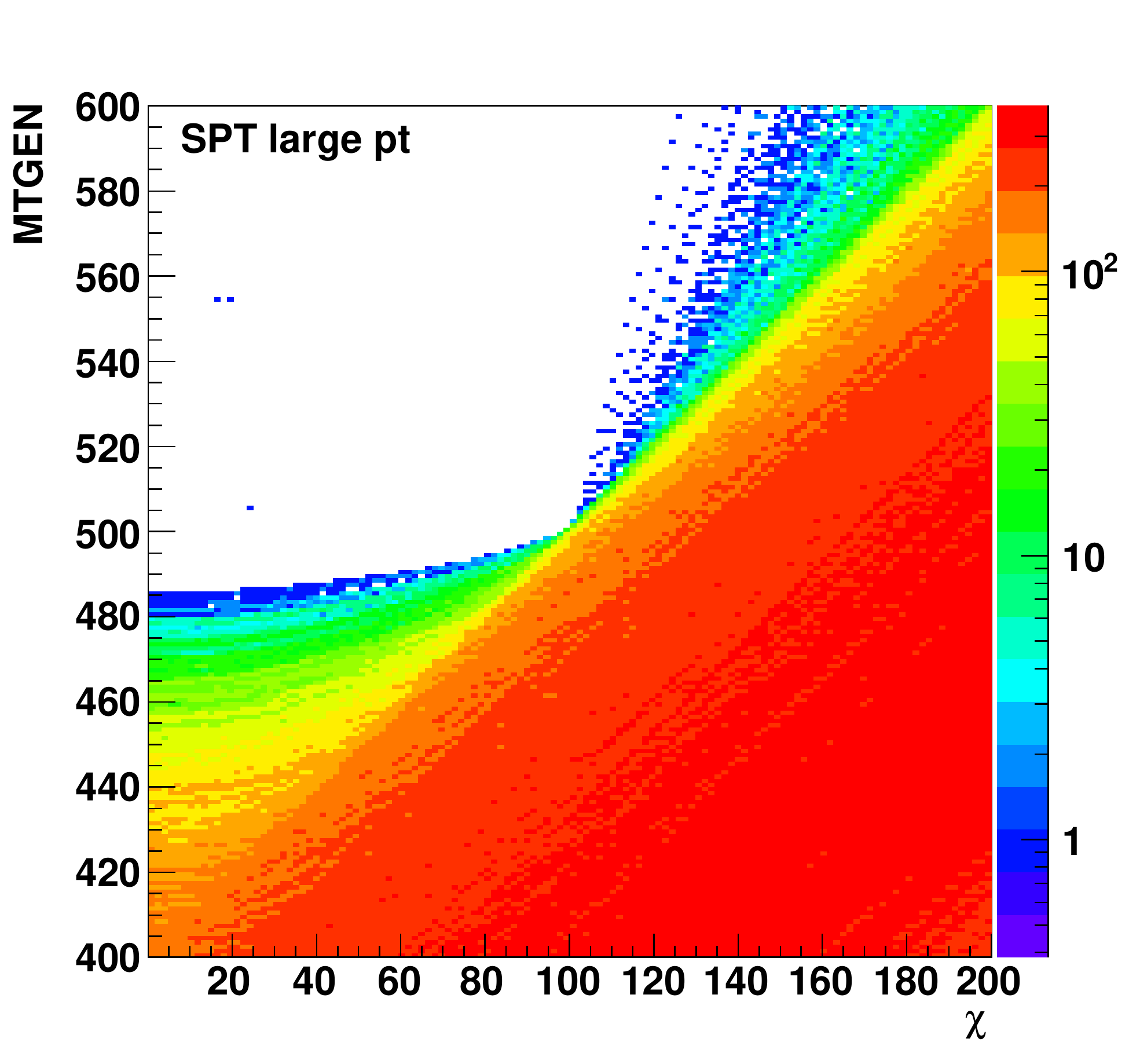}
        {\bf (d)}
     \end{center}
   \end{minipage}
\end{center}
\caption{ (\casesixv\ \MTGEN)
As for \figref{fig:six-v}, but for the kinematic variable 
\MTGEN, rather than \MTTWO. $m_{0,5} = 500$, $m_{1,6} = 100$ and $m_{2,3,7,8}=0$.
\label{fig:six-v-mtgen}}
\end{figure}

Note that the generation of \figref{fig:six-v} (and \figref{fig:six-s})
required the Monte Carlo truth record to be inspected in order that
the quarks could be combined together only in the {\em correct
pairings}.  In reality, the pedigree of each jet will be a
mystery, and so it is reasonable to wonder whether the kink structure
of \figref{fig:six-v} will remain in a plot where we are blinded
with respect to the ``truth''.  In \cite{Cho:2007qv} it was claimed
that in at least two examples a kink was seen survive the imposition of a set
of cuts designed to place jets heuristically into the appropriate
pairings.  Here, however, we note
an alternative approach which may be followed.  Rather than attempt to
assign quarks to the appropriate parings, one can instead use the
\MTGEN\ variable \cite{Lester:2007fq} in place of \MTTWO.  The \MTGEN\
variable does not need to be told which momenta come from which side
of the event -- it treats all visible momenta on an equal footing.
Internally, \MTGEN\ performs a minimisation over a (possibly large)
number of \MTTWO\ evaluations, with an evaluation for each 
distinct partition of the visible particles into two sets (one for each parent).  
It was shown in \cite{Lester:2007fq}
that the endpoint of \MTGEN\ shares the same properties as the
endpoint of \MTTWO\ at $\chi=m_1$.  In \figref{fig:six-v-mtgen} 
therefore, we look for a kink in the maximal values of \MTGEN\ as a
function of $\chi$, and indeed the kink is found to be present at both ZPT and SPT.  
Moreover, it is interesting to note that variation with $\chi$ of the \MTGEN\
maximum seen in \figref{fig:six-v-mtgen} is surprisingly similar to the
variation with $\chi$ of the \MTTWO\ maximum seen in
\figref{fig:six-v}.   This is remarkable as it suggests that the endpoint
of the \MTGEN\ variable appears to be able to capture almost the same information as the endpoint
of the \MTTWO\ variable, even though the former is always blind to the correct quark
assignments, and the latter expects to be given them.  Further
study is needed to determine which of these two variables is the
better to pursue.

\subsection{Pair production with three-daughter cascade decays
  (\casesixs) \label{sec:mc6s}}
\begin{figure}
   \begin{center}
    \begin{minipage}[b]{.32\linewidth}
      \begin{center}
         \includegraphics[width=0.99\linewidth]{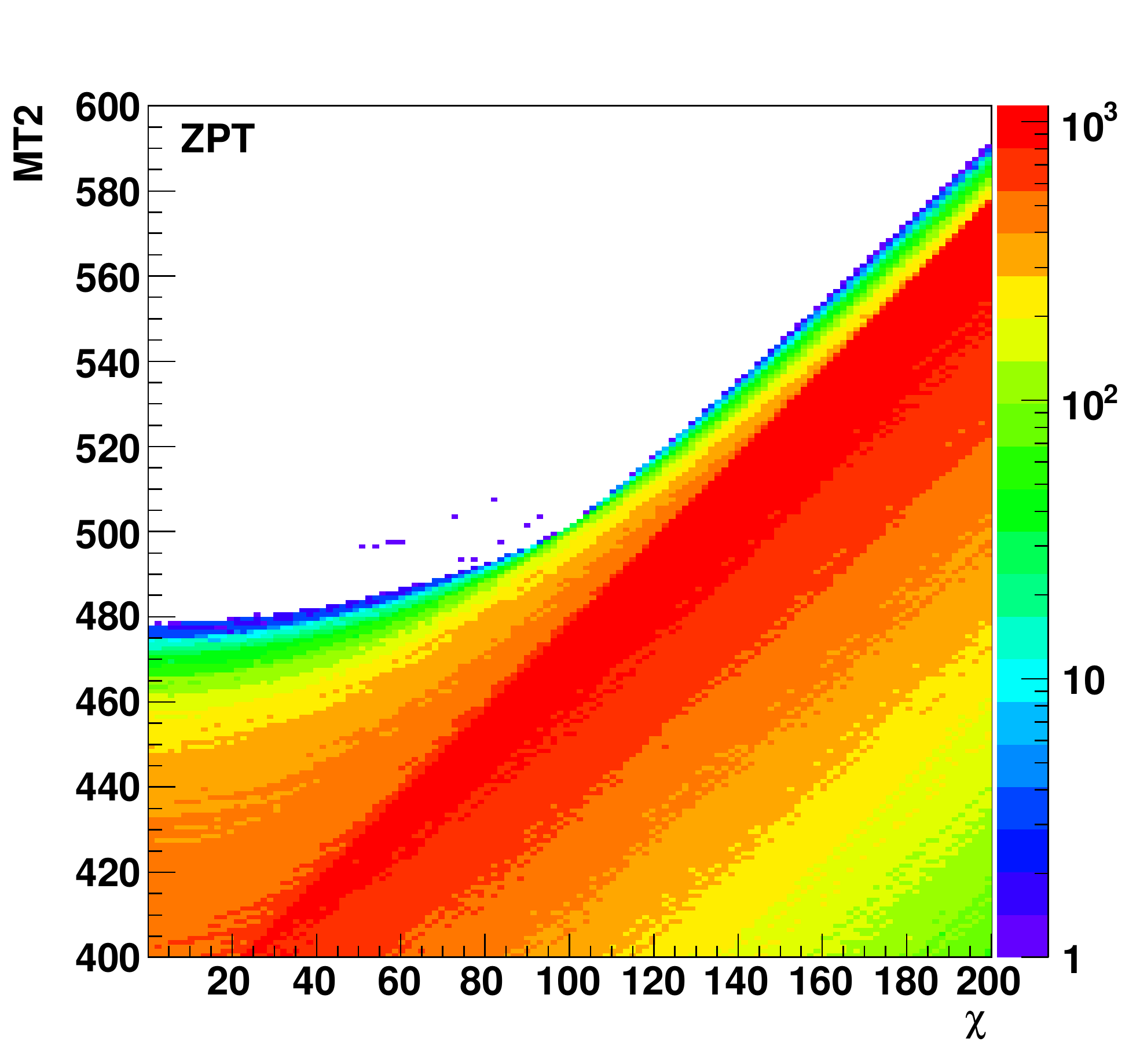}
 	{\bf (a)}
      \end{center}
    \end{minipage}
   \begin{minipage}[b]{.32\linewidth}
     \begin{center}
        \includegraphics[width=0.99\linewidth]{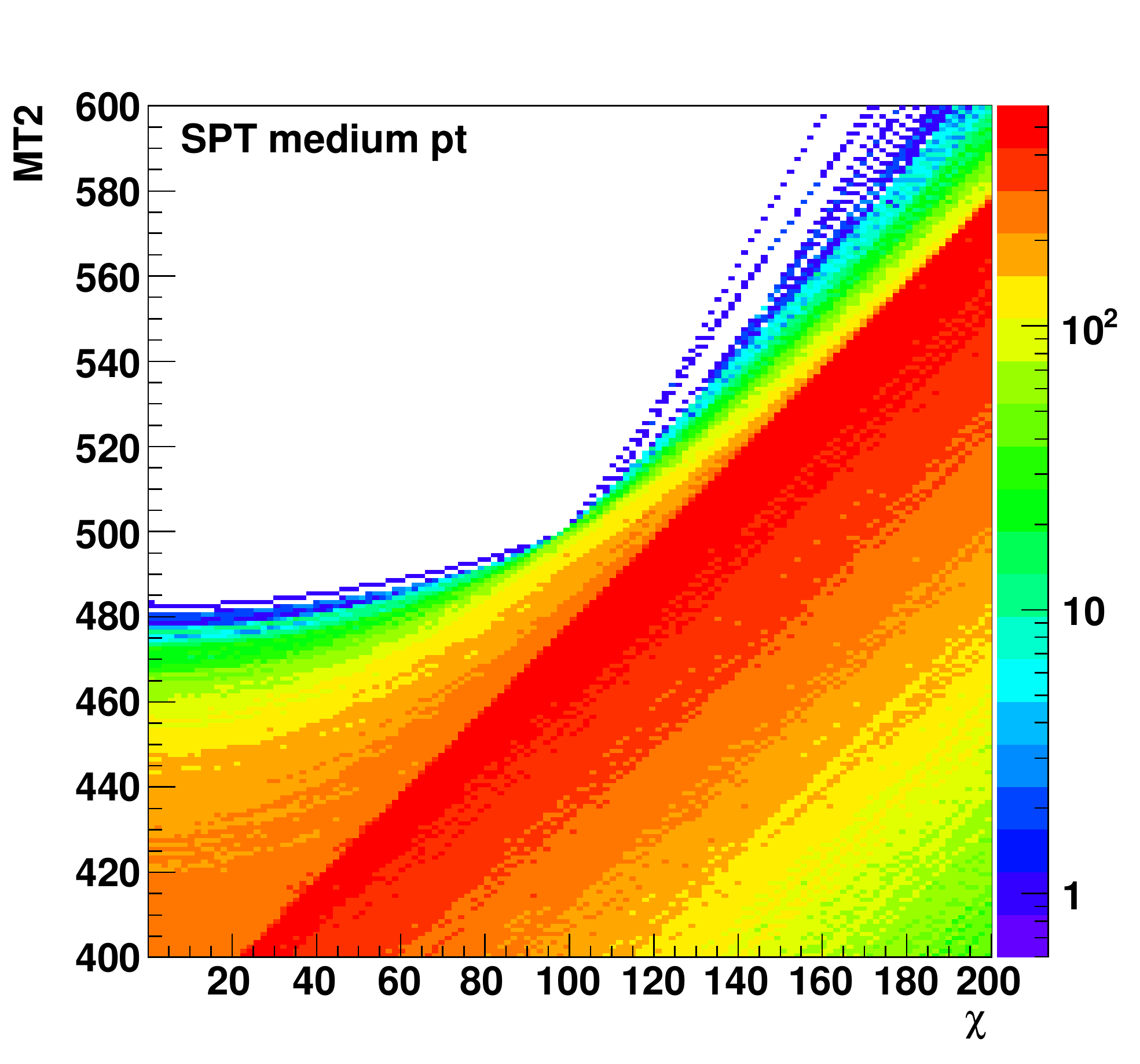}
        {\bf (b)}
     \end{center}
   \end{minipage}
   \begin{minipage}[b]{.32\linewidth}
     \begin{center}
        \includegraphics[width=0.99\linewidth]{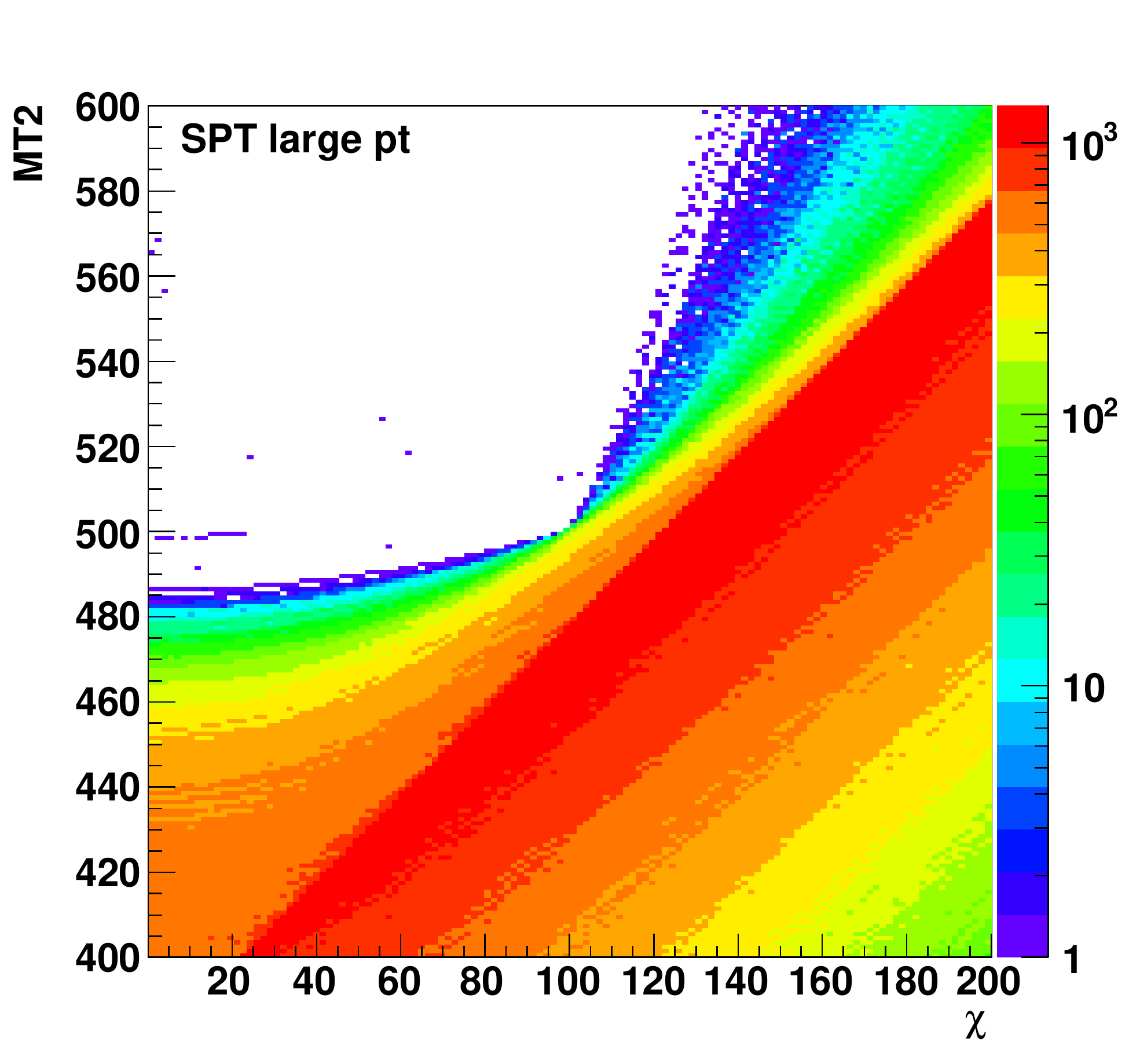}
        {\bf (c)}
     \end{center}
   \end{minipage}
\caption{(\casesixs) \label{fig:six-s}
As for \figref{fig:three-s} but plotting \MTTWO\ for pairs of decays.
$m_{0,5}=500, m_{1,6}=100, m_{2,3,7,8}=0, m_{4,9}=300$.
}
  \end{center}
\end{figure}

One may conduct the same distributions as in Section \ref{sec:mc6v}
but for pairs of cascade decays into three particles
(\casesixs) rather than for pairs of point-like three-body decays.
The results for \casesixs\ (\figref{fig:six-s}) are qualitatively the same as those for
\casesixv: a kink is seen in each of the SPT {\em and} the ZPT
distributions.  As before, the kink becomes more pronounced as the
amount of $p_T$ increases: ZPT $\rightarrow$ SPT (medium pt)
$\rightarrow$ SPT (large pt). 

\section{Conclusions}\label{sec:conclusions}
We have examined a wide range of cases in which heavy particles are produced
singly or in pairs at hadron colliders and decay to heavy invisible daughters 
(as exemplified by the diagrams in \figref{fig:general_idea} and \figref{fig:cases}).
We have examined how the endpoints of suitable kinematic variables
depend on the {\em a priori} unknown mass of the invisible daughter, paying particular attention
to the behaviour around its true mass.

We found that, unless the mass of the invisible particle is known exactly, 
the global maxima ({\em i.e} the true endpoints) for the general case (SPT) always correspond to events 
in which the transverse momenta of the particles are asymptotically-large compared to their masses. 
For $\chi<m_1$, the global maximum occurs when the invisible daughters are emitted in the same direction 
as the velocity vectors of their parents in the lab frame. For $\chi>m_1$, the global maximum occurs when the invisible daughters 
are emitted backwards relative to their parents' lab velocities.
We have shown that, because  asymptotically large energies are required, 
it is rather unlikely that significant numbers of maximal (or rather, near-maximal) 
events will be obtained in experiments at energy scales not far above the masses of the particles involved. 
But we have also demonstrated that non-maximal events still generate a kink.

We found that the maximum is different if the extra ZPT condition is applied.
The ZPT global maximum for $\chi>m_1$ occurs in events in which the invariant mass, 
$m_N$, of the visible $N$-daughter system takes its maximum value.
The global maximum for $\chi<m_1$ occurs when $m_N$ takes its minimum value. 
For \casetwo\ , the invariant mass is fixed, and so in the absence of ISR (i.e.\ in the ZPT case), 
this is the one case in which the kink does not appear even in principle.
In {\em all} other single-parent decays, the global maxima generate a kink both at SPT and ZPT.

We note that the ZPT kink has a different origin to the SPT kink. 
The former is related to a degree of freedom possessed by the invariant mass(es) 
of the set(s) of visible daughters of each decaying particle.
The latter is an artifact caused by invariant masses being evaluated in a frame 
which is {\em not} the parent's rest frame when the invisible particle's 
mass is hypothesized at incorrect values. This second effect
is only significant when the parent's velocity is comparable
to the speed of light.
\begin{table}
\begin{center}
\begin{tabular}{|c|c|c|c|}
\hline
Case  & Some PT (SPT)  & Zero PT (ZPT) &  Figures\\
\hline
2  & Kinky & Smooth & \ref{fig:two} \\
3v & Kinky & Kinky & \ref{fig:three-v} \\
3s & Kinky & Kinky & \ref{fig:three-s} \\
4  & Kinky & Smooth & \ref{fig:four} \\
6v & Kinky & Kinky & \ref{fig:six-v}, \ref{fig:six-v-mtgen} \\
6s & Kinky & Kinky & \ref{fig:six-s} \\
\hline
\end{tabular}
\caption{Summary of whether or not we expect (and observe) kinks in the different cases, 
(as shown in \figref{fig:general_idea} and \figref{fig:cases}).
``Kinky'' indicates cases in which we observe a discontinuity in the gradient of the 
maximum value of an appropriate kinematic transverse mass variable at $\chi=m_1$,
whereas ``Smooth'' indicates the lack of such a discontinuity.
\label{tab:summary_of_kinks} }
\end{center}
\end{table}
Now, it seems likely that the ISR in events of this type at the LHC is likely to be small, but non-vanishing  \cite{Plehn:2005cq}. 
But since the maxima obtained with vanishing ISR are not maxima (not even local maxima) with non-vanishing ISR, it is clear that the behaviour with small, but non-vanishing ISR is important.

To assess the implications of this, we considered the values of $m_T$ that result from a finite distribution of events in $\beta$ and $\cos \theta$, as defined in Section \ref{boost}.
In this formalism, ZPT events have $\beta=0$. As well as reproducing our previous results, and showing that the global maximum of the SPT case is unlikely to be reached, we showed that a kink still arises at $(m_0,m_1)$, even if we do not reach the global maximum.
In order to generate a kink, 
one simply needs events that 
have small relative rapidity, and are kinematically distinct {\em either} in their values of $m_N$, {\em or} in the values of $\beta \cos \theta$. So a kink will always be generated in a single-parent decay, unless all events have the same value of the invariant mass and the same value of $\beta \cos \theta$. Similar conclusions are reached in the case of identical pair decays. A full analysis of pair decays, including non-identical decays, remains to be done.

If the bulk of events are SPT (are
produced with enough transverse recoil), then kinks are ubiquitous,
occurring in every case shown in \figref{fig:cases}, and in any
topology that can be represented as \figref{fig:general_idea}(a) or
\figref{fig:general_idea}(b).  By contrast, if the bulk of events are
ZPT (produced with little or no transverse momentum), we found that
for \casetwo\ and \casefour, where the invariant mass is fixed, the maximal and minimal gradients are the same, and no kink is found.
These results are summarized for our exemplar cases in Table \ref{tab:summary_of_kinks}.

We conclude by making some remarks about the experimental viability of the method.
Firstly we recognise that backgrounds, both from Standard Model and competing new-physics processes
will need to be considered. 
Secondly we point out that, in real physics events, the finite widths of the unstable 
particles will have the effect of smearing the end-points.

When backgrounds, detector smearing and finite width effects are included, 
it will clearly become impossible to determine the position of the endpoint from individual maximal events. 
Unless the shape of the distribution near the endpoint is known then 
one must try to infer its position by looking for a region of rapidly decreasing numbers of events.
The presence of a significant fraction of events near the kinematic limit therefore
becomes much more important when such effects are considered.
Since we expect that cases in which kinks only exist {\em only} at SPT (i.e.\ \casetwo\ and \casefour) 
will have a small fraction of events potentially contributing to the kink, 
we have less confidence in these cases being experimentally accessible with this method compared
to cases which {\em also} have kinks at ZPT (i.e.\ all the other cases).

One might have reasonable confidence that, even after the inclusion 
of backgrounds, measurement imprecision, and finite widths, the method presented would still be very useful.
After all, as remarked in Section \ref{end-point}, the Tevatron $W$-mass measurements
demonstrate that, with adequate Monte Carlo modelling of these complicating physical effects,
endpoint techniques can be extremely useful at extracting mass information.

\begin{acknowledgments}
BMG thanks H.\ D.\ Kim, B.\ McElrath, R.\ Rattazzi, and G. G. Ross for
useful discussions.
\end{acknowledgments}

\bibliography{kink}
\end{document}